%% file: thesis-driver.tex
\documentstyle[12pt,mystyle]{report}
\input psfig

\begin{document}
\pagenumbering{roman}
\pagestyle{plain}

\tableofcontents
\newpage
\input ack_0.tex

\newpage
\addcontentsline{toc}{chapter}{List of Figures}
\listoffigures
\newpage
\addcontentsline{toc}{chapter}{List of Tables}
\listoftables
\newpage

\setcounter{page}{0}
\pagenumbering{arabic}
\pagestyle{myheadings}
\markright{}

\input intro1_qgp.tex
\newpage
\input intro2_rhic.tex

\newpage
\input intro3_signal.tex
\newpage
\input sec4-initial.tex
\newpage
\input sec5-pre.tex
\newpage
\input sec6-ratio.tex

\newpage
\input sec7-bg.tex

\newpage
\input outlook_8.tex

\newpage
\input appendix_9.tex
\newpage

\input ref_10.tex
\end{document}

%% file: ack_0.tex
\addcontentsline{toc}{chapter}{Acknowledgments}
\begin{center}
{\Large \bf Acknowledgments}
\end{center}

This thesis would not have been possible without the help from my advisor,
Professor Miklos Gyulassy.  I wish to give my deepest gratitude to him for his
patient guidance and heart-warming care.  I also wish to express my sincerest
respect for his hardworking spirit, which has always been inspiring to me.  

I would like to thank Professor Allan S. Blaer for his assistance to my
graduate life.  I would like to thank Mrs. Anne Billups and Irene Tramm for
their help when I was in the theoretical physics group.  I would also like to 
thank Mrs. Della Jean Dumlao and Mr. David Yan for their kindly assistances. 
I would like to thank especially Mrs. Sarah Goldstein and Stephen Vance for 
correcting the English of this thesis.

Working with colleagues in the nuclear theory group, Dirk H. Rischke, Masayuki
Asakawa, and Bin Zhang has been very harmonious. 
Last but not least, I thank all my fellow graduate friends, who have made my
five years at Columbia University very enjoyable.

%% file: intro1_qgp.tex
\hmychapter{Introduction}

The quark-gluon plasma phase of matter \cite{free,kislinger}, a state
consisting of weakly interacting deconfined quarks and gluons, was predicted
to exist soon after the discovery \cite{asym} of the asymptotic
freedom property of Quantum Chromodynamics (QCD). That novel
phase of matter was predicted to occur perhaps deep in the cores of neutron 
stars \cite{baym} and during the first few moments (microseconds) of 
the Big Bang \cite{applegate}.
It was also suggested that perhaps the quark-gluon plasma could be formed in
the laboratory in very high energy collisions of hadrons or nuclei \cite{qgp}. 
Since the mid 1970's the experimental search for new phases of nuclear matter
\cite{leewick,migdal} using heavy ion reactions has been underway. 
Until 1984 only light ion reactions were possible to study at energies below
$2$AGeV (GeV per incident baryon) \cite{nagamiyamg}.
In the past ten years, experiments with heavy ion with energies up
to 200 AGeV have been investigated \cite{qm88,qm95,qm96}. 
In 1999 a new era of experiments with heavy ion beams in the
collider mode at the Relativistic Heavy Ion Collider (RHIC) at 
the Brookhaven National Laboratory (BNL) with center of mass energies
up to $100$AGeV will begin. 

In this chapter, we will review key elements of QCD thermodynamics and the
dynamics of heavy ion reactions in order to put the thesis work described in
the later chapters into context. 
The work on open charm that forms this thesis is related to
the search for the QGP state of matter at RHIC energies. 
It deals with a probe of the earliest moments of heavy ion reactions and is
expected to be one of the key diagnostic tools of that plasma.
In Chapter~\ref{sec-qgp} we will review, from both the perturbative QCD and
lattice QCD calculations, the basic idea about the 
phase transition from hadronic matter to the quark-gluon plasma.  
In Chapter~\ref{sec-nucl} we will introduce nuclear collisions as a tool of
studying the quark-gluon plasma, and estimate thermalization and
chemical equilibration processes in heavy ion collisions.  
Then in Chapter~\ref{sec-signals} we will review several of the important
proposed observables including Heavy Quarkonia, dileptons, direct photons,  
disoriented chiral condensates (DCC), transverse flow, and Hanbury-Brown Twiss
(HBT).   
Finally in Chapter~\ref{sec-goal} we will give the introduction on open charm
signal, which is the main subject of this thesis. 

\hmysection{Quark-Gluon Plasma}
\label{sec-qgp}

One of the most remarkable properties of the Standard Model of strong
interactions, QCD, is that the fundamental color degrees of freedom
carried by quarks and gluons can not be directly observed. 
That property, called confinement, restricts physical observables to
composite color neutral objects called hadrons and nuclei. 
It is thought that this property is caused by the very complex,
non-perturbative nature of the physical vacuum.  That vacuum structure is also
thought to account for the observed breaking of the dynamical chiral symmetry
of QCD \cite{shuryak_report}. The naive perturbative QCD vacuum is not the
ground state, but has an energy density $B\approx 200$GeV/fm$^3$ above the 
physical vacuum that is filled with gluon and quark condensates.
The rich hadron spectroscopy and complex strong interaction phenomenology
is caused by that non-trivial vacuum structure. 
As T.D. Lee has repeatedly emphasized over the past 20 years \cite{vacuum},
understanding that structure is one of the fundamental problems of modern
physics. 

One of the few untested and key predictions of the Standard Model
is that the structure of the physical vacuum could drastically change
at high temperature and baryon density. 
Due to asymptotic freedom, as discussed below, the condensates could melt at
sufficiently high temperature, and thus hadronic or nuclear matter is expected
to transform into a weakly interacting gas of partons.

\hmysubsection{Asymptotic Freedom}

From perturbative QCD it is known that the coupling constant
depends on the momentum scale $Q$ \cite{asym}.  
For $SU(N)$ gauge theory with $n_f$ flavours of fermions, this is
expressed as the renormalization group equation:
\begin{eqnarray}
\frac {d \alpha_s}{dt} = b_0 \alpha_s^2 + b_1 \alpha_s^3 + O(\alpha_s^4) \;\; ,
\end{eqnarray}
where $t \equiv \ln (Q^2/\Lambda^2)$, and
\begin{eqnarray}
b_0 &=& \frac {\beta_0}{4\pi} 
= \frac {1}{4\pi} \left ( \frac {11N}{3} - \frac {2n_f}{3} \right ) \;\; ,
\\[2ex] 
b_1 &=& \frac {\beta_1}{16\pi^2} 
= \frac {1}{16\pi^2} \left [ \frac {34N^2}{3} 
- \left ( \frac {10N}{3}+\frac {N^2-1}{N} \right )n_f \right ] \;\; .
\end{eqnarray}
For example, for QCD $SU(3)$ theory the lowest order expression of the coupling
constant is 
\begin{eqnarray}
\alpha^{LO}_s (Q^2) = \frac {4\pi} {\left ( 11-\frac {2n_f}{3} \right )
\ln \left ( \frac {Q^2}{\Lambda^2} \right )} \;\; ,
\end{eqnarray}
where $Q$ is the momentum transfer and $\Lambda \sim 200$MeV.

Therefore QCD has asymptotic freedom in that for $n_f \leq 6$ the running
coupling constant approaches zero at the limit of $Q \rightarrow \infty$.
This raises the possibility that we will have a free gas of quark
and gluon plasma when the nuclear matter is at high temperature or at high
density \cite{free}.  

\hmysubsection{Deconfinement and Chiral Restoration}

From the asymptotic freedom of QCD one expects to have an ideal gas of quarks
and gluons at high temperature. 
However at low temperature quarks and gluons are subject to color confinement,
and chiral symmetry is spontaneously broken.
Therefore at some high temperature we expect to have deconfinement and chiral
restoration.

The QCD Lagrangian can be written as
\begin{eqnarray}
{\cal L}=\frac {1}{4} F^a_{\mu \nu} F^a_{\mu \nu} + \sum_i \bar {\psi_i} 
\left ( D+m_i \right ) \psi_i \;\; ,
\label{EQ:qcdl}
\end{eqnarray}
where following quantities expressed in terms of the gauge field $A^a_\mu$ are:
\begin{eqnarray}
F^a_{\mu \nu}&=&\partial_\mu A^a_{\nu} - \partial_\nu A^a_{\mu} + gf^{abc}
A^b_{\mu} A^c_{\nu} \;\; , \\[2ex]
D&=&\gamma_\mu \left ( i \partial_\mu+\frac {1}{2} g A^a_\mu \lambda^a \right )
\;\; , \\[2ex]
A_\mu &\equiv& \frac {1}{2} A^a_\mu \lambda^a \;\; ,
\end{eqnarray}
where $\lambda^a$ are the generators of the $SU(N)$ group in the
fundamental representation.  The $\lambda^a$ satisfy the following relations
with $f^{abc}$ being the antisymmetric $SU(N)$ structure constants:
\begin{eqnarray}
\left [ \lambda^a, \lambda^b \right ]&=&2if^{abc} \lambda^c \;\; , \\[2ex] 
Tr \left (\lambda^a \lambda^b \right )&=&2 \delta^{ab} \;\; .
\end{eqnarray}

The above Lagrangian in eq.(~\ref{EQ:qcdl}) has exact symmetry only when the
quark masses are infinity or zero.  In the former case we have the pure gauge
theory with $Z(N)$ symmetry, and in the latter case we have chiral symmetry. 

When the quarks are infinitely heavy, we can study the deconfinement phase
transition from the breaking of the global $Z(N)$ symmetry in the pure gauge
theory.  The order parameter is found to be the Polyakov loop operator
\begin{eqnarray}
L(\vec x)=Tr {\cal P} \exp \left ( i \int_0^{1/T} dt A_0 (\vec x,t) \right )
\;\; , 
\end{eqnarray}
which takes the trace of the path integral of an exponential function.

The thermal expectation value of the Polyakov loop operator is related to the
free energy of inserting a static quark in the vacuum:
\begin{eqnarray}
<L(\vec x)> = \exp \left ( -{ \frac {F_q-F_0}{T}} \right ) \;\; ,
\end{eqnarray}
where $F_0$ stands for the free energy of the vacuum.

In the confining phase the field can not be screened in the pure gauge
theory, and the string potential grows linearly with distance, 
it therefore takes infinite free energy to insert a quark and $<L(\vec x)>=0$. 
However in the deconfining phase the field can be Debye screened and this
results in $<L(\vec x)> \neq 0$.  Thus $<L(\vec x)>$ is the order
parameter for the deconfinement phase transition
and provides information on the 
breaking of $Z(N)$ symmetry in pure gauge $SU(N)$ theory.

It was pointed out by Svetitsky and Yaffe \cite{zn} that the four dimensional
$SU(N)$ pure gauge theory at finite temperature is in the same universality
class as the three dimensional $Z(N)$ spin theory.  Numerical studies on the
effective models give strong evidence that $SU(3)$ pure gauge theory goes
through first order deconfinement phase transition \cite{deconf}.  

The situation with dynamical quarks is quite different.  The $Z(N)$ global
symmetry is explicitly broken by the finite quark mass.  In general it is
expected that with dynamical quarks the confinement phase transition weakens,
since the field can be screened by $q\bar q$ pair productions and the strings
can be broken even in the confining phase \cite{deconf_q}.

On the other hand, when the quarks are massless, the QCD Lagrangian is
invariant under the following transformations on the fermion fields $\psi_i$:
\begin{eqnarray}
\psi_i &\rightarrow& e^{i\alpha} \psi_i \;\; , \\[2ex]
\psi_i &\rightarrow& e^{i\alpha \gamma_5} \psi_i \;\; ,\\[2ex]
\psi^R_i &\rightarrow& U^R \psi^R_i \;\; , \\[2ex]
\psi^L_i &\rightarrow& U^L \psi^L_i \;\; ,
\end{eqnarray}
where $\gamma_5$ is the Dirac matrix, and
\begin{eqnarray} 
U^R, U^L &\in& SU(n_f)  \;\; , \\[2ex]
\psi^R_i &=& \frac {1+\gamma_5}{2} \psi_i \;\; , \\[2ex]
\psi^L_i &=& \frac {1-\gamma_5}{2} \psi_i \;\; .
\end{eqnarray}
Hence the Lagrangian has the following chiral symmetry
\begin{eqnarray}
U_V(1) \times U_A(1) \times SU_R(n_f) \times SU_L(n_f) \;\; .
\end{eqnarray}
However, $U_A(1)$ symmetry is broken by quantum effects, and the corresponding
axial current \cite{thooft}
\begin{eqnarray}
j^5_{\mu} = \sum_i \bar {\psi_i} \gamma_{\mu} \gamma_5 \psi_i
\end{eqnarray}
is not conserved.  Instead we have the Adler-Bell-Jackiw anomaly \cite{anomaly}
as
\begin{eqnarray}
\partial^{\mu} j^5_{\mu} = \frac {n_f g^2}{32\pi^2} 
Tr \left (\epsilon^{\mu \nu \lambda \sigma} F_{\mu \nu} F_{\lambda \sigma}
\right ) \;\;.
\end{eqnarray}
As a result the chiral symmetry is \cite{zanf}
\begin{eqnarray}
Z_A(n_f) \times SU_R(n_f) \times SU_L(n_f)\;\;.
\end{eqnarray}
Pisarski and Wilczek \cite{chiral1} studied the following Lagrangian with the
above chiral symmetry
\begin{eqnarray}
{\cal L}_{ch}&=& 
\frac {1}{2} Tr \left ( \partial_{\mu} \Phi^\dagger \partial^{\mu}\Phi \right )
-\frac {1}{2} m_{\Phi}^2 Tr \left ( \Phi^\dagger \Phi \right ) 
+ c \left ( Det \Phi + Det \Phi^\dagger \right ) \nonumber \\
&& -\frac {\pi^2}{3} g_1 \left [ Tr \left ( \Phi^\dagger \Phi
\right ) \right ]^2 
-\frac {\pi^2}{3} g_2 Tr \left [ \left ( \Phi^\dagger \Phi \right )^2
\right ] \;\; ,
\end{eqnarray}
where the fields are
\begin{eqnarray}
\Phi_{i,j} \propto <\bar {\psi_i} \left ( 1+\gamma_5 \right ) \psi_j> \;\; ,
\end{eqnarray}
and where $g_1,g_2$ are related to $\beta_1,\beta_2$ in the renormalization
group. 

The order of the chiral phase transition from the above general Lagrangian is
found to depend strongly on $n_f$.  The transition is first order for $n_f \geq
3$.  For $n_f=1$ the transition does not even exist.  For the interesting
$n_f=2$ case, the transition depends sensitively on the parameter $c$, the
magnitude of the mass term.  The phase transition is first order if $c=0$,
weakens when $c$ becomes bigger, and even disappears if $c$ is too large.  The
exact nature of the phase transition has also been studied with mean field
analysis and lattice Monte Carlo calculations. 

\hmysubsection{QGP Perturbative Equation of State}

For QCD with massless quarks, the thermodynamic potential was calculated as 
a function of temperature
\begin{eqnarray}
\Omega &\equiv& \frac {-1}{VT^3} \ln Z \nonumber \\
&=& -\left [ a_0 + a_2 g^2 +a_3 g^3 + \left (a_{4l} \ln g^2 + a_4 \right ) g^4
\right . \nonumber \\ 
&& \left . + \left (a_{5l} \ln g^2 + a_5 \right ) g^5 \right ] + O(g^6) \;\; ,
\label{EQ:partition}
\end{eqnarray}
where $Z$ is the partition function.  The first four coefficients in the above
equation are derived in terms of number of colors and flavours \cite{omega} 
\begin{eqnarray}
a_0&=&\frac {\pi^2}{45} \left ( N^2-1 +\frac {7}{4} N n_f \right ) \;\; , \\[2ex]
a_2&=&\frac {-1}{144} \left ( N^2-1 \right ) \left ( N+ \frac {5}{4}n_f \right
) \;\; , \\[2ex]
a_3&=&\frac {1}{12\pi} \left ( N^2-1 \right ) \left ( \frac {N}{3}+ \frac
{n_f}{6} \right )^{3/2} \;\; , \\[2ex] 
a_{4l}&=&\frac {1}{32\pi^2} N \left ( N^2-1 \right ) \left ( \frac {N}{3}+
\frac {n_f}{6} \right ) \;\; .
\end{eqnarray}

Then other thermodynamic quantities such as energy density, pressure and
entropy density are expressed in terms of the thermodynamic potential $\Omega$:
\begin{eqnarray}
p&=& T \partial_V \ln Z = -T^4 \Omega \;\; , \\[2ex]
\epsilon &=& \frac {T^2}{V} \partial_T  \ln Z = 3p-T^5 \frac {d\Omega}{dT}
\;\; , \\[2ex] 
s &=& \frac {\epsilon +  p}{T} = \frac {4p}{T} - T^4 \frac {d\Omega}{dT} \;\; .
\end{eqnarray}
We expect there to be a free gas of quarks and gluons at the weak coupling
limit $g \rightarrow 0$, so the 
leading term $a_0$ in eq.~(\ref{EQ:partition}) just gives the Stefan-Boltzmann
quantities for massless bosons and fermions:
\begin{eqnarray}
p&=& a_0 T^4 = \frac {\pi^2}{90} \left [ 2(N^2-1) +\frac {7}{8} N \cdot 2 \cdot
2 n_f \right ] \;\; , \\[2ex]
\epsilon &=& 3a_0 T^4 = 3p \label{EQ:e_sb} \;\; , \\[2ex]
s &=& 4a_0 T^3 \;\; .
\end{eqnarray}

When one considers finite baryon density, one introduces the quark chemical
potential $\mu$, and then the equation of state (EOS) is given by \cite{eos_mu}
\begin{eqnarray}
p&=& \frac {\pi^2}{90} \left [ 2(N^2-1) +\frac {7}{8} N \cdot 2 \cdot
2 n_f \right ] + \frac {N n_f}{6} \left ( T^2 \mu^2 + \frac {\mu^4}{2\pi^2}
\right ) \label{EQ:pmu_sb} \;\; , \\[2ex]
\epsilon &=& 3p \;\; , \\[2ex]
n &=& \frac {N n_f}{3\pi^2} \left ( \mu^3 + \pi^2 T^2 \mu \right ) \;\; , 
\\[2ex]
s &=& \frac {\epsilon + p - \mu n}{T} \;\; , 
\end{eqnarray}
where $n$ is quark number density.  We mostly
concentrate our discussions on the physics for zero baryon density case, since
small baryon density is expected for very high energy heavy ion collisions.

Perturbative methods provide us with a powerful tool to study QCD \cite{pqcd}.
However, the result is an expansion in the coupling $g$, which is of the order
$1$ at any practical energies despite the fact the it is a decreasing function
of energy.  Therefore the higher order terms in eq.~(\ref{EQ:partition}) are 
comparable with the lower terms.
Furthermore the expansion often encounters technical difficulties such as
the summation of infinite terms with the same order.  

We take eq.(~\ref{EQ:partition}) as an example.  In order to eliminate the
infra-red divergence at $O(g^8)$, one has to introduce a magnetic mass in the
gluon propagator \cite{m_m}
\begin{eqnarray}
m_M \propto g^2 T \;\; .
\end{eqnarray}
Then there are infinite number of diagrams contributing at $O(g^6)$, since a
diagram with $n$ number of four gluon vertices gives a term which is
proportional to
\begin{eqnarray}
g^6 \left (\frac {g^2 T}{m_M} \right )^{n-3} \;\; .
\end{eqnarray}

Although eq.(~\ref{EQ:partition}) is a perturbative result, it includes
infinite resummation of ring diagrams that take into account the polarization
and screening effects of the plasma.  
Similar problems occur at all higher orders, and it seems impossible to solve
the problem, which is called ``Linde's disease'', analytically via
perturbative expansion.
Thus even when $g$ is small, it seems that we still can not solve the problem
analytically.  
Non-perturbative calculations starting from first principle are therefore
urgently needed.

\hmysubsection{QGP Non-perturbative Equation of State}

In order to overcome the difficulties encountered in perturbative QCD
calculations, lattice QCD was proposed by Wilson in 1974 \cite{wilson}.
Lattice QCD regularizes the QCD Lagrangian by fixing the lattice
spacing.  Then thermodynamic properties of the QCD 
can be studied by Monte Carlo simulation \cite{mc}.  First one needs to
discretize the QCD action  
\begin{eqnarray}
S=\int_0^{1/T} dx_0 \int_V {\cal L} \;\; , 
\end{eqnarray}
where the QCD Lagrangian ${\cal L}$ is given by eq.(~\ref{EQ:qcdl}).

Wilson proposed a way to map the gluon part of the action to lattice
\cite{wilson}.  First a link variable is defined to be
\begin{eqnarray} 
U_{n,\mu} = \exp \left [ -\frac {iga}{2} \sum_b \lambda^b
A_\mu^b \left (x=n\cdot a \right ) \right ] \;\; , 
\end{eqnarray}
where $a$ is the lattice spacing, $n^\lambda$ is a point in the
four-dimensional Euclidean lattice, and $\mu$ labels the direction of the link.
Then the gluon part of the action is
\begin{eqnarray}
S_G = \beta \sum_{n;\mu<\nu} P_{n,\mu\nu} 
= \beta \sum_{n;\mu<\nu} \left [ 1- \frac {1}{N} 
Re Tr \left ( U_{n,\mu} U_{n+\mu,\nu} U^\dagger_{n+\nu,\mu} U^\dagger_{n,\nu}
\right ) \right ] \;\; , 
\label{EQ:sg}
\end{eqnarray}
where $Re$ takes the real part of a function, and
\begin{eqnarray}
\beta = \frac {2N}{g^2} \;\; .
\end{eqnarray}

The above gluon action is equivalent to that in the QCD Lagrangian
eq.(~\ref{EQ:qcdl}) at the continuum limit $a \rightarrow 0$.  
We expand the trace term in eq.(~\ref{EQ:sg}) to $O(a^3)$:
\begin{eqnarray}
U_{n,\mu} U_{n+\mu,\nu}&=&e^{-igaA_\mu} e^{-igaA_\nu(x+a_\mu)}
\simeq e^{-igaA_\mu} e^{-iga(A_\nu+a\partial_\mu A_\nu)} \nonumber \\
&\simeq&e^{-iga(A_\mu+A_\nu)-iga^2\partial_\mu A_\nu
+(iga)^2 [A_\mu,A_\nu]/2} \;\; , \\[2ex]
U^\dagger_{n+\nu,\mu} U^\dagger_{n,\nu}
&\simeq&e^{iga(A_\mu+A_\nu)+iga^2\partial_\nu A_\mu
+(iga)^2 [A_\mu,A_\nu]/2} \;\; , \\[2ex]
\Rightarrow 
U_{n,\mu} U_{n+\mu,\nu} U^\dagger_{n+\nu,\mu} U^\dagger_{n,\nu}&\simeq&
e^{-iga^2 \left (\partial_\mu A_\nu - \partial_\nu A_\mu -ig [A_\mu,A_\nu]
\right ) + O(a^3)} \nonumber \\
&=&e^{-iga^2 F^a_{\mu\nu} \lambda^a/2 + O(a^3)} \;\; , 
\end{eqnarray}
\begin{eqnarray}
\Rightarrow P_{n,\mu\nu} &=&1-
\frac {1}{N} ReTr \left ( e^{-iga^2 F^a_{\mu\nu} \lambda^a/2 + O(a^3)} \right )
\nonumber \\
&=&\frac {g^2 a^4}{4N} F^a_{\mu\nu} F^a_{\mu\nu} + O(a^6) \;\; , \\[2ex]
\Rightarrow
S_G &=& \frac {2N}{g^2} \sum_{n;\mu<\nu} P_{n,\mu\nu} = \frac {1}{4} V_4
F^a_{\mu\nu} F^a_{\mu\nu} + O(a^6) \;\; , 
\end{eqnarray}
where
\begin{eqnarray}
V_4 \equiv (N_t a) (N_s a)^3 =\frac {1}{T} L^3
\end{eqnarray}
is the volume of the four dimensional lattice.

The lattice regularization of fermion action turns out to have fundamental
difficulties.  The naive regularization produces 16 poles in the fermion
propagator, thus results in 16 fermions from the every one original fermion.
This fermion doubling problem can be solved in at least two ways.
Wilson \cite{wilson} 
proposed to add a large mass term to eliminate all the other 15 unwanted
fermions.  
However, the mass term explicitly breaks the chiral symmetry, which
is very important for the study of QCD phase transitions.  Kogut and
Susskind \cite{staggered} proposed a way to reduce the 16 fermions to 4
(staggered fermion formulation), and 
this method has the advantage that part of the chiral symmetry still remains.
Wilson's fermion method and staggered fermion method will agree and both
produce the continuum fermionic Lagrangian in the limit $a \rightarrow 0$.

In the staggered fermion formulation, the fermion part of the action is given
in terms of the fermion fields $\chi (\bar \chi)$, which are anticommuting
Grassmann variables
\begin{eqnarray}
S_F=\sum_i {\bar \chi}_{n,i} Q^i_{n,m} \chi_{m,i} \;\; .
\end{eqnarray}
Here the fermion matrix $Q$ includes the usual derivative term and the mass
term. 

Once one fixes the formulation of the fermion action $S_F$, the partition
function on lattice becomes
\begin{eqnarray}
Z=\int \prod_{n,\mu} dU_{n,\mu} \prod_{m,i} d\chi_{m,i} d{\bar \chi}_{m,i}
e^{-S_G-S_F} \;\; , 
\end{eqnarray}
where $\chi_{m,i}$ are the fermion fields on lattice.  
Then the expectation value of an operator $\hat O$ is
\begin{eqnarray}
<\hat O>=Z^{-1} \int \prod_{n,\mu} dU_{n,\mu} \prod_{m,i} d\chi_{m,i} d{\bar
\chi}_{m,i} \hat O e^{-S_G-S_F} \;\; .
\end{eqnarray}
Since the fermion action $S_F$ is bilinear in the fermion fields, it can be
integrated out and only the gluonic degrees of freedom remain:
\begin{eqnarray}
<\hat O>=Z^{-1} \int \prod_{n,\mu} dU_{n,\mu} \hat O 
\prod_i \det Q^i e^{-S_G} \;\; .
\end{eqnarray}

\begin{figure}[p]
\psfig{figure=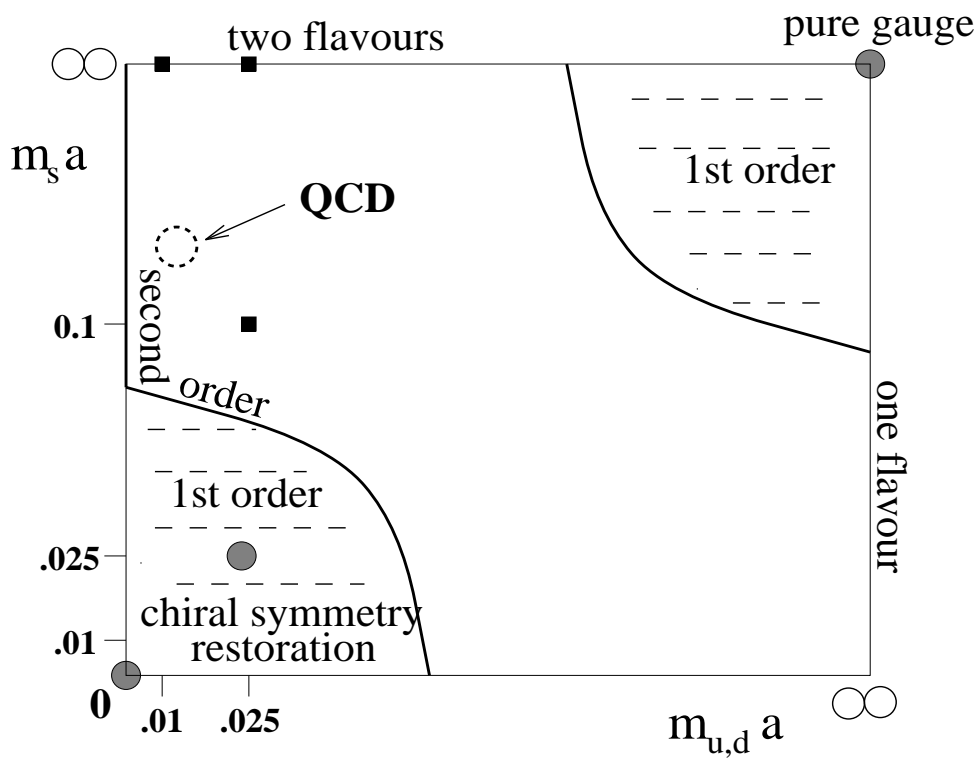,height=4.in,width=5.in,angle=-90} 
\caption{Phase transition from lattice QCD at finite temperature as a function
of $m_{u,d}a$ and $m_sa$ from ref.~\protect{\cite{prl65}}.
}
\label{fig:phase_diagram}
\end{figure}

The lattice QCD results on phase transitions are summarized in
Figure~\ref{fig:phase_diagram} from ref.\cite{prl65}.  
The order of the phase transition is plotted as a function of the light quark
mass $m_{u,d}$ and the strange quark mass $m_s$ measured in
units of $a^{-1}$.  We proceed anti-clockwise from the upper-right corner.
In pure gauge theory where all quarks are infinitely heavy, there is a first
order phase deconfinement transition \cite{prl67}.  With two flavours of quarks
with small but finite masses, one has the two solid boxes on the upper border
of Figure~\ref{fig:phase_diagram} and there is no phase transition
\cite{prl65}, but there is a rapid crossover \cite{crossover}.
For two massless quarks there is a second order transition.  When we have three
(or more) massless quarks, the transition is known to be first order; and it
remains first order even for the case of three light quarks \cite{prl65}, which
is represented by the lower solid circle.
For QCD with one flavour of massless quark, there is no phase transition.
In lattice QCD physical quantities are expressed in terms of the lattice
spacing $a$.  To have a feeling of the magnitude of $a$, $m_\pi a \sim 0.4$ in
calculations of ref. \cite{prl65}.  Therefore the real world where one has two
nearly massless quarks and one heavier strange quark lies in the area
represented by the virtual circle in Figure\ref{fig:phase_diagram}.  To probe
near the real QCD, the lattice calculation with $m_{u,d}a=0.025, m_s a=0.1$
(solid box in the middle of the figure) was performed but showed no phase
transition.
However, the current lattice calculations have 
uncertainties due to finite size effects.  
For example, in ref. \cite{prl65} $m_K/m_\rho=0.46$ instead of the physical
value of 0.64.  
Much more work needs to be done to determine the exact QCD phase diagram.

\begin{figure}[p]
\psfig{figure=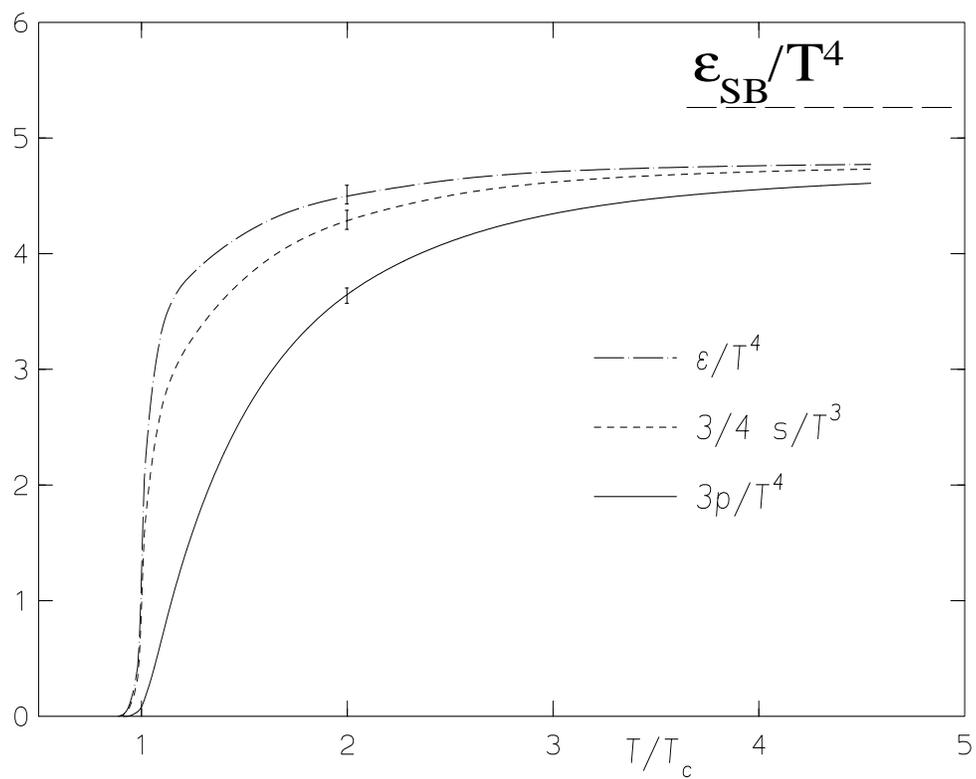,height=4.in,width=5.in,angle=180} 
\caption{Energy density, entropy density and pressure for $SU(3)$ gauge theory
as a function of $T/T_c$ from ref.~\protect{\cite{eos_su3}}.
}
\label{fig:eos_su3}
\end{figure}

The lattice QCD results on the equation of state of $SU(3)$ pure gauge theory
is shown in Figure~\ref{fig:eos_su3} \cite{eos_su3}.  The upper dashed straight
line is the Stefan-Boltzmann value for the energy density according to
eq.~(\ref{EQ:e_sb}), and the other plotted physical quantities are the values
extrapolated to the continuum limit.
One notices that the energy density quickly approaches the ideal gas value
after $T_c$, however the pressure approaches slowly and thus gives the biggest
deviation from the ideal gas equation of state, $\epsilon=3p$, 
at $T \sim (1-2)T_c$. 
From the perturbative QCD result given by eq.~(\ref{EQ:partition}), the
deviation from the ideal gas equation of state is
\begin{eqnarray}
\frac {\epsilon-3p}{T^4} = -T \frac {d\Omega}{dT} 
\simeq c_1 gT\frac {dg}{dT} = c_2 g^4 \;\; , 
\label{EQ:pqcd_eos}
\end{eqnarray}
where
\begin{eqnarray} 
c_1=-\frac {N(N^2-1)}{72},\;\;\;c_2 > 0 \;\; . 
\end{eqnarray}
The deviation can also be connected to the gluon condensate
\begin{eqnarray}
\epsilon-3p = <G^2>_0 - <G^2>_T \;\; , 
\end{eqnarray}
where
\begin{eqnarray}
<G^2>_0 \sim 2\;{\rm GeV/fm^3} \;\; . 
\end{eqnarray}
The zero temperature gluon condensate is important for the deviation of
the quantity $(\epsilon-3p)/T^4$ when close to $T_c$, and its effect decreases
rapidly at high 
temperature.  The deviation at high temperature $T/T_c \geq 2$ could be
attributed to the leading order perturbative result in eq.~(\ref{EQ:pqcd_eos})
assuming a large running coupling constant $g^2(T) \sim 2$. 
However the perturbative expansion can not be expected to provide us a
quantitative result in this case.

As seen above, the lattice calculation gives us valuable insight into
the non-perturbative aspects of QCD.  
The critical temperature for the QCD phase transition appears to be about
$150$MeV \cite{crossover}, which means a critical energy density of about
$1$GeV/fm$^3$ according to eq.~(\ref{EQ:e_sb}) for three flavour QGP.
However, one needs higher temperature (and thus a much higher energy
density) for the onset of the perturbative behaviour, as can be seen from
Figure~\ref{fig:eos_su3}.

\begin{figure}[p]
\hspace{1.in}
\psfig{figure=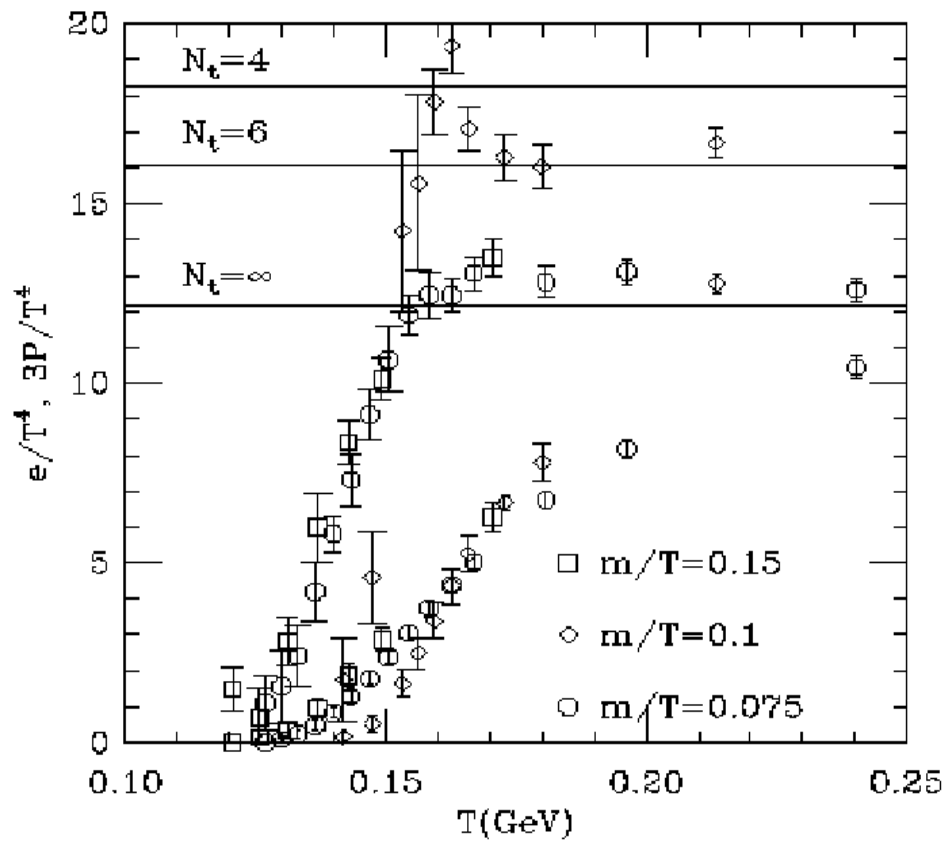,height=4.5in,width=5.in,angle=0} 
\caption{
Equation of state for two flavour QCD on $N_t=4$ and $N_t=6$ lattices as a
function of $T$ from ref.~\protect{\cite{finite_size}}. 
}
\label{fig:finite_size}
\end{figure}

Nevertheless, one has to be careful when drawing conclusions from lattice QCD
results before one understands uncertainties such as finite size effects. 
To give an example, we show the results of the equation of state for two
flavour QCD.  
In Figure\ref{fig:finite_size} \cite{finite_size} energy density and pressure
are plotted for two flavour QCD.  The diamonds are the results on $N_t=4$
lattices, while the circles and squares are those on $N_t=6$ lattices.  The
bottom straight line indicates the Stefan-Boltzmann value for energy density,
and the other two straight lines indicate the Stefan-Boltzmann values corrected
for finite size effects.  

By comparing $N_t=4$ results with $N_t=6$ results in
Figure~\ref{fig:finite_size} one finds large finite size effects.  This is not
unexpected since one has significant finite size effect in the free theory on
lattices already.  For example, the finite temporal size effect for the energy
density of an ideal gluon gas can be expressed approximately by
\cite{nt_correction} 
\begin{eqnarray}
\frac {\epsilon}{T^4} = \frac {3p}{T^4}
\simeq (N^2-1) \frac {\pi^2}{15} \left (1+\frac {30}{63} \frac {\pi^2}{N_t^2} +
\frac {1}{3} \frac {\pi^4}{N_t^4} \right ) \;\; .
\end{eqnarray}
The finite temporal size effect can therefore be as large as $50\%$ on $N_t=4$
lattices for pure gauge theory.
Improved actions on both gluon \cite{improved} and fermion part could help to
limit the sensitivity to the finite size.

In order to improve the results from lattice QCD, larger machines are
definitely needed.  At present a 0.4 Teraflops supercomputer is under
construction by N. Christ et al. at Columbia University \cite{teraflop}.
That machine will be optimized for full QCD calculations, and will be one
of the most promising lattice QCD projects to approach the continuum limit.

%% file: intro2_rhic.tex
\hmysection{Why Do We Need Heavy Ion Collisions?}
\label{sec-nucl}

In order to experimentally study the phase transition of the quark-gluon
plasma, we need to achieve very high energy density in the lab.  It is also
important to have a large interaction volume in order to approach the
thermodynamic limit of the new phase.  
It is hoped that high energy heavy ion collisions can provide us with these
necessary conditions.  Since the initial interaction volume and parton
multiplicity turn out to be large, the dense parton system may go through both
thermalization and chemical equilibration processes after its formation.  
At extremely high energies one may study an almost fully-equilibrated plasma of
quarks and gluons with zero baryon density, and at currently accessible
energies matter with high baryon density may be studies as well.  
The QCD phase transition is also expected to occur at high baryon density 
\cite{kislinger}.

\hmysubsection{Glauber Geometry}

In nuclear collisions many $pp$ collisions occur almost simultaneously in a
small nuclear volume.   For hard perturbative QCD processes with small cross
sections, the nuclear geometry and the QCD dynamics factorize to lowest order.
Then it is assumed that they are independent two body reactions.
For $A+B$ collisions at a fixed impact parameter $\vec b$, 
the number of binary collisions is given by \cite{eskola3}  
\begin{eqnarray}
N^{AB} (\vec b) &=&
\int dp_\perp^2 dy_1 dy_2 d^2\vec {b_1} d^2\vec {b_2} \delta^{(2)} (\vec b
-\vec {b_1} - \vec {b_2}) \nonumber \\
&& \times \; x_1 \Gamma_{a/A}(x_1, Q^2,\vec {b_1})
x_2 \Gamma_{b/B}(x_2, Q^2,\vec {b_2}) \frac {d\hat \sigma^{ab}}{d\hat t} \;\; .
\end{eqnarray}
For simplicity we do not consider nuclear shadowing effects in this section.
Therefore the nuclear parton density in this case is simply
\begin{eqnarray}
\Gamma_{a/A}(x, Q^2,\vec s)=T_A(\vec s)f_{a/N}(x,Q^2) \;\; ,
\end{eqnarray}
where $f_{a/N}(x,Q^2)$ is the structure function of parton $a$ in the nucleon,
and $T_A(\vec s)$ is the nuclear thickness function
\begin{eqnarray}
T_A(\vec s)=\int_{-\infty}^{\infty} dz n_A(\sqrt {z^2+{|s|}^2}) \;\; , 
\;\;\;\int d^2\vec s T_A(\vec s)=A \;\; .
\end{eqnarray}

To show $A$ scaling of hard processes, we take the simplest hard sphere
model with radius $R_A$ for the nuclear density function
\begin{eqnarray}
R_A &=& r_0 A^{1/3}, \;\;\;r_0 \simeq 1.2\;{\rm fm} \;\; . 
\end{eqnarray}
Therefore
\begin{eqnarray}
T_A(\vec s) &=& \frac {3A}{2\pi R_A^2} \sqrt {1-\frac {{|s|}^2}{R_A^2}}
\;\; , \\[2ex] 
\Rightarrow N^{AB} (\vec b) &=&
\int d^2\vec {b_1} d^2\vec {b_2} \delta^{(2)} (\vec b -\vec
{b_1} - \vec {b_2}) T_A(\vec {b_1}) T_B(\vec {b_2}) \nonumber \\
&& \times \; \int dp_\perp^2 dy_1 dy_2 x_1 f_{a/N}(x_1, Q^2)
x_2 f_{b/N}(x_2, Q^2) \frac {d\hat \sigma^{ab}}{d\hat t} \\[2ex]
&=& T_{AB}(\vec b) \sigma^{pp} \;\; , 
\end{eqnarray}
where
\begin{eqnarray}
T_{AB}(\vec b) &=&
\int d^2\vec {b_1} d^2\vec {b_2} \delta^{(2)} (\vec b -\vec
{b_1} - \vec {b_2}) T_A(\vec {b_1}) T_B(\vec {b_2}) \\[2ex]
&=&\int d^2 \vec s T_A(\vec s) T_B(\vec b- \vec s) \;\; .
\end{eqnarray}
The scaling with atomic number $A$ in central $pA$ and $AA$ collisions is,
for example, given by
\begin{eqnarray}
T_{pA}(0)&=& \int d^2 \vec s T_p(\vec s) T_A(\vec s) \simeq \frac
{3A^{1/3}}{2\pi r_0^2} \;\; , \\[2ex]
\Rightarrow
N^{pA}/N^{pp} &\propto & A^{1/3} \;\; . \\[2ex]
T_{AA}(0)&=& \int d^2 \vec s T_A^2(\vec s) = \frac {9A^{4/3}}{8\pi r_0^2}
\;\; , \\[2ex] 
\Rightarrow N^{AA}/N^{pp} &\propto & A^{4/3} \;\; .
\end{eqnarray}

An estimate shows that minijet gluons may be produced with a cross section
$\sim 10$mb in $pp$ collisions at $\sqrt s=200$AGeV \cite{eskola1}.  This
estimate was obtained with a conservative transverse momentum cutoff $p_0=2$GeV
for the gluon scattering process.  Thus, according to the above scaling
with atomic number $A$, the number of minijet gluons may be on the order of
thousands at RHIC. 

\hmysubsection{Global Features of Heavy Ion Collisions}

In this section we review the standard calculations for certain
global features of heavy ion collisions such as energy density, temperature and
multiplicity.  

At present there are two major facilities for heavy ion experiments: the
Alternating Gradient Synchrotron at Brookhaven (AGS), and the Super Proton
Synchrotron at CERN (SPS).  They are both fixed target machines.  The AGS
accelerates nuclear beams from proton to gold up to the energy of $29(Z/A)$GeV,
and the SPS accelerates beams from proton to lead up to the energy of
$400(Z/A)$GeV.  
Two larger colliders are under construction:
the Relativistic Heavy Ion Collider at Brookhaven (RHIC) \cite{cdr}, and the
Large Hadron Collider at CERN (LHC) \cite{lhc}.  RHIC is scheduled to be
completed in 1999 with the center of mass energy of $200$AGeV for $Au+Au$
collisions. 
Heavy ion physics is one important part for the LHC program, and
the center of mass energy for $Pb+Pb$ collisions at LHC will be $5.4$ATeV.

The total multiplicity rapidity density for $pp$ collisions from SPS to
Tevatron energies \cite{dndy_exp} has been fit with
\begin{eqnarray}
\left ( \frac {dN}{dy} \right )^{pp} \simeq 0.9 \ln \left ( \frac {\sqrt
s}{2m_p} \right ) \;\; .
\end{eqnarray}
For $AA$ collisions, that density can be parameterized approximately as
\begin{eqnarray}
\left ( \frac {dN}{dy} \right )^{AA} \simeq A^\alpha \left ( \frac {dN}{dy}
\right )^{pp} \;\; , 
\label{EQ:dndy_aa}
\end{eqnarray}
where $\alpha$ can be determined at least from lower energy $pA$ data
\begin{eqnarray}
\alpha \simeq 1.1 \;\; . 
\end{eqnarray}

In the standard scenario based on Bjorken's longitudinal boost-invariant
boundary conditions \cite{bj}, the initial volume is assumed to be 
\begin{eqnarray}
V_0=\pi r_0^2 A^{2/3} \tau_0,\;\;\; \tau_0 \simeq 1{\rm fm} \;\; , 
\end{eqnarray}
and is furthermore assumed to grow linearly with proper time.
Given the average transverse momentum of final particles,
\begin{eqnarray}
<k_\perp> &\simeq& 400{\rm MeV} \;\; , \\[2ex]
\Rightarrow <m_\perp> &\simeq& 500{\rm MeV} \;\; .
\end{eqnarray}
Hence the initial energy density is
\begin{eqnarray}
\epsilon_0=\frac {<m_\perp> \left ( dN/dy \right )^{AA}}{V_0}
\simeq 0.1 A^{\alpha-2/3} \ln \left ( \frac {\sqrt s}{2m_p} \right )
\;{\rm GeV/fm^3} \;\; . 
\label{EQ:ed_s}
\end{eqnarray}
From this equation, we realize that the initial energy density grows as a
power of $A$, the atomic number of the heavy ion beam, but only grow
logarithmically with the center of mass energy.
Therefore it is much more efficient to reach a higher initial energy density
by using heavy ion collisions than increasing the beam energy alone.

According to eq.~(\ref{EQ:e_sb}) we get the initial temperature assuming 
that the initial state is an ideal gas of gluons and three flavor of massless
quarks:
\begin{eqnarray}
T_0=\left ( \frac {30\epsilon_0}{47.5\pi^2} \right )^{1/4}
\simeq \epsilon_0^{1/4} \times 150\;{\rm MeV} \;\; , 
\label{EQ:t_s}
\end{eqnarray}
where $\epsilon_0$ is expressed in the unit of GeV/fm$^3$.

\begin{figure}[p]
\vspace{1cm}
\psfig{figure=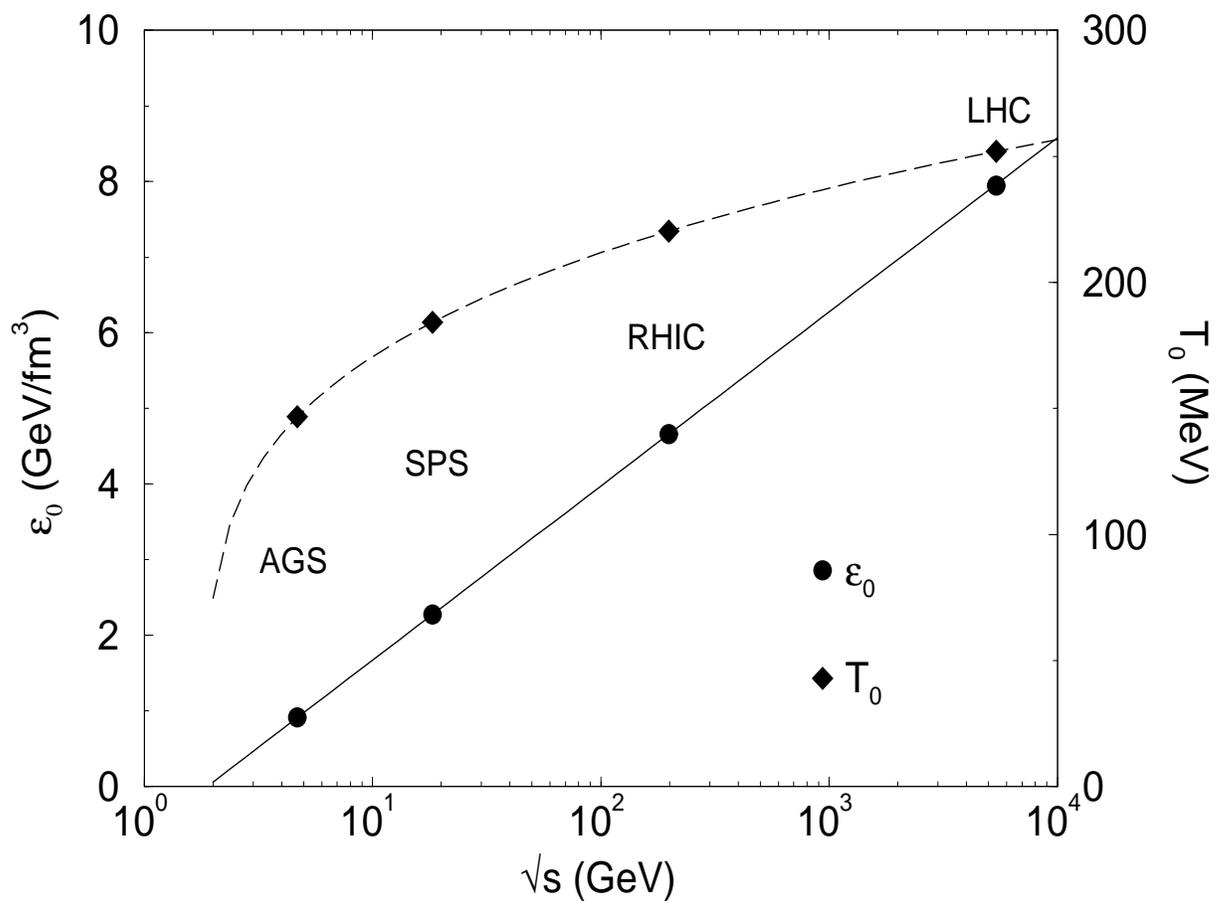,height=5.in,width=3.5in,angle=-90} 
\caption{
Initial energy density and temperature as a function of center of mass energy
of heavy ion collisions.
}
\label{fig:energy_density}
\end{figure}

In Figure~\ref{fig:energy_density} we plot the initial energy density and
temperature as a function of the center of mass energy.  We use
eq.~(\ref{EQ:ed_s}) and eq.~(\ref{EQ:t_s}) from the standard scenario, and
assume $AA$ collisions with $A \simeq 200$.

Note that the multiplicity density $dN/dy$ is about 300 times the energy
density (in the unit of GeV/fm$^3$) according to eq.~(\ref{EQ:ed_s}).
In Figure~\ref{fig:energy_density} we see that almost all four experimental
facilities can achieve an initial energy density above the critical $\epsilon_c
\simeq 1$GeV/fm$^3$ for the QCD phase transition.  
However at the currently available AGS and SPS energies, the baryon stopping
power is sufficiently large that matter is formed with high baryon density.  
We recall from the discussion of QCD thermodynamics that much less is
known about the high baryon density case than the zero baryon density case, and
we expect to gain valuable knowledge from these experiments.
At future collider facilities, matter is expected to be formed at much
lower baryon densities.

We have to stress that the above estimates for the global quantities in this
section are based on a standard scenario, which only provides an order of
magnitude estimate.  
Much higher energy densities may be achieved by taking multiple minijet
production into account, as we discuss later in Chapter \ref{subsec-minijets}. 
Factors such as a bigger $<k_\perp>$ \cite{biro}, a larger $\alpha$
from more secondary scatterings \cite{geiger} as found in cascade models, 
and a smaller $\tau_0$ as in the hot glue scenario \cite{hotglue} can all
result in higher temperatures for the initial QGP system.
However, we note that even the conservative estimate \cite{bj} stated above
gives a high enough initial energy density to motivate the search for
quark-gluon plasma in heavy ion experiments.

\hmysubsection{Thermalization}
\label{subsec-therm}

The quark-gluon plasma is defined to be an asymptotically free gas of quarks
and gluons with both thermal and chemical equilibration.  
After the dense system of quarks and gluons is formed in high energy heavy
ion collisions, it will approach thermalization via rescattering, and approach
chemical equilibration via creation and annihilation processes.   
However, the parton system only has a finite lifetime.
Thus it is important to know the time scales that the system
requires to reach thermalization and equilibration, and whether the parton
system can achieve both stages in its finite lifetime.  

In this section we first follow Shuryak's estimate of the above elastic
scattering cross sections \cite{hotglue}, and remark on the limitations of
that approach, and then estimate the thermalization time.  Finally we introduce
new results on parton energy loss \cite{bdmps}, which imply that faster
thermalization may be possible via inelastic processes.

The cross sections for the relevant elastic processes are
\begin{eqnarray}
\frac {d\sigma}{dt} &=& \frac {\pi \alpha_s^2}{s^2} M^2 \;\; , \\[2ex]
M^2_{gg \rightarrow gg} &=& \frac {9}{2} \left ( 3-\frac {ut}{s^2}-\frac
{us}{t^2} -\frac {st}{u^2} \right ) \;\; , \\[2ex]
M^2_{gq \rightarrow gq} &=& -\frac {9}{4} \frac {u^2+s^2}{us}+\frac
{u^2+s^2}{t^2} \;\; , \\[2ex] 
M^2_{qq \rightarrow qq} &=& \frac {4}{9} \frac {u^2+s^2}{t^2} \;\; .
\label{EQ:scatterings}
\end{eqnarray}

One may estimate the above cross sections by separating them into large-angle 
and small-angle scattering parts \cite{hotglue}.  
Since most entropy in high energy heavy ion collisions is in
gluons \cite{eskola1}, let us consider gluons.

For an ideal gas of gluons at a temperature $T$, 
\begin{eqnarray}
n_g&=&\int 2(N^2-1) \frac {d^3p}{e^{p/T}-1}
=\frac {(N^2-1) T^3}{\pi^2} \int_0^\infty \frac {z^2 dz}{e^z-1} \nonumber
\\[2ex] 
&=&\frac {2(N^2-1) \zeta (3)}{\pi^2} T^3 \equiv a_1 T^3 \;\; , \\[2ex]
\label{EQ:nd_gluon}
\zeta(3) &\simeq& 1.202 \;\; , \\[2ex]
\epsilon_g&=&\frac {2(N^2-1) \pi^2}{30} T^4 \equiv a_2 T^4 \;\; , \\[2ex]
\label{EQ:ed_gluon}
\Rightarrow p_g &=& \frac {\epsilon_g}{n_g} \simeq 3T \;\; .
\end{eqnarray}
For an ideal gas of quarks, equations become
\begin{eqnarray}
n_q&=&b_1 T^3, \;\; b_1= \frac {9n_f \zeta (3)}{\pi^2} \;\; , \\[2ex]
\label{EQ:nd_quark}
\epsilon_q&=&b_2 T^4, \;\; b_2= \frac {7\pi^2 n_f}{20} \;\; .
\label{EQ:ed_quark}
\end{eqnarray}
For large angle scatterings, the matrix elements in eq.~(\ref{EQ:scatterings})
are finite.  Therefore one might estimate the corresponding relaxation time
using the average values of the Mandelstam variables for an ideal gas of
gluons:  
\begin{eqnarray}
&&<s>/2 = -<t> = -<u> \simeq (3T)^2  \;\; , \\[2ex]
&\Rightarrow& <M^2_{gg \rightarrow gg}> \sim 30.4  \;\; , \\[2ex]
&\Rightarrow& \frac {1}{\tau _g^{large-angle}}=n_g <\sigma> 
\sim n_g \frac {\pi \alpha_s^2}{<s>^2} <M^2> <t> \sim 5\alpha_s^2 T .
\end{eqnarray}
For small angle scatterings, the matrix elements are divergent at $t,u
\rightarrow 0$. 
If one just use the finite Debye mass \cite{debye} to regulate the divergence
\cite{hotglue}, then the contribution from small angle scattering is 
\begin{eqnarray}
t_{min}&=&\mu_D^2=g^2 T^2  \;\; , \\[2ex]
\Rightarrow <M^2_{gg \rightarrow gg}> &\sim& \frac {9}{2} \frac {s^2}{t_{min}}
\;\; ,  \\[2ex] 
\Rightarrow \frac {1}{\tau _g^{small-angle}} 
&\sim& \frac {9\pi \alpha_s^2}{t_{min}} \sim 2.2 \alpha_s T \;\; .
\end{eqnarray}

Therefore the estimate of the rate for gluon elastic scatterings is
\begin{eqnarray}
\frac {1}{\tau_g} \sim 5\alpha_s^2 T + 2.2 \alpha_s T \;\; .
\label{EQ:t_g_hotglue}
\end{eqnarray}

The above method \cite{hotglue} of treating the small angle divergence is far
from exact.   
Although Debye screening eliminates the divergence from longitudinal exchange
of gluons, the transverse interactions, i.e. the color-magnetic interactions,
can not be simply screened.
This problem was discussed earlier by Danielewicz and Gyulassy \cite{damp1}.  
Later Baym et al. \cite{damp2} found that the Landau damping of transverse
gluons provides an effective cutoff at a distance $\sim 1/\mu_D$ for the
infrared divergence of transverse interactions.  The result for the gluon
relaxation time considering the viscosity of a quark-gluon plasma in the
weak-coupling limit is given by
\begin{eqnarray}
\frac {1}{\tau_g} \simeq 4 \alpha_s^2 \ln(1/\alpha) T \;\; .
\label{EQ:t_g_baym}
\end{eqnarray}
Comparing eq.~(\ref{EQ:t_g_hotglue}) to eq.~(\ref{EQ:t_g_baym}),  we note that
the more precise dependence of the relaxation time $\tau_g$ on the coupling
constant $\alpha$ in eq.~(\ref{EQ:t_g_baym}) is hinted by the combination
of linear and quadratic $\alpha$ dependent terms in the crude estimate found in
eq.~(\ref{EQ:t_g_hotglue}).

Now we take eq.~(\ref{EQ:t_g_baym}) and find that for a high
temperature $T \sim 0.5$GeV, the thermalization time for the gluons is large:
\begin{eqnarray}
\alpha &\sim& 0.3 \;\; , \\[2ex]
\Rightarrow \tau_g &\simeq& 1{\rm fm} \;\; .
\end{eqnarray}
Simila ranalysis gives a quark relaxation time which is about twice larger than
the gluon relaxation time.  Thus, via elastic scatterings it is difficult
to quickly approach thermalization.

However, thermalization may be accelerated by inelastic processes such as
parton radiations from multiple scatterings in the dense medium and/or gluon
multiplication processes ($2\;g \rightarrow n\; g$, $n\geq 3$)
\cite{xiong_2n}.  
The radiative energy loss due to multiple scatterings was first considered in
QED by Landau, Pomeranchuk and Migdal (the LPM effect) \cite{lpm_qed}.  Similar
effects were considered in QCD by Gyulassy and Wang \cite{lpm_qcd}, and
recently new calculations from Baier, Dokshitser, Mueller, Peign\'e and Schiff
(BDMPS) \cite{bdmps} yield unexpected results.
Instead of being finite, the energy loss per unit length in QCD for a fast
parton grows with the energy of the parton as $\sqrt E$ when the medium is
infinite, or grows with the medium length $L$ when the parton is extremely
energetic.  
The BDMPS result for the spectrum of radiative gluon energy loss per
unit length is given by \cite{bdmps}
\begin{eqnarray}
\omega \frac {dI}{d\omega dz} = \frac {3\alpha_s C_R}{2\pi \lambda_g}
\sqrt {\kappa \ln \frac {1}{\kappa}} \;\; , 
\end{eqnarray}
where $\kappa=\lambda_g \mu^2/\omega$, $w$ is the frequency of
the radiation, $\mu$ is the Debye mass induced by the medium, $\lambda_g$
is the mean free path for the gluon, and $C_R$ is the Casimir operator for the
parton.

In terms of radiation frequency, the LPM effect in QCD occurs for
\begin{eqnarray}
\omega_{BH} \sim \lambda \mu^2 < \omega < \omega_{Fact} \sim \frac {\mu^2
L^2}{\lambda} \;\; ,
\end{eqnarray}
while the Bethe-Heitler regime lies below $\omega_{BH}$, and the factorization
regime lies above the frequency $\omega_{Fact}$ and up to the energy of the
parton, $E$. 

In the case of an infinite medium, the LPM effect occurs for frequencies up to
the order of $E$, and therefore
\begin{eqnarray}
-\frac {dE}{dz} = \int d\omega \;\left (\omega \frac {dI}{d\omega dz} \right )
\;\;\propto\;\; \int d\omega \sqrt {\frac {1}{\omega} \ln \omega}
\;\;\propto\;\; \sqrt E \ln E \;\; . 
\end{eqnarray}

While in the case of an infinite energy $E$, the LPM effect stops at
$\omega_{Fact}$, thus
\begin{eqnarray}
-\frac {dE}{dz} \propto L \ln L \;\; . 
\end{eqnarray}
This nonlinear behaviour may imply a huge radiative energy loss
for fast partons in a dense system.  The strong radiation may accelerate the
process of thermalization, which is probably slow  when one only includes the
elastic scatterings according to the previous estimate.

The previous estimate of thermalization time, 
e.g. eq.~(\ref{EQ:t_g_hotglue}), assumed at the initial stage a chemically
equilibrated gluon gas.  With a high initial temperature,
the corresponding gluon density is expected to be large,
eq.~(\ref{EQ:nd_gluon}).  However, the initial gluon density produced via
minijets and soft mechanisms may still not be high enough to populate the
whole phase space.  
In order to study the chemical equilibration of the dense parton system, as
will be discussed in the next section, one needs the initial condition
including thermalization time scale, temperature and chemical
compositions for quarks and gluons.  
Therefore one has to take the non-zero chemical potential into account.  

\hmysubsection{Chemical Equilibration}
\label{subsec-chem}

Chemical equilibration is approached via flavour creation and annihilation
processes.  The evolution of chemical composition can be described
by a set of master equations \cite{master}.  In order to solve the equations
one usually assume that the partons have reached approximate thermal
distribution.  As we will see, the time scale for chemical equilibration is
much longer than that for thermalization, and therefore the above assumption is
supported.  

As discussed in the previous section, gluons and quarks in general are not
in chemical equilibrium.  With non-zero chemical potential, one can introduce
fugacity parameters $\lambda_g$ and $\lambda_q$ \cite{biro}:
\begin{eqnarray}
f_i=\sum_n \lambda_i^n e^{-n \beta u \cdot k}
=\frac {\lambda_i}{e^{-\beta u \cdot k} \pm \lambda_i}
\simeq \frac {\lambda_i}{e^{-\beta u \cdot k} \pm 1} \;\; , 
\end{eqnarray}
where $i=g,q$, $\beta=1/T$, and $u^\mu$ is the velocity of the local co-moving
frame.  

We study the following dominant processes for parton creation and
annihilation
\begin{eqnarray}
gg \leftrightarrow ggg, \;\;gg \leftrightarrow q\bar q \;\; .
\end{eqnarray}
For a baryon symmetric system where $n_q=n_{\bar q}$, the evolutions of the
parton densities are 
\begin{eqnarray} 
\partial_\mu \left ( n_g u^\mu \right )
&=&\frac {1}{2} \sigma_3 n_g^2 \left (1-\frac {n_g}{\tilde{n_g}} \right )
- \sigma_2 n_g^2 \left (1-\frac {n_q {\tilde{n_g}}^2}
{{\tilde{n_q}}^2 n_g^2} \right ) 
\label{EQ:ev_g} \;\; , \\[2ex]
\partial_\mu \left ( n_q u^\mu \right )
&=&\frac {1}{2} \sigma_2 n_g^2 \left (1-\frac {n_g}{\tilde{n_g}} \right )
\;\; , \label{EQ:ev_q}
\end{eqnarray}
where
\begin{eqnarray}
\sigma_3&=&<\sigma \left (gg \rightarrow ggg \right ) v>, \;\;\;
\sigma_2=<\sigma \left (gg \rightarrow q\bar q \right ) v> \;\; , \\[2ex]
\tilde {n_g}&=&n_g/\lambda_g, \;\;\;
\tilde {n_q}=n_q/\lambda_q \;\; .
\end{eqnarray}
The temperature evolution is
\begin{eqnarray}
\partial_\mu \left (\epsilon u^\mu \right ) + p\partial_\mu u^\mu =0 \;\; . 
\label{EQ:bj_t}
\end{eqnarray}
During early times one can neglect transverse expansion, and assume the Bjorken
longitudinal expansion, hence the above equation becomes
\begin{eqnarray}
\frac {d\epsilon}{d\tau} + \frac {\epsilon + p}{\tau} =0 \;\; .
\end{eqnarray}
For a thermally equilibrated quark and gluon gas, according
to eqs.~(\ref{EQ:ed_gluon}) and (\ref{EQ:ed_quark}), the energy density
becomes 
\begin{eqnarray}
\epsilon=3p=\left (\lambda_g a_2 + \lambda_q b_2 \right ) T^4 \;\; .
\label{EQ:ed_lambda}
\end{eqnarray}

Hence the solution of eq.~(\ref{EQ:bj_t}) is just
\begin{eqnarray}
\epsilon \tau^{4/3} = \left (\lambda_g a_2 + \lambda_q b_2 \right ) T^4
\tau^{4/3} = {\rm constant} \;\; .
\label{EQ:chem1} 
\end{eqnarray}

\begin{figure}[p]
\vspace{1cm}
\psfig{figure=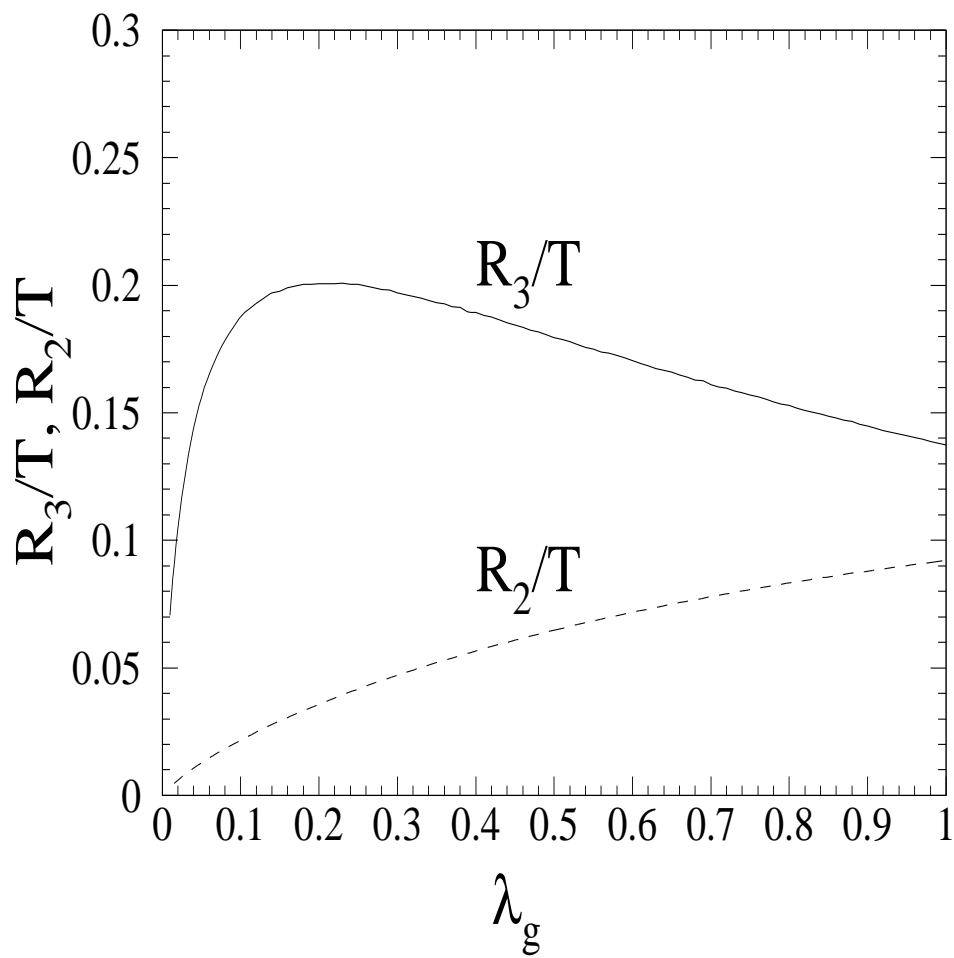,height=5.in,width=5.in,angle=0} 
\caption{
The gluon and quark production rates from ref.~\protect{\cite{xn_report}}. 
}
\label{fig:rates}
\end{figure}

With the assumption of pure longitudinal expansion, one has
\begin{eqnarray}
\partial_\mu \left ( n_i u^\mu \right )
=\frac {\partial n_i}{\partial \tau} + \frac {n_i}{\tau} \;\; .
\end{eqnarray}
Therefore one can write eqs.~(\ref{EQ:ev_g}) and (\ref{EQ:ev_q}) as
\begin{eqnarray}
\frac {\dot \lambda_g}{\lambda_g} + 3 \frac {\dot T}{T} + \frac {1}{\tau}
&=& R_3 \left ( 1-\lambda_g \right ) - 2R_2 \left ( 1-\frac {\lambda_q^2}
{\lambda_g^2} \right ) 
\;\; , \label{EQ:chem2} \\[2ex]
\frac {\dot \lambda_q}{\lambda_q} + 3 \frac {\dot T}{T} + \frac {1}{\tau}
&=& R_2 \frac {a_1}{b_1} \left ( \frac {\lambda_g} {\lambda_q} - \frac
{\lambda_q} {\lambda_g} \right ) 
\;\; , \label{EQ:chem3} 
\end{eqnarray}
where the overdot represents the derivative with respect to $\tau$, 
$a_1$ and $b_1$ are from eqs.~(\ref{EQ:nd_gluon}) and (\ref{EQ:nd_quark}), and
\begin{eqnarray}
R_3=\frac {1}{2} \sigma_3 n_g, \;\;\; R_2=\frac {1}{2} \sigma_2 n_g.
\end{eqnarray}
The above rates $R_3$ and $R_2$ contain infrared divergences.  One can 
estimate the rates using the Debye screening mass for gluons and the thermal
quark mass to regulate the divergences \cite{biro,lmw,xn_report}.
Numerical results for the estimate are shown in Figure.~\ref{fig:rates}
\cite{xn_report}. 

The initial conditions for above differential equations as in 
the case of $Au+Au$ collisions at RHIC energy $\sqrt s=200$AGeV is \cite{biro}:
\begin{eqnarray}
\tau_{iso} =0.7 {\rm fm}, \;\; n_0=3.2 {\rm fm}^{-3}, 
\;\;<k_\perp>=1.17 {\rm GeV} \;\; .
\label{EQ:initial_iso}
\end{eqnarray}
Using eq.~(\ref{EQ:ed_lambda}), $\epsilon_0=4n_0<k_\perp>/\pi$,
and $\lambda^0_q=0.16 \lambda^0_g$ \cite{xn_report}
\begin{eqnarray}
n=\left (\lambda_g a_1 + \lambda_q b_1 \right ) T^3 \;\; .
\end{eqnarray}
One can obtain initial values of fugacities and temperature
\begin{eqnarray}
\lambda^0_g = 0.05, \;\;\; \lambda^0_q = 0.008, \;\;\; 
T_0=0.57 {\rm GeV} \;\; . 
\end{eqnarray}

\begin{figure}[p]
\vspace{1cm}
\psfig{figure=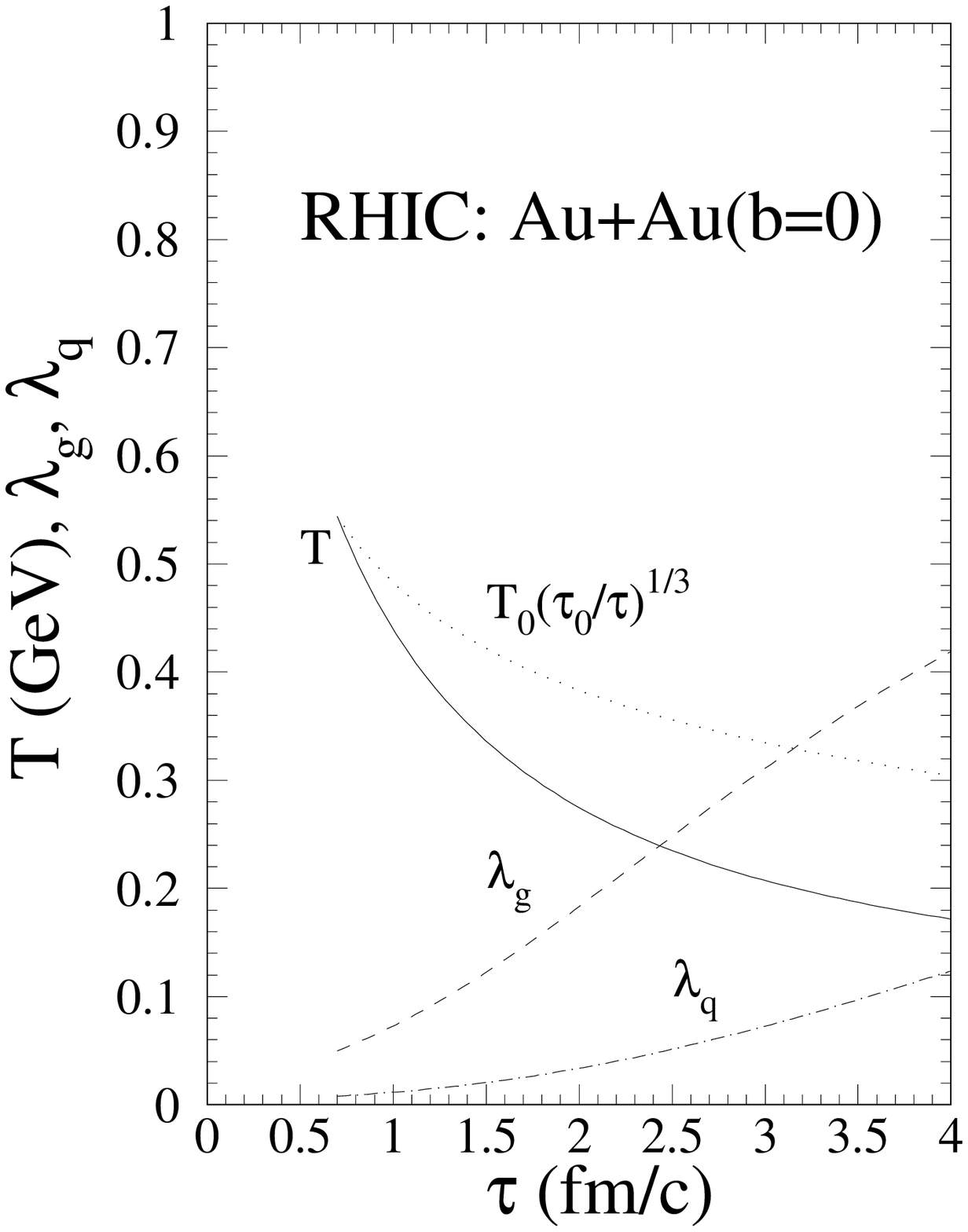,height=5.in,width=5.in,angle=0} 
\caption{
Chemical equilibration in $Au+Au$ collisions at RHIC
from ref.~\protect{\cite{xn_report}}. 
}
\label{fig:chem_evolution}
\end{figure}

Given the above initial conditions, the chemical equilibration
eqs.~(\ref{EQ:chem1}), (\ref{EQ:chem2}) and (\ref{EQ:chem3}) can be solved
numerically, and results of which are shown in
Figure~\ref{fig:chem_evolution} \cite{xn_report}.

In Figure~\ref{fig:chem_evolution} the temperature in this calculation drops
faster than in Bjorken's free streaming case, since some energy is used to
produce more partons in order to approach chemical equilibration.  
Gluons are much more likely to approach the chemical equilibrium than quarks.
However, both of them can hardly reach chemical equilibrium within
the lifetime of the plasma state.

One can see that the result in this study depends strongly on initial
conditions.    
This particular set of initial conditions may be conservative.  As discussed in
Chapter~\ref{subsec-therm}, if fast thermalization could be achieved via
inelastic processes, then one would begin with high initial parton fugacities,
and the lifetime of the plasma would become longer.  All of these contributions
would accelerate the chemical equilibration.
However, initial parameters are, unfortunately, quite uncertain since we have
insufficient understanding of the soft physics. 
There are uncertainties in cutoff scales, rescattering cross sections, $2g
\leftrightarrow ng$ process \cite{xiong_2n}, and small-$x$ gluon distributions.
Therefore careful investigations are still needed to determine whether more
equilibrated plasma can be formed at RHIC. 

\hmysubsection{Phase Diagram}

\begin{figure}[p]
\psfig{figure=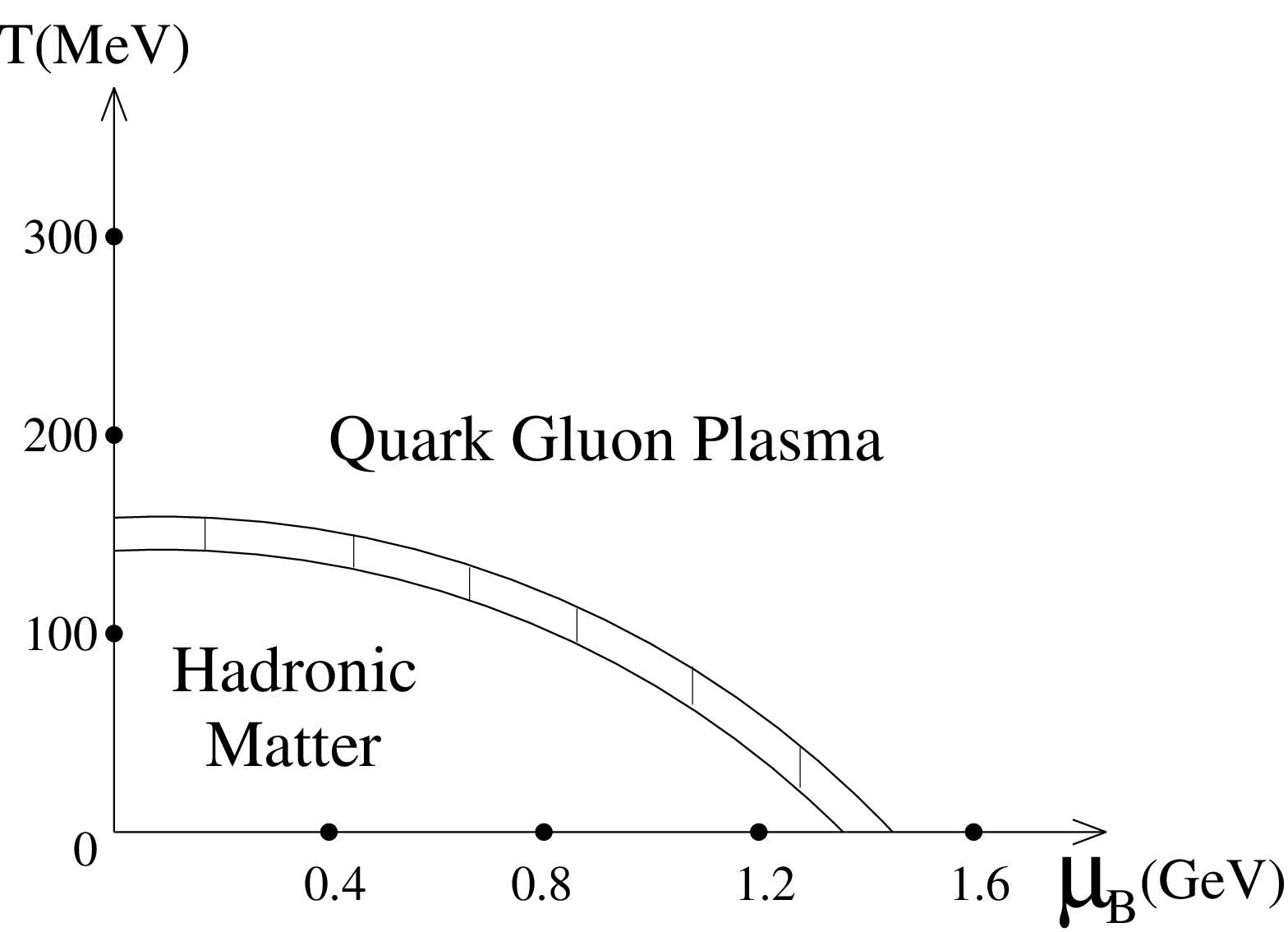,height=2.7in,width=5.5in,angle=0} 
\caption{
Schematic QCD phase transition diagram as a function of $T$ and $\mu_B$.
}
\label{fig:phase_mub}
\vspace{1cm}
\psfig{figure=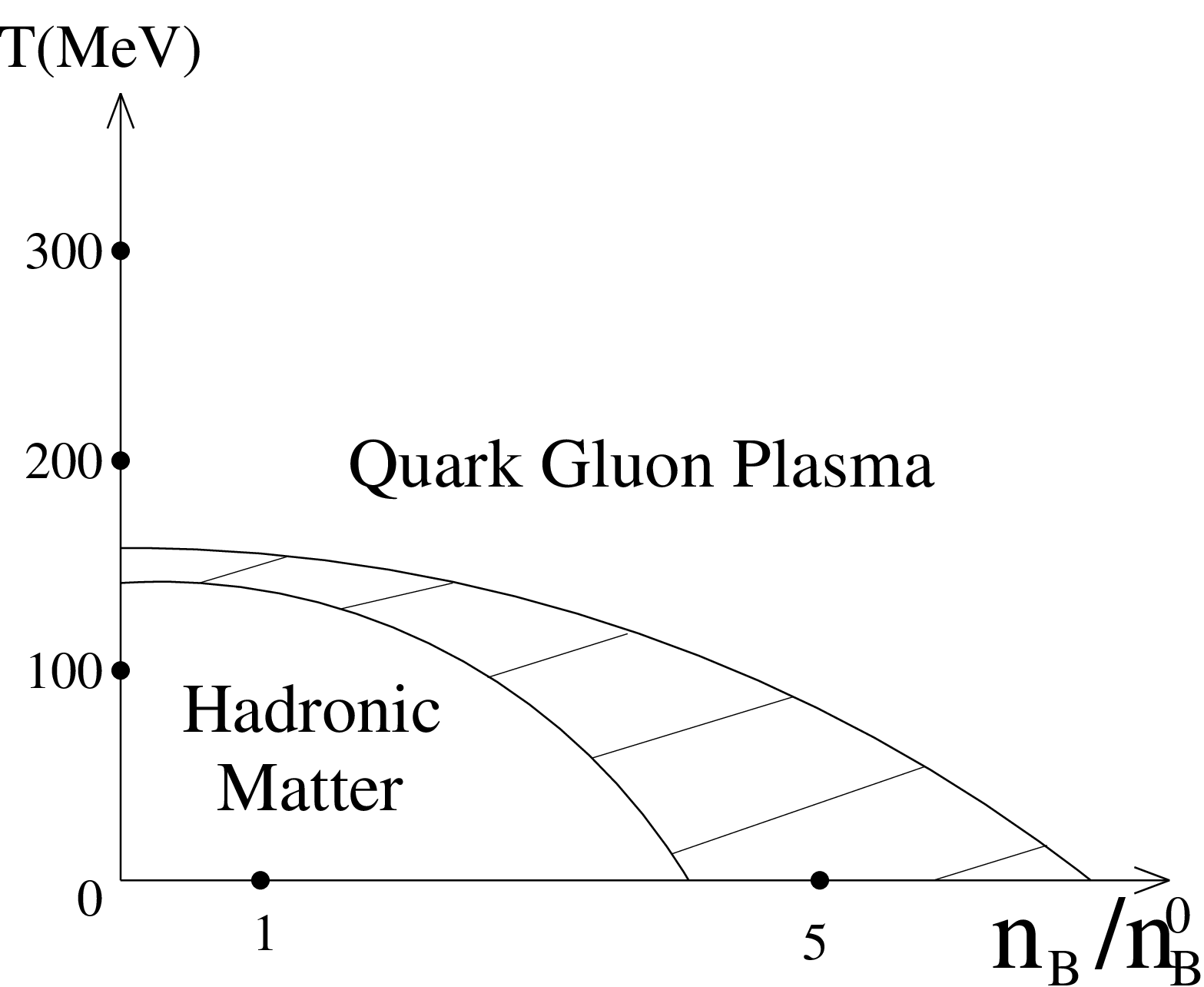,height=2.7in,width=5.in,angle=0} 
\caption{
Schematic QCD phase transition diagram as a function of $T$ and $n_B/n^0_B$.
}
\label{fig:phase_nb}
\end{figure}

From lattice QCD numerical simulations, the QCD phase transition happens at
$T_c=150 \pm 10$MeV \cite{crossover} at zero net baryon density. 
The QCD phase transition is also expected to occur at some high baryon density
$n_c^B$ at zero temperature, when the hadrons are so dense that the partons
from different nucleons start to overlap.  At present, there is no lattice QCD
calculation on phase transitions with finite baryon density, and therefore
there is uncertainty with the critical baryon density $n_c^B$.  In general one
expects that the QCD phase transition occurs on a smooth curve on the plane of 
temperature $T$ and the baryon chemical potential $\mu_B$.  In terms of 
baryon number densities, there is a jump from the hadronic phase to the QGP
phase, just like the jump in the energy densities at zero baryon
density.  The phase diagram for the transition from the hadronic
phase to the QGP phase is roughly plotted in Figure~\ref{fig:phase_mub} as a
function of the temperature $T$ and the baryon chemical potential $\mu_B$.
Figure~\ref{fig:phase_nb} shows the phase diagram as a function of the
temperature $T$ and the scaled baryon number density $n_B/n^0_B$, where
$n^0_B\simeq 0.14$/fm$^3$ is the normal density in nuclear matter. 
The shaded area in Figure~\ref{fig:phase_mub} represents the error
from our current knowledge about the phase transition, and the shaded area
in Figure~\ref{fig:phase_nb} mostly represents the mixed phase.

Let us estimate the critical chemical potential at $T=0$.
For high density quark matter consisting of $u,d$ quarks at
zero temperature:
\begin{eqnarray}
n_q&=&\frac {12}{(2\pi)^3} \int_0^{\mu_q} d^3 p=\frac {2}{\pi^2} \mu_q^3
\;\; , \\[2ex] 
\Rightarrow \epsilon&=&\frac {3}{2\pi^2} \mu_q^4 \;\; .
\end{eqnarray}
If we assume that the QCD phase transition at zero temperature needs
approximately the same critical energy density as in the zero baryon potential
case, then 
\begin{eqnarray}
\epsilon_c &\simeq& 1{\rm GeV/fm^3} \;\; , \\[2ex]
\Rightarrow \mu^c_q&=&\left ( \frac {2\pi^2 \epsilon_c}{3} \right )^{1/4}
\simeq 480 {\rm MeV} \;\; , \\[2ex] 
\Rightarrow \mu^c_B &\simeq& 3\mu^c_q =1.4 {\rm GeV} \;\; , \\[2ex]
n^c_B &\simeq& \frac {n^c_q}{3} = 0.93{\rm /fm^3} \simeq 7 n^0_B \;\;{\rm in\;
quark\; matter\; phase.}
\end{eqnarray}

These estimates are consistent with the zero temperature values in
Figure~\ref{fig:phase_mub} and Figure~\ref{fig:phase_nb}. 
In order to obtain a general estimate on the phase transition boundary, 
expressions for the pressures in the two phases need to be known.
The pressure $P_p(T, \mu)$ in the plasma phase is derived perturbatively and
expressed in eq.~(\ref{EQ:pmu_sb}).
The hadronic equation of state can not be obtained from perturbative
calculations, but can be parameterized for different hadron species.  
In this simple case one gets the parameterization of the EOS in the hadronic
phase by adding the parameterizations for nucleons, pions and other resonances
together.  
The energy density for nucleons, for example, can be parameterized as
\begin{eqnarray}
\epsilon=n \left [ m - W_0 + \frac {K(n/n_0-1)^2}{18} + \frac {3T}{2} \right ]
\;\; , 
\end{eqnarray}
where $n$ is the nucleon density, $m$ is the nucleon rest mass, $W_0$ is the
binding energy, $K$ represents the compressional energy density, and the last
term is the thermal energy 
\cite{csernai}. 
One then take Gibbs' criteria of phase equilibrium $P_h(T,\mu)=P_p(T,\mu)$ 
to determine the phase boundary curve $T_c(\mu)$, which is schematically shown
in Figure~\ref{fig:phase_mub} \cite{boundary}.

%% file: intro3_signal.tex
\hmysection{Signals of the Quark-Gluon Plasma}
\label{sec-signals}

\begin{figure}[p]
\psfig{figure=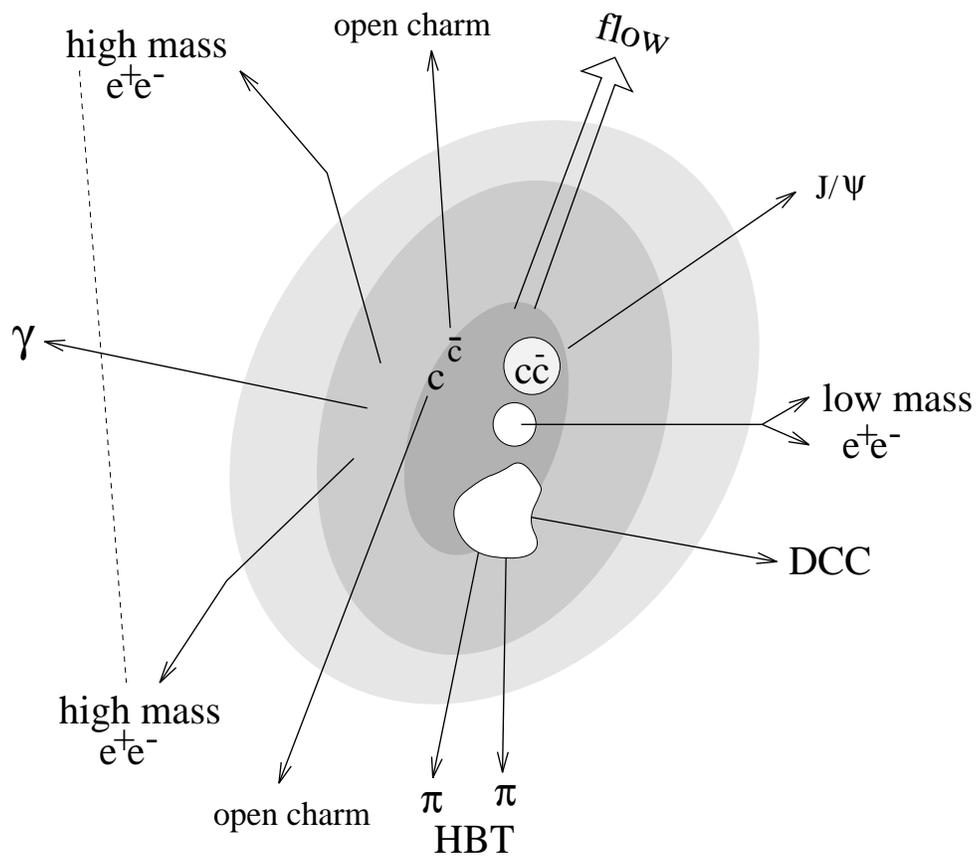,height=4.5in,width=5.in,angle=0} 
\caption{
Several key observables of the quark-gluon plasma.
}
\label{fig:signal}
\end{figure}

In this section we discuss several key observables of the quark-gluon plasma,
as sketched in Figure~\ref{fig:signal}.
Among the probes of the early formed quark-gluon plasma, electromagnetic probes
such as direct photons \cite{hotglue,photon,sx} and dileptons
\cite{qgp,dilepton} are the most direct signals, since they do not experience 
the subsequent thermalization and final state interactions.   
Heavy quarkonia \cite{jpsi} and heavy quark \cite{geiger,mw} productions are
also proposed to be the sensitive probes of the properties of the QGP.
Recently new interesting experimental results such as low mass
dilepton enhancement \cite{ceres,lowmass_95} and $J/\psi$ suppression in $A+B$
collisions \cite{jpsi_lead} were presented.  These results were heavily
discussed at the Quark Matter'96 meeting \cite{qm96}. 

\hmysubsection{Heavy Quarkonia}

The production and suppression of heavy quarkonia bound states, such as
$J/\psi$, was proposed by Matsui and Satz \cite{jpsi} 
to be an ideal test of the quark-gluon plasma production.
The authors claimed that 
"there appears to be no mechanism for $J/\psi$ suppression in nuclear
collisions except the formation of a deconfining plasma, and if such a plasma
is produced, there seems to be no way to avoid $J/\psi$ suppression."  
The proposed suppression signature of QGP formation
was first observed in 200 AGeV $O+U$ reactions at the CERN/SPS
in 1987 \cite{na38}. However, far from settling the matter, the debate on the
interpretation of those data and on the uniqueness of the QGP mechanism for
suppression has only intensified since then.

The nuclear suppression of charmonium was observed in $p+A$ reaction and was
found to depend linearly on $A^{1/3}$. 
In addition, the suppression was also observed for $\psi^\prime$.
Since the QGP production is {\em not} expected in $p+A$ collisions,
alternate mechanisms for suppression of quarkonia production were proposed.
One attempt to explain the $p+A$ data was to ascribe the suppression to nuclear
gluon shadowing \cite{satz_shad}. 
(We will discuss more fully the physics behind this effect later.)
However,  this explanation was later excluded because the lowest energy data at
$150$GeV did not fit into the universal shadowing curve.  The most satisfactory
explanation thus far \cite{satz_octet} of $J/\psi$ suppressions in both $p+A$
and {\em light} ion $O+U$ and $S+U$ reactions \cite{jpsi_su} is that it is due
to pre-formation inelastic final state interactions of the color-octet $c\bar c
g$ component which is produced in the initial hard glue-glue fusion event,
with nucleons in the target with a cross section $\sigma_{\psi N} \sim 6mb$.
In this {\bf nucleon absorption model}, the $J/\psi$ survival probability in
nucleus $A$ can be estimated as
\begin{eqnarray}
S_A=\exp \left [ {-\int_{z_0}^\infty dz \rho_A(\vec b,z) \sigma_{\psi N}}
\right ] \;\;\sim\; \exp {\left (-\sigma_{\psi N} \rho_0 L_A \right )} \;\; , 
\end{eqnarray}
where $z$ is the longitudinal coordinate along the beam axis, and $\vec b$ is
the impact parameter of the initial $gg \rightarrow c\bar{c}$ hard pQCD event
in the target nucleus with density profile, $\rho_A$, normalized to $A$.  
The survival probability is the probability of having no final state
interactions in the remaining part of the nucleus. 
It depends exponentially on the {\em effective} path length, $L_A$, travelled
by the $c\bar c$ pair in a nucleus of mean density $\rho_0 \approx 0.14/{\rm
fm}^3$. 
We note that typically the average rapidity of the produced $c\bar{c}$ pair is
high in the target frame and that is why the straight line (Eikonal)
approximation can be used. 

In $A+B$ collisions, the data on quarkonia production are usually sorted in
terms of an external calorimetric transverse energy trigger, $E_T$. 
The transverse energy is usually measured in a window with restricted rapidity
and azimuthal angle coverages, and depends sensitively on the particular
experimental device and configuration. 
Generally only a detailed Monte Carlo nuclear event generator such as HIJING
\cite{hijing} or VENUS \cite{venus} can provide a way to calculate the
dependence of the observed transverse energy on the impact parameter of the
collision.  
With such a generator, the conditional transverse energy probability density,
$P^{AB}(E_T;\vec{b})$, that an $A+B$ reaction at impact parameter $\vec{b}$
will result in a measured $E_T$, can be estimated.  

The survival probability of $J/\psi$ in an $A+B$ reaction triggered with a
given $E_T$ can then be calculated as
\begin{eqnarray}
S_{AB}(E_T)&\propto&
\frac {\sigma_{\psi}^{AB}(E_T)}{\sigma_{DY}^{AB}(E_T)}\nonumber \\[2ex]
&\propto&\frac{1}{AB}
\int d^2 b P^{AB}(E_T; \vec{b}) \; dz_A d^2 b_A\; dz_B d^2 b_B\;\delta^2(
\vec{b}-\vec{b}_A-\vec{b}_B) \nonumber\\
&&\;\;\;\;\times \; \rho_A(z_A,\vec{b}_A) \rho_B(z_B,\vec{b}_B)
e^{ -\sigma_{\psi N} ( \int_{z_A}^\infty dz \rho_A(\vec{b}_A,z)
+\int_{z_B}^\infty dz \rho_B(\vec{b}_B,z))} \nonumber\\[2ex]
&=& e^{-\sigma_{\psi N} \rho_0 L_{AB}(E_T)} \;\;\; .
\end{eqnarray}
In the first line the Drell-Yan cross section \cite{drell-yan} is used to
divide the quarkonium cross section, since the absence of final state
interactions in $q\bar q \rightarrow \mu^+\mu^-$ makes the atomic number
scaling of that cross section to be particularly simple.  Integrated over
transverse energy triggers and hence impact parameter,
$\sigma_{DY}^{AB}=AB\sigma_{DY}^{NN}$.  
For the most central collision (high $E_T$) trigger $N_{DY}^{AA} \approx
A^{4/3} N_{DY}^{NN}$ up to nuclear shadowing corrections.
We note that both the above quarkonia and Drell Yan cross sections are
understood to refer to comparable rapidity and transverse momentum kinematic
regions. 

\begin{figure}[p]
\psfig{figure=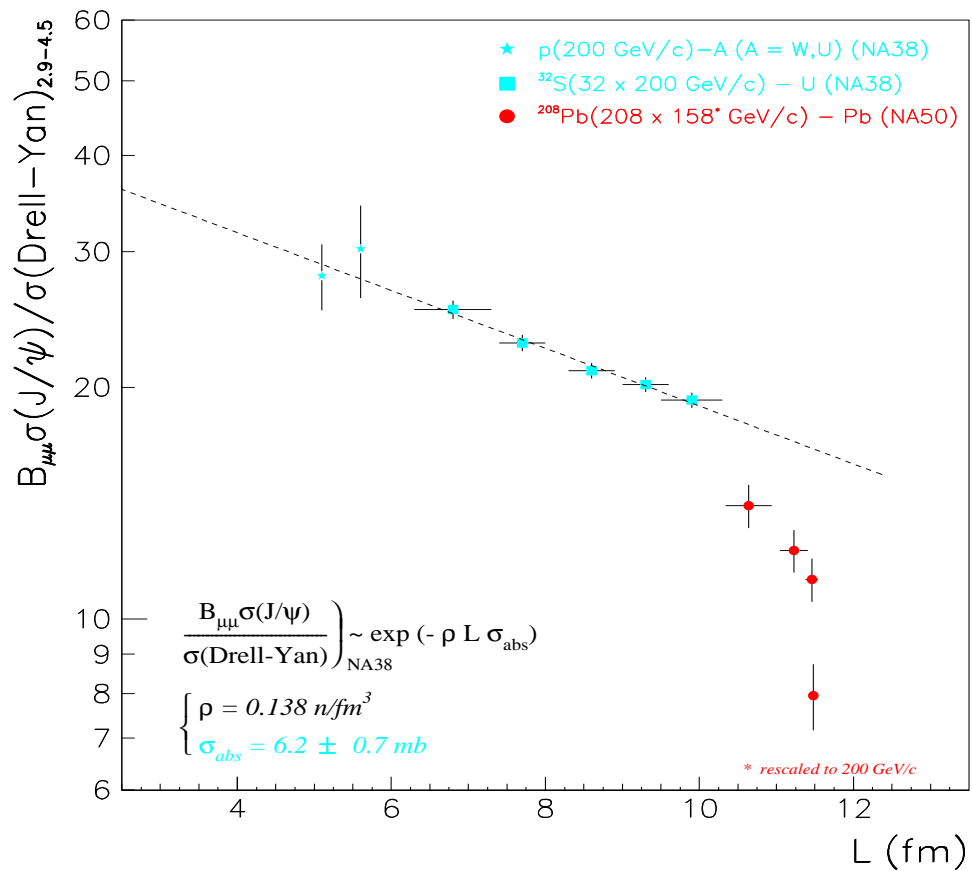,height=4.5in,width=5.in,angle=0} 
\caption{
$J/\psi$ to Drell Yan ratio in $p+A$, $S+U$ and $Pb+Pb$ collisions as a
function of the average path length $L$ from ref.~\protect{\cite{na50_www}}. 
}
\label{fig:na50}
\end{figure}

In Figure~\ref{fig:na50} the ratio of $J/\psi$ to Drell-Yan cross sections is
plotted against the effective nuclear absorption length $L \equiv \langle
L_{AB}(E_T) \rangle$. 
Unfortunately, the $L$ dependence on the experimental observable $E_T$ is
rather model dependent \cite{l_et}.  
Nevertheless, the exponential fit shows that both $p+A$ and $S+U$ data can be
well explained taking $\sigma_{\psi N}\approx 6 $ mb. 

A caveat in the above nucleon absorption explanation of suppression in light
ion reactions is that a larger suppression of $\psi^\prime$ is observed in
$S+U$ collisions than for $\psi$ \cite{psi_prime}. This can only be understood 
in terms of an additional suppression mechanism that only applies to the larger
sized $\psi^\prime$ states in the co-moving high density hadronic matter formed
along side the quarkonium state. 

The most provocative discrepancy seen in Fig.~\ref{fig:na50}, however, is the
failure of the nucleon absorption model to explain the latest data  showing
enhanced suppression of $J/\psi$ in NA50 $Pb+Pb$ experiment \cite{jpsi_lead}.  
The observed enhanced extra suppression of $J/\psi$ in $Pb+Pb$ data has been
interpreted \cite{jpsi_qgp} as the first direct evidence  for the formation of
quark gluon plasma.

The above claim has however been challenged by results of a recent study by
Gavin and Vogt based on a more careful analysis of the co-mover suppression
model \cite{comover_lead}. 
In an earlier study of NA38 $S+U$ data, it is argued \cite{comover_su} that
nucleon absorption alone (even with a cross section of $7mb$) can not explain
the $E_T$ distribution of $S+U$ data, while co-mover dissociation cross section
of $3mb$ together with nucleon absorption of $4.8mb$ could. 
Using the same parameters and updating the kinematical variables appropriate to
the NA50 experiment, this model predicted the extra suppression of $\psi$ 
in $Pb+Pb$ in quantitative accordance with the recent data.  

The $J/\psi$ survival probability due to co-mover scatterings can be simply 
estimated according to \cite{co_gg}
\begin{eqnarray}
S_{co}=\exp {\left (-\int_{\tau_0}^{\tau_F} d\tau n(\tau) \sigma_{co} v_{rel}
\right ) } \;\; , 
\end{eqnarray}
where due to Bjorken longitudinal expansion, the local co-moving
density decreases with proper time as 
\begin{eqnarray}
n(\tau)= n_0 \frac {\tau_0}{\tau} \;\; .
\end{eqnarray}
Here, $\tau_0$ is the formation proper time of the quarkonium state in the 
frame where its longitudinal rapidity vanishes. The initial co-moving density
is found to scale approximately linearly with transverse energy, $n_0=\tilde
{n_0} E_T/\tilde {E_T}$ \cite{comover_lead}.  
The final proper time, $\tau_F$, marking the time that the $\psi$ exits the
interaction area as it propagates in the transverse direction is approximately
given by 
\begin{eqnarray}
\tau_F \approx \frac {R_A}{v_{rel}} \;\; .
\end{eqnarray}
The survival probability after the co-mover scattering with cross section
$\sigma_{co}$ introduces another exponentially decreasing factor with
increasing $E_T$: 
\begin{eqnarray}
S_{co} \simeq 
\exp {\left [ - \sigma_{co} v_{rel} \tilde {n_0} \tau_0 
\ln \left ( \frac {R_A} {v_{rel}\tau_0} \right ) E_T /\tilde {E_T} \right ] }
=\exp { \left (-\beta E_T \right )} \;\; . 
\end{eqnarray}
Therefore, together with the nucleon absorption factor the survival probability
including co-mover $\psi$ dissociation processes is given schematically by
\begin{eqnarray}
S_{AB}(E_T)
\;\;\propto\; <S_A S_B S_{co}> \sim e^{-\sigma_{\psi N} \rho_0 L_{AB}(E_T)}
e^{-\beta E_T} \;\; . 
\end{eqnarray}

\begin{figure}[p]
\psfig{figure=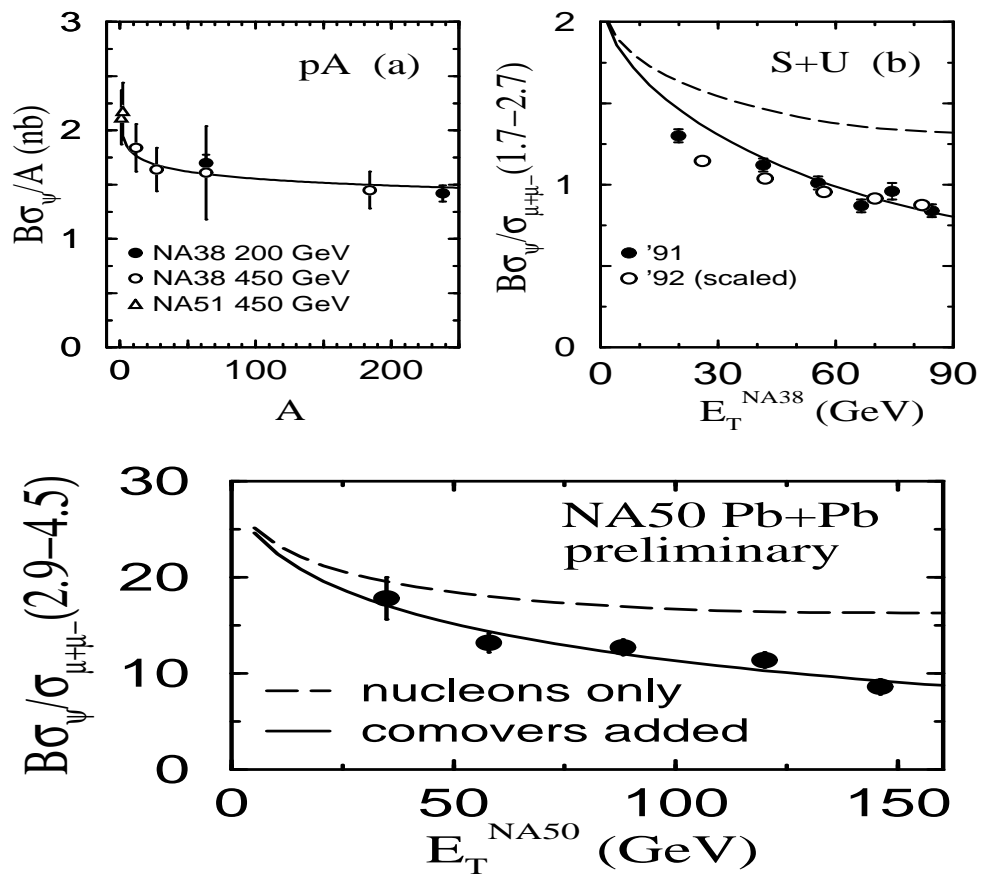,height=4.5in,width=5.in,angle=0} 
\caption{
Co-mover $J/\psi$ suppression vs $S+U$ and $Pb+Pb$ data from
ref.~\protect{\cite{comover_lead}}.  
}
\label{fig:comover}
\end{figure}

Although the average nuclear absorption length $L_{AB}(E_T)$ grows with $E_T$, 
it has a geometrical upper limit $\sim (R_A+R_B)$ in central collisions.
Therefore, nucleon absorption mechanism saturates at high $E_T$.  
However in central collisions $E_T$ may fluctuate to higher and higher values
as the local rapidity density of hadrons increases.  Co-mover dissociation can
therefore increase indefinitely with increasing $E_T$.  
The result of the detailed analysis \cite{comover_lead} is shown in
Figure~\ref{fig:comover}. 
The suppression of $J/\psi$ production as a function of the actual
experimentally observed $E_T$ is plotted for NA38 $S+U$
\cite{jpsi_su,psi_prime} and NA50 $Pb+Pb$ data \cite{jpsi_lead}.  
Furthermore, the co-mover model could also be adjusted to fit  well the
observed $\psi^\prime$ data in $p+A$, $S+U$ and $Pb+Pb$ collisions. This
agreement indicates no compelling need to invoke the QGP explanation at this
time from purely the phenomenological point of view. 
Theoretically, the validity of the large co-moving dissociation cross section
has been questioned \cite{kharzeev}, but from the phenomenological side all the
available data seem to be well accounted for by this conventional mechanism. 

Another ambiguity associated with the interpretation of quarkonium data in
terms of QGP formation is due to uncertainties in the basic production
mechanism of such bound heavy quark state even at the $pp$ level. 
In the generalized version of the color evaporation model \cite{evaporation}, 
it is assumed that charmonium cross sections are proportional to the 
sub-threshold charm cross section.   With a fit proportionality constant of
3\% \cite{evap_rev}, the model seems to give a good description (with a factor
of 2) of the energy dependence of the total $J/\psi$ production with pion and
proton beams.  It is also consistent with $\psi^\prime/(J/\psi)$ ratio data.   
This model has only one parameter for any bound state.
However, it remains to be justified that this factor is independent of
processes. In particular, the factor for pion and proton beams differs somewhat
and the color evaporation mechanism can easily depend on the environment and
hence on $A$ in nuclear reactions. 

In the color-singlet model \cite{singlet}, on the other hand, it is assumed
that charmonium cross sections are proportional to the cross section of forming
a color-singlet $c\bar c$ state with the corresponding quantum numbers.   
Then the non-perturbative probability for the color-singlet state to convert to
charmonium is related to the derivative of the wavefunction of the singlet 
state in the origin. 
This model was able to give absolutely normalized production cross section, and
the ratio of different charmonium states.  However, it completely fails to
reproduce the high $p_\perp$ $J/\psi$ data by orders of magnitude. 
Formally, its problem can be seen \cite{divergence},e.g. via the infra-red
divergence of the $P$-wave charmonium cross sections.

Recently, it was proposed that in some region of the phase space color-octet
components might be dominant \cite{jpsi_octet,braaten}.  
Contrary to the color-singlet model, the short-distance part of the interaction
could also involve a color-octet $c \bar c$. 
Although the color-octet matrix elements are suppressed by powers of $v$, the
relative velocity of the heavy quark in the quarkonia, color-octet mechanism
might dominate in the processes where the leading term in $v$ is suppressed by
other small parameters such as strong coupling constant $\alpha_s$ or
$m_Q/p_{\perp}$ at large transverse momentum.
The conversion of the color-octet state to a color-singlet charmonium is
described by  non-perturbative matrix elements.  
With those color-octet matrix elements fixed by the data, the model gives good
agreement with the shape of the $p_\perp$ distribution.  
Although the magnitudes of the unknown matrix elements are consistent with the
non-relativistic QCD model (NRQCD) \cite{nrqcd}, more theoretical work is
needed.

Given the above uncertainties in the production dynamics at the $pp$ level as
well as the ability for conventional nucleon absorption and co-mover
dissociation mechanisms to explain the quarkonium data, it is important to
consider a much wider class of observables in the search for QGP formation.

\hmysubsection{Dileptons}

Dileptons are one of the direct electromagnetic probes.  Thermal dileptons, for
example, can tell us about the thermal profile of the dense plasma.  In
\cite{sx} it was proposed that in the hot-glue scenario, where the plasma is
less chemically equilibrated but hotter than in the chemically equilibrated
plasmas, dileptons of a few GeV energy will be enhanced over the standard
scenario by a factor of 2.  

However, in order to observe the interesting thermal signals, one has to deal
with a large dilepton background.  One such background is the dileptons decayed
from heavy flavor meson pairs.  
In a study by Vogt et al. \cite{vogt_charmdilepton}, it was suggested that
dileptons from open charm decay would be about an order of magnitude above the
thermal and Drell-Yan signals in the few GeV region, and therefore it would be
very difficult to subtract this large background and observe the desired plasma
signal. 
However, Shuryak \cite{loss} suggested that one important effect was missing in
the arguments of that study \cite{vogt_charmdilepton}, which was the energy
loss of the charm quark in the dense medium formed after high energy $A+A$
collisions.  Once one assumes that the charm quark would lose a finite amount
of energy along its way (typically $2$GeV/fm), then the dilepton spectrum from
open charm will be pushed to the low mass region, and dileptons with a few GeV
energy would be suppressed by over an order of magnitude.  More theoretical
work on the energy loss in a finite nucleus needs to be done before we know the
magnitude of the suppression on high mass dileptons in $A+A$ collisions.

\begin{figure}[p]
\hspace{-0.5in}
\vspace{9.0in}
\psfig{figure=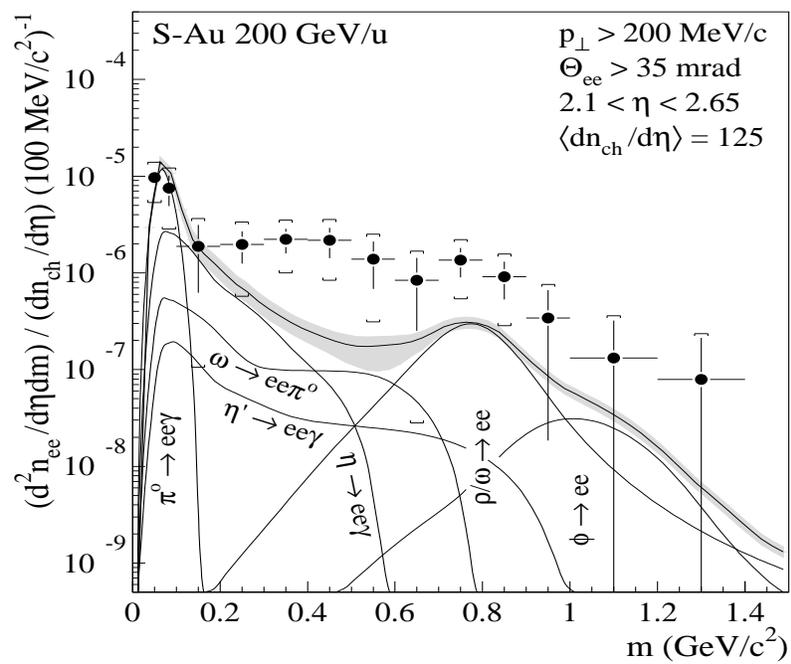,height=8.in,width=5.7in,angle=0} 
\vspace{-10.5in}
\caption{
Low mass dilepton enhancement from CERES experiment from
ref.~\protect{\cite{lowmass_95}}.
}
\label{fig:lowmass_95}
\end{figure}

Very interesting data on low mass dilepton enhancement was reported at the
Quark Matter'95 meeting \cite{lowmass_95,emprobes}. 
It was found that dileptons in the low mass region $0.2-1.5$GeV are enhanced
by a factor of 5 in $S+Au$ collisions \cite{ceres}.  
In Figure~\ref{fig:lowmass_95} the CERES $S+Au$ data is presented
\cite{lowmass_95}.    
Similar effects were also observed in $S+W$ collisions \cite{helios}. 

Dileptons from vector mesons are of interest \cite{rho,phi} because they can
provide information on the possible variation of hadron masses due to chiral
symmetry restoration.   
The low mass dilepton enhancement was also one of the focal points of the Quark
Matter'96 meeting \cite{qm96}.

\begin{figure}[p]
\psfig{figure=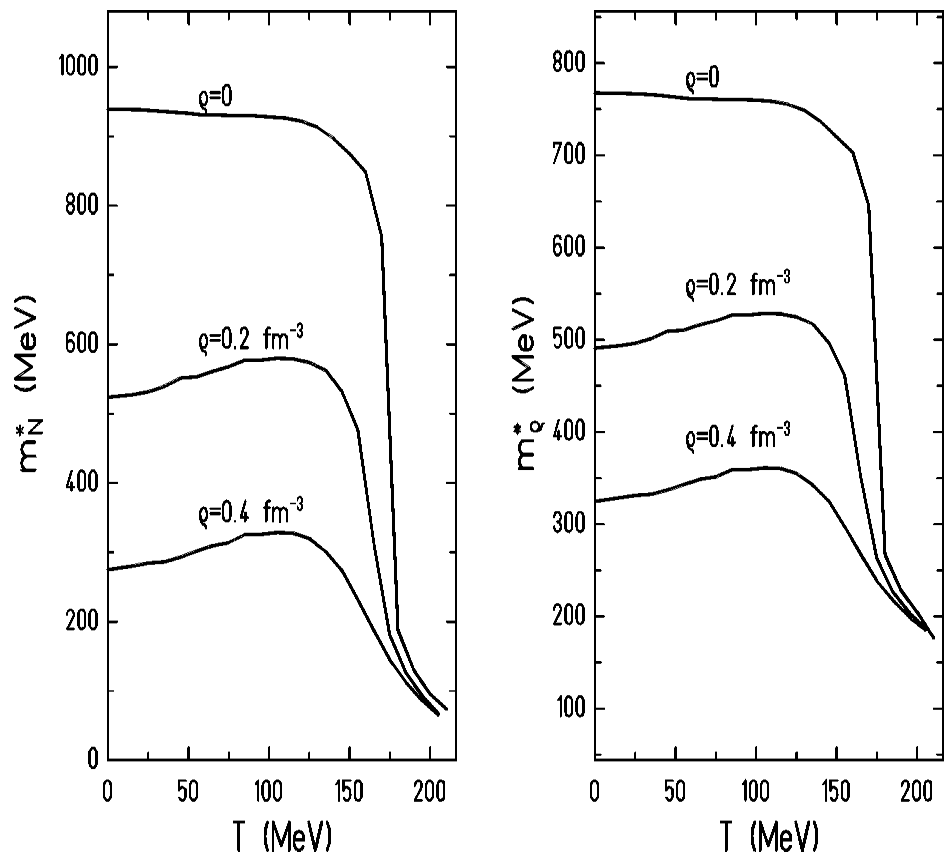,height=4.5in,width=5.in,angle=0} 
\vspace{0.5in}
\caption{
Assumption of in-medium nucleon and $\rho$ meson masses as a function of
temperature from ref.~\protect{\cite{ko_lowmass2}}.  
}
\label{fig:rhomass}
\end{figure}

It was proposed \cite{ko_lowmass,ko_lowmass2} that the CERES data \cite{ceres}
could be explained by a drop of $\rho$ meson mass in nuclear medium.
In Figure~\ref{fig:rhomass} the assumption \cite{ko_lowmass2} of in-medium mass
reductions of nucleons and $\rho$ mesons is plotted for several baryon
densities.  The $\rho$ mass was assumed to have dropped to $270$MeV at the
density $0.4$/fm$^3$ and temperature $165$MeV.  However, this assumption of 
$\rho$ mass drop is too large, and is inconsistent with QCD sum rules
\cite{hatsuda}. 

\begin{figure}[p]
\psfig{figure=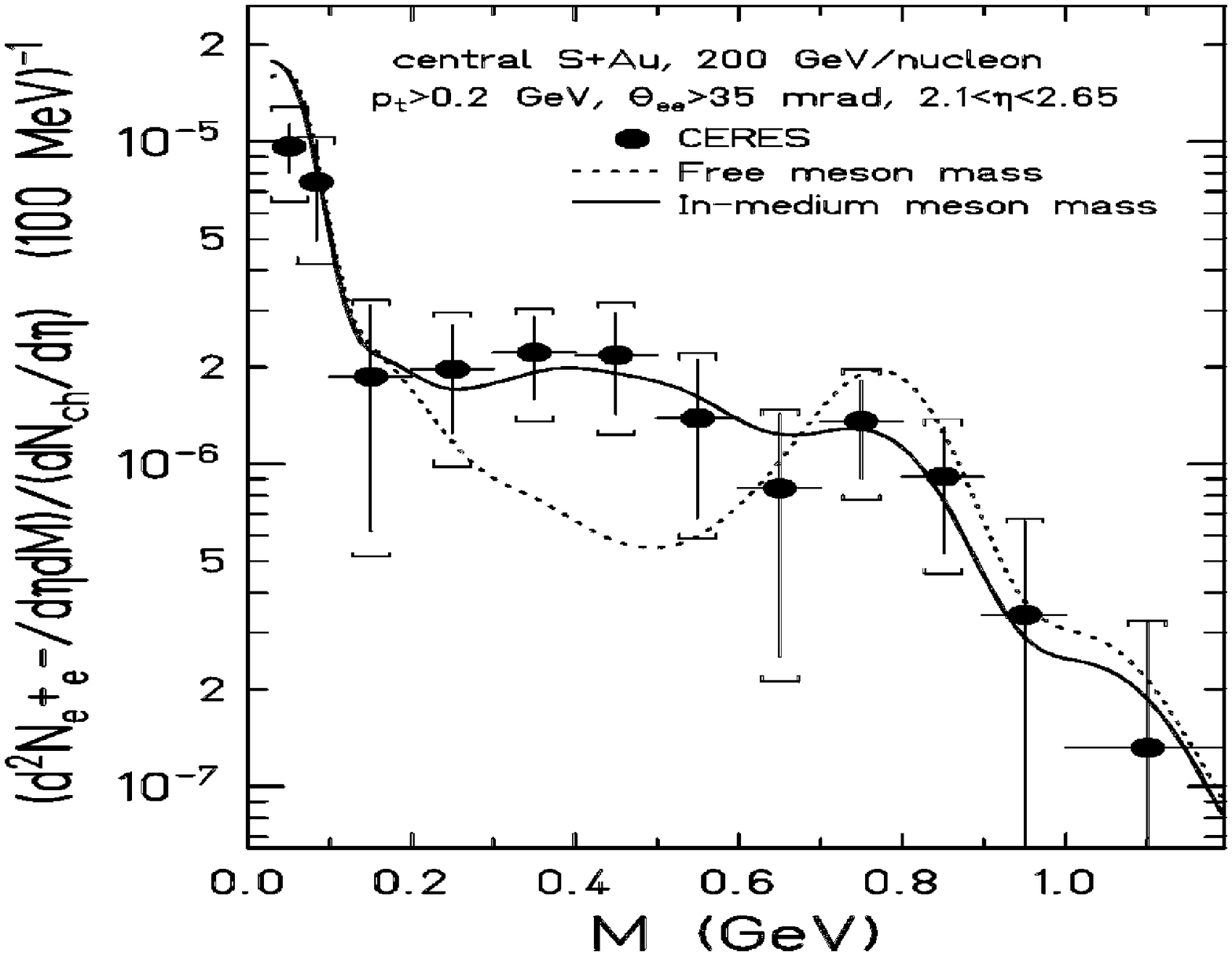,height=4.in,width=5.in,angle=0} 
\vspace{0.5in}
\caption{
CERES data on low mass dileptons vs the calculation from
ref.~\protect{\cite{ko_lowmass2}} assuming in-medium
meson masses.
}
\label{fig:lowee_medium}
\end{figure}

In Figure~\ref{fig:lowee_medium} the CERES data on low mass dilepton invariant
mass spectra is plotted vs calculations from ref.\cite{ko_lowmass2} assuming
the in-medium meson masses assumed in Figure~\ref{fig:rhomass}. 
With the significant $\rho$ mass reduction in the dense hadronic matter, the
agreement with the CERES data is greatly improved in the invariant mass region
from $2m_\pi$ to $m_\rho$. 

In calculations from another group \cite{cassing1,cassing2} a smaller $\rho$
mass dropping was assumed:
\begin{eqnarray}
m^*_\rho \simeq m^0_\rho \left ( 1-0.18 \rho_B/\rho_0 \right ) \;\; , 
\end{eqnarray}
and these calculations also explained the CERES $S+Au$ data, as well as the
HELIOS $S+W$ data \cite{helios}.

We see that the explanations of the observed low mass dilepton enhancement are
not unique.  
Different scenarios of $\rho$ meson mass shifts in medium were assumed.  
In addition, the conventional scenario of the broadening of the $\rho$ meson
spectral function in medium \cite{herrmann} could enhance the low mass
dileptons.  The present experimental error bars are still too large to
pin down different models, and therefore more precise experimental data and
further theoretical studies are needed to extract the physics from the CERES
and HELIOS data.

\hmysubsection{Direct Photons}

Direct photons are also one of the direct probes which could provide pristine
information on the quark-gluon plasma.  
In addition, since the direct photon production in nucleon-nucleon collisions 
\cite{photon_nucleon} is understood, its production in nucleus-nucleus
collisions could provide constraints on the little-known gluon structure
functions in the nucleus via the Compton process $gq \rightarrow q \gamma$.

Data on single photons in $S+Au$ collisions  \cite{wa80} was claimed to be
understood only if the quark-gluon plasma is formed, because an overly
simplified hadronic fireball model led to yields an order of magnitude
\cite{photon_qgp,photon_hydro} above the data. 
The problem with this interpretation is that the fireball model is inconsistent
with the observed rapidity and transverse momentum distributions.

A Large background is another problem with the direct photon measurements.
Compared to dileptons, the photon background, which mainly comes from $\pi_0$
and $\eta$ decays, is usually much higher \cite{emprobes}.  
Direct photon signatures are therefore rather difficult to disentangle
\cite{photon_jet}. 

\hmysubsection{Disoriented Chiral Condensate}

The restoration of chiral symmetry at high temperatures is one of the most
basic features of the QCD matter.  According to lattice QCD calculations the
critical temperature for chiral restorations coincides with that for the
deconfinement transition.
Thus the chiral phase transition in QCD is expected from both the theory study
and lattice numerical simulations \cite{chiral1,chiral2}.  A possible signal
for the chiral transition is the formation of a coherent pion field called 
the disoriented chiral condensate (DCC) \cite{early_dcc,dcc_equil,quench}.  
In equilibrium, a DCC can not grow larger than $\sim 1/T_c \sim 1$fm
\cite{dcc_equil}.  
However, in non-equilibrium it is possible to grow larger. 
A large domain of DCC will manifest itself by the striking fluctuation in the
spectrum of neutral and charged pions in terms of the ratio
$n_{\pi ^0}/n_{\pi}^{total}$.  There also exists some cosmic ray data
\cite{centauro} with that ratio far from the average value $1/3$.  

Most studies of this phenomena are based on the linear sigma model, with the
effective potential as 
\begin{eqnarray}
V_{eff}=\lambda ({\vec \pi}^2 + \sigma^2 -v^2)^2/4 - H \sigma \;\; , 
\end{eqnarray}
where $m_{\sigma} = \sqrt {2\lambda v^2}$ as the sigma mass at $T=0$ if $H=0$, 
The explicit chiral symmetry breaking term $H \sigma$ takes into account
the non-zero mass of the quarks in the real world.
In the so-called quench scenario \cite{quench}, the initial $(\sigma, \vec
\pi)$ field is assumed to be symmetric under chiral transformations
\begin{eqnarray}
\sigma \rightarrow \sigma + \vec \epsilon \cdot \vec \pi \;\; , \\[2ex]
\vec \pi \rightarrow \vec \pi - \vec \epsilon \sigma \;\; , 
\end{eqnarray}
with $<\sigma>=<\pi_i>=0$.
The evolution of the field is then described by 
zero-temperature equations of motion.  
Long wavelength modes of the pion field are amplified according to
\begin{eqnarray}
\frac {d^2}{dt^2} \vec \pi_k = (\lambda v^2 - k^2)\vec \pi_k \;\; , 
\end{eqnarray}
because the Mexican-hat form of the effective potential makes the
$\sigma=\pi_i=0$ point unstable.
In this quenched scenario the domain size could grow as large as $4-5$fm
\cite{both}.  In the annealing scenario,  $(\sigma, \vec \pi)$ mean
fields evolve almost synchronously with the effective potential, thus the
system just oscillates close to the equilibrium point.  Long wavelength modes
are less amplified and domain size is therefore smaller \cite{both}.

DCC growth depends strongly on the initial condition, and it is one main
uncertainty in present DCC studies \cite{both,anneal}.
Furthermore, most studies are based on the linear sigma model, and this model
is not dependable at the starting chiral symmetric phase where the temperature
is above $T_c$.
To verify that some pions come from the coherent source of the DCC domain, 
one can seek secondary signals such as pion pair correlations.
However, Asakawa probed that the pion pair correlations are very small and
mainly at $\theta=180^\circ$ \cite{yuki_pair}.
Theories must be improved in order to give the size of the DCC domain in heavy
ion collisions and/or the possibility that we would observe a strong pion
number imbalance.  Experiments to detect DCC formation could be 
done at RHIC, or as proposed in \cite{bj_dccexp}.

\hmysubsection{Directed Transverse Flow}

Transverse flow from azimuthal anisotropy was observed recently in $Au+Au$
collisions at $10.8$AGeV at the AGS \cite{e877_flow}.
It was first discovered in 1984 at much lower energies at Bevalac
\cite{bevalac_flow}, and has been the subject of intense studies since then.
In general, flow can provide information on the compressibility of the dense
baryonic system and on the equation of state.
It can also assist us in studying the role of the mean field in nuclear
collisions. 

The azimuthal distribution of energy and multiplicity would be isotropic if
there were no collective transverse flow.  
Transverse flow manifests itself via the anisotropy in the azimuthal
distribution on an event-by-event basis.  To study the anisotropy, one can
Fourier-transform the distribution $f(\phi)$ \cite{fourier_flow}:  
\begin{eqnarray}
f(\phi) = \sum_i f_i e^{in \phi} = \sum_i v_i e^{-i \psi_n} e^{in \phi} \;\; .
\end{eqnarray}
For real functions, the above equation can be written as
\begin{eqnarray}
f(\phi)=\sum_i v_i \left [ \cos (n\psi_n)\cos (n\phi)+\sin (n\psi_n) \sin
(n\phi) \right ]  \;\; .
\end{eqnarray}
Anisotropy gives non-zero values of $v_n$ ($n \geq 1$). 
Variables $v_1$ and $\psi_1$ for the first harmonic represent a shift of the
center of the isotropy circle, with $\psi_1$ being the azimuthal angle of the
reaction plane, which is the plane formed by the beam axis and the impact
parameter vector.  The variable $v_2$ represents the eccentricity of the 
ellipse in the reaction plane. 

The above analysis is complicated by the fluctuation resulting from finite
multiplicity in a nuclear collision \cite{multiplicity}.  
Therefore the reaction plane can not be exactly determined event by event, and
the values of $\psi_n$ suffer from this uncertainty.
More importantly, the above variables for the harmonics for an ensemble of
events are described not by a constant, but by a Gaussian distribution with a
finite width.  

The two-dimensional Gaussian would be centered at zero if there were no
transverse flow: 
\begin{eqnarray}
\frac {d^2 w}{v_n dv_n d\psi_n}=\frac{1}{2\pi\sigma_n^2}
e^ {\frac {-v_n^2}{2\sigma_n^2}} \;\; .
\end{eqnarray}
Assuming that the fluctuation changes little with transverse flow, for
\begin{eqnarray}
f(\phi)=\sum_i f_i \delta (\phi - \phi_i)
\end{eqnarray}
we have
\begin{eqnarray}
v_n^2 &=& \frac {1}{\pi^2} <\sum_i f_i^2> =\frac {N}{\pi^2} <f_1^2> \;\; , \\
\Rightarrow \sigma_n &=& \frac {N}{\pi^2} \sigma_1 \;\; . 
\end{eqnarray}
Note that $\sigma_n$ is independent of $n$, and we now denote it as $\sigma$.

With transverse flow, the Gaussian is centered at non-zero $(\tilde
{v_n},\tilde {\psi_n})$:
\begin{eqnarray}
\frac {d^2 w}{v_n dv_n d\psi_n}
&=&\frac{1}{2\pi\sigma^2}
e^ {-\frac {(\vec v_n - \tilde {\vec v_n})^2}{2\sigma^2}}
=\frac{1}{2\pi\sigma^2}
e^ {-\frac {v_n^2 +{\tilde {v_n}}^2 - 2v_n \tilde {v_n}\cos (\psi_n-\tilde
{\psi_n})} {2\sigma^2}}, \\[2ex]
\Rightarrow 
\frac {d^2 w}{v_n dv_n} &=& \frac{1}{\sigma^2} 
e^ {-\frac {v_n^2 +{\tilde {v_n}}^2}{2\sigma^2}} 
I_0(\frac {v_n \tilde {v_n}}{\sigma^2}) \;\; , 
\end{eqnarray}
where $I_0(x)$ is the modified Bessel function.  

Redefining the dimensionless variable as
\begin{eqnarray}
x_n \equiv \frac {v_n}{\sigma} \;\; ,
\tilde{x_n} \equiv \frac {\tilde {v_n}}{\sigma} \;\; ,
\end{eqnarray}
the new distribution becomes
\begin{eqnarray}
\frac {d^2 w}{x_n dx_n} = \frac{1}{\sigma^2} 
e^ {-(x_n^2 +{\tilde x_n}^2)} I_0(x_n \tilde x_n) \;\; .
\end{eqnarray}
The above distribution $\frac {d^2 w}{x_n dx_n}$ resembles a Gaussian function
when $\tilde {x_n}<1$ and then is hard to be distinguished from the finite
multiplicity fluctuation.  However, when the transverse flow is strong enough
so that $\tilde {x_n}>\sqrt 2$, it has a minimum at $x_n=0$, which is a
distinct signature for the transverse flow. 

From the experiment E877 at AGS \cite{hemmick}, clear transverse flow signal
is observed.  In mid-central events, a clear minimum in the $\frac {d^2 w}{v_1
dv_1}$ distribution is seen.  The $v_1$ coefficients extracted from transverse
energy anisotropy and multiplicity anisotropy are quite different, and the
reason is that nucleons and pions have different flow patterns.  For protons,
transverse flow grows linearly as a function of $p_T$, and reaches 20\% of the
total transverse momentum at $p_T=1$GeV.  From the comparison with different
transport models such as RQMD \cite{rqmd_flow} and ARC \cite{arc_flow}, one can
study the effect of the mean field in the nuclear collisions. 
The most striking collective flow signature of a QGP formation has been
predicted to be a local minimum of the flow excitation function at the
so-called ``softest point'' of the nuclear matter \cite{dirk}.  This proposed
signature is being looked for at AGS.

\hmysubsection{Hanbury-Brown Twiss}

Correlation functions for identical particles probe the source size and
freeze-out time of the dense medium \cite{hbt_gkw,hbt_pratt}. 
The two-particle correlation function in momentum space is defined as
\begin{eqnarray}
C_2(\vec p_1,\vec p_2) \equiv \frac {P(\vec p_1,\vec p_2)} 
{P(\vec p_1) P(\vec p_2)} \;\; . 
\end{eqnarray}
If the particles are classical without dynamical correlations, the above
function is always equal to $1$.  For identical quanta with bosonic statistics 
\cite{hbt}, 
symmetrization introduces a dependence on two momenta and $C_2(\vec p_1,\vec
p_2)=2$ without dynamical correlations. 
More generally, for a source described by seven-dimensional phase space density
function $S(x,\vec p)$, the semi-classical expression for the correlation
function is given by \cite{hbt_pratt,padula}
\begin{eqnarray}
C_2(\vec K,\vec q)=1+\frac {\int d^4x d^4{x^\prime} S(x,\vec K) S(x^\prime,\vec
K) e^{iq(x-x^\prime)}}{\int d^4x S(x,\vec p_1) \int d^4{x^\prime}
S(x^\prime,\vec p_2)}  \;\; , 
\end{eqnarray}
where $\vec K=(\vec p_1+\vec p_2)/2$, $q=(E_1-E_1,\vec p_1-\vec p_2)$, and
$\int d^4x S(x,\vec p) \equiv P(\vec p)$.  

For an azimuthally symmetric source \cite{hbt_cnh}
\begin{eqnarray}
S(x,\vec p) \propto \exp \left [-\frac {x^2+y^2} {2 R_\perp^2}
- \frac {z^2} {2 R_z^2} - \frac {(t-t_0)^2} {2 \Delta^2} \right ] \;\; , 
\end{eqnarray}
using
\begin{eqnarray}
q_0=E_1-E_2=\frac {{\vec p_1}^2 - {\vec p_2}^2} {E_1+E_2}
\simeq \frac {\vec K \cdot \vec q}{E_K} \equiv \vec \beta \cdot \vec q
=\beta_{out} q_{out} + \beta_{long} q_{long} \;\; , 
\end{eqnarray}
where $q_{long}$ is defined to be the component of $\vec q$ along the beam
direction, $q_{out}$ is the transverse component which is in the plane with the
beam direction and the $K$ vector, and $q_{side}$ is the other orthogonal
transverse direction.
The approximation sign is valid for $|\vec q| \ll |\vec K|$, and $\beta_{side}
\propto K_{side} = 0$ by definition.  Thus 
\begin{eqnarray}
C_2(\vec K,\vec q) &=& 1+\exp \left [-R_\perp^2 (q_x^2+ q_y^2) - 
R_z^2 q_z^2 - \Delta ^2 q_0^2 \right ] \\[2ex]
\label{EQ:hbt_1}
& \simeq & 1+\exp \left [ -q_{side}^2 R_\perp^2 
-q_{out}^2 (R_\perp^2 + \beta_{out}^2 \Delta^2) \right . \nonumber \\
&& \left . -q_{long}^2 (R_z^2 + \beta_{long}^2 \Delta^2)
-2 \beta_{out} \beta_{long} q_{out} q_{long} \Delta^2 \right ] .
\label{EQ:hbt_2}
\end{eqnarray}

In the longitudinally co-moving system (LCMS) where $K_{long} = 0$, it becomes
\begin{eqnarray}
C_2(\vec K,\vec q)
= 1+\exp { \left [ -q_{side}^2 R_\perp^2 
-q_{out}^2 (R_\perp^2 + \beta_{out}^2 \Delta^2)
-q_{long}^2 R_z^2 \right ]}.
\label{EQ:hbt_3}
\end{eqnarray}

Experimentally one usually uses the parameterization \cite{hbt_q}
\begin{eqnarray}
C_2(\vec K,\vec q) = 1+\lambda \exp \left [-q_{side}^2 R_{side}^2 - q_{out}^2
R_{out}^2 - q_{long}^2 R_{long}^2 \right ] \;\; . 
\label{EQ:hbt_4}
\end{eqnarray}
Thus one can obtain information about the space-time dimension and expansion
velocity of the emitting source from the fitted radius parameters. 
The chaoticity parameter $\lambda$ is different from $1$ due to other
correlations and resonance decays.  For charged particle correlations such as
pion and kaon correlations, the Gamov correction due to Coulomb interactions is
important and must be included.
Data from $p+Pb$, $S+Pb$ and $Pb+Pb$ collisions \cite{hbt_nadata}
show that the fitted radii grow with the size of the system, and are 
generally larger than the projectile size.  Pions have larger radii and smaller
chaoticity than kaons due to more resonance decays. 
Because of collective flow phenomena associated with transverse expansion, the
above simple formulae (\ref{EQ:hbt_1}-\ref{EQ:hbt_4}) are modified.  The fitted
radii do not correspond directly to the source size but must be interpreted in
terms of a detailed transport model.  Pion interferometry can nevertheless
serve as a powerful signal of the QGP formation.  Given a rapid crossover
region around $T_c$, the speed of sound $C_s^2$ acquires a local minimum near
$T_c$.  As matter cools and crosses that mixed phase region, the minimum of
$C_s^2$ causes the transverse expansion to stall for some time.  This leads to
the time delay signature of the phase transition that could be seen in 
detailed systematics of the $q_{out},q_{side}$ dependences of the function
$C_2(q_{out},q_{side})$ \cite{dirk}.

\hmysection{Open Charm}
\label{sec-goal}

One of the main uncertainties in heavy ion physics is the initial conditions
of the dense parton system.  As shown in section\ref{subsec-chem}, for example,
the initial momentum distribution, energy density and chemical composition
determine the fate of the QGP.  
If the assumed initial conditions in eq.~(\ref{EQ:initial_iso}) \cite{biro}
are true, we may not have fully equilibrated quark-gluon plasma at RHIC.  
Recently it is found that minijets lead to initial conditions characterized
by large fluctuations of the local energy density and of the collective flow
field.  The ``hot spots'' \cite{hotspots} corresponding to higher local energy
density may change the usual picture from cylindrically symmetric, homogeneous
quark-gluon plasma to turbulent, volcanic plasma.  Certain signals of QGP
may be sensitive to this change of the initial condition.
Initial conditions are themselves determined by the parton structure
functions from the two colliding nuclei, mostly by the gluon structure
function.  Parton distributions in nuclei are depleted due to nuclear
shadowing effects \cite{emc,muellerq}.   
As we go to higher energies in search of the quark-gluon plasma, we can see
softer (small-x) partons deeper in the shadowing regime.
It is therefore important to understand nuclear parton distributions in order
to calculate reliably all the other observables in $A+A$ collisions.

Open charm production is well suited to measure the initial gluon structure
function and its nuclear modifications.
At collider energies $\sqrt{s}>200$ AGeV the initial minijet plasma is mostly
gluonic \cite{hotglue,eskola1} with a quark content far below its chemical
equilibrium value.  
Because open charm is produced mainly through gluon fusion, it can provide
information on both the incoming gluon structure functions and the later-formed
dense minijet plasma.  

This thesis on open charm production \cite{lm_charm} in high energy nuclear
collisions is motivated by two recent studies \cite{mw,geiger} which predicted
widely different rates.
In \cite{mw} the pre-equilibrium contribution was found to be almost equal
to the initial gluon fusion rate, similar to the suggestion of the charm
enhancement from thermal productions in the hot-glue scenario \cite{hotglue}.
In \cite{geiger}, a more provocative claim was made that open charm may
even be enhanced by over an order of magnitude above the initial perturbative
QCD (pQCD) rate. 

The proposed method to measure charm meson ($D/\bar D$) cross section at the
Relativistic Heavy ion Collider (RHIC) is $e\mu$ coincidence measurements, or
dilepton measurements. 
Traditionally, dileptons from charm decay are regarded as background, while
thermal dileptons or Drell-Yan dileptons are considered as signals. 
Dileptons from Drell-Yan process probe the quark degrees of freedom, and
thermal dileptons tell us about characteristics such as temperature and phase
transition in the dense plasma.   
However, those signals will probably be overshadowed by the large background of
dileptons from charm meson ($D/\bar D$) decay in $Au+Au$ collisions at
RHIC \cite{vogt_charmdilepton}, and it is easier to measure the dileptons from
open charm decay.
In the QCD factorization limit, open charm mainly comes from initial gluon
fusion, and it thus depends strongly on the gluon distribution in the nuclei.

The thesis discussed at length in the next chapters consists of two parts.
First, it is shown that the pre-equilibrium charm production is small compared
to the initial fusion rate \cite{lm_charm,lmw}.  
Second, it is shown that this dominance of the initial gluon fusion mechanism
can be used to probe the gluon shadowing in nuclei \cite{lm_dilepton}.   
While the first result of the thesis work is pessimistic in the sense that it
indicates, contrary to an earlier claim \cite{geiger}, that open charm is not a
good probe of the QGP formation, the second result is optimistic as it provides
a new tool to measure the gluon shadowing via $p+A$ reactions.

Perturbative QCD analysis \cite{muellerq,eskolaq} predicts that the gluon
structure will be depleted at $x<0.01$ due to gluon recombination processes.  
Shadowing of the quark nuclear structure functions, $q_A(x,Q^2)$, is well
established from deep inelastic $\ell A\rightarrow \ell X$ reactions
\cite{emc}.  For heavy nuclei, the quark structure functions are shadowed by a
factor, $R_{q/A}(x\ll 0.1,Q^2)\approx 1.1 - 0.1 \;A^{1/3}$, nearly independent
of $Q^2$.  It is far less known about the gluon shadowing.  In high energy
nuclear collisions, the gluon nuclear structure is of central importance since
it controls the rate of minijet production that determines the total entropy
produced at RHIC and higher energies \cite{hijingprl}.
The understanding of the gluon shadowing is therefore of fundamental
interest in this field.

In Chapter~\ref{sec-initial} the dependence of the direct pQCD rates for charm
production on structure functions, momentum scale $Q^2$ is reviewed, and we
compare our results with a previous one \cite{geiger}.
Chapter~\ref{sec-pre} presents the results of part one of this thesis on the
pre-equilibrium charm production.  
The important role of space-time and momentum correlations 
in suppressing that component of the charm production is emphasized.
Chapter~\ref{sec-ratio} contains the second part of this thesis.
The dilepton spectra from open charm decay in $p+Au$ collision at 200 GeV/A are
calculated and its sensitivity to the nuclear shadowing is quantitatively
estimated.  We pay particular attention to possible backgrounds of the dilepton
spectra for the PHENIX detector geometry at RHIC.
Discussion and summary are given  at the end of each chapter.
Chapter~\ref{sec-outlook} presents an outlook of some open questions on
charm productions.

%% file: sec4-initial.tex
\hmychapter{Initial Charm Production}
\label{sec-initial}

In this chapter we will study how the direct pQCD rates of charm production
depend on structure functions, momentum scale $Q^2$, and $K$-factor. 
We will then compare our results with results from the Parton Cascade Model
\cite{geiger}. 

\hmysection{Production Mechanisms}

Heavy quark production in $pp$ reactions was calculated long ago in
\cite{combridge} including both fusion and heavy flavor excitation processes in
the leading order pQCD.   
It was proposed that the flavor excitation processes were dominant at high
energies because a small $Q^2$ exchange can easily liberate any charm component
in the nucleon while gluon fusion was suppressed because $Q^2 \geq 4M_c^2$.  
In the Parton Cascade Model \cite{geiger}, both mechanisms are incorporated to
calculate $s,c,b$ quark production in nuclear collisions.  Results from this
model suggested that the flavor excitation of the charm quark of nuclear
structure functions would be the dominant source of charm production in nuclear
collisions as well. 

However, it is pointed out \cite{collins} that the original flavor excitation
rates in \cite{combridge} were too high in the $x_f\sim 0$ region due to
neglected interference with other pQCD amplitudes of the same order.  
When all diagrams were added together, a large destructive interference was
found to suppress the flavor excitation rates by powers of $\Lambda/M_q$, where
$\Lambda \sim 300MeV$ is a typical QCD scale, and $M_q$ is the heavy quark
mass.  The suppression factor appears in the process $g + c(\bar c)$ where the
charm is evolved from the structure functions using {\em perturbative} QCD, as
also shown in \cite{ellis}.   

We note that there is a possible non-perturbative charm component (intrinsic
charm) in the nucleon.  However, there are experimental constraints on the
amount of this non-perturbative charm component \cite{hoffmann,ingelman}. 
The total contribution of the intrinsic charm was shown in
 \cite{vogt_delta,brodsky} to be small (about 10\%) in the midrapidity
region where most of the charm is made. Although the contribution of the
intrinsic charm component appears important at large $x_f$, its contribution to
the total cross section is small and well within the uncertainties from other
sources which we will discuss.  

\hmysection{Cross Section}

We therefore only include fusion processes for the parton level cross sections
as in \cite{mw}.  
The cross sections are the following \cite{combridge}:
\begin{eqnarray} 
d\sigma=d x_1 d x_2 f_1(x_1,Q^2) f_2(x_2,Q^2) \hat \sigma(\hat s) \;\; , 
\end{eqnarray} 
where $\hat s=x_1 x_2 s$.  

Let $x_L=x_1-x_2$, and $F=xf$ as the fractional-momentum parton
distribution function, then  
\begin{eqnarray}
{d\sigma}
& = &d\hat s dx_L \frac{F_1(x_1,Q^2)F_2(x_2,Q^2)\hat\sigma(\hat s)} 
{x_1 x_2 (x_1+x_2) s} \nonumber \\
& = &d\hat s dx_L \frac{F_1(x_1,Q^2)F_2(x_2,Q^2)\hat\sigma(\hat s)} 
{\hat s \sqrt{x_L^2+4 \hat s/s}} \;\; , 
\label{EQ:initial}
\end{eqnarray}
where 
\begin{eqnarray}
x_1 = (x_L+\sqrt{x_L^2+4 \hat s/s})/2,
x_2 = (-x_L+\sqrt{x_L^2+4 \hat s/s})/2 \;\; . 
\end{eqnarray}

By assuming that $Q^2$ only depends on $\hat s$, we can explicitly obtain the
pQCD differential cross sections for $a+b \rightarrow c\bar c + X$ as the
following \cite{combridge}.
\begin{eqnarray}
\hat \sigma_{q\bar q \rightarrow c\bar c}
=\frac{8 \pi \alpha _s^2(Q^2) }{27 \hat s^2} (\hat s + 2 M_c^2)~\chi \;\; , 
\end{eqnarray}
\begin{eqnarray}
\hat \sigma_{g g \rightarrow c\bar c}
=\frac{\pi \alpha _s^2(Q^2)}{3 \hat s} \left[ - (7 + \frac{31 M_c^2}{\hat s})
\frac{1}{4} \chi + (1 + \frac{4 M_c^2}{\hat s} + \frac{M_c^4}{\hat s^2})
\log{\frac{1 + \chi}{1 - \chi}} \right] ,
\label{EQ:ggfusion}
\end{eqnarray}
where $\chi=\sqrt{1 - 4 M_c^2/\hat s}$.

\hmysection{Parton Structure Functions, Momentum Scale, and K-Factor}

\begin{figure}[p]
\psfig{figure=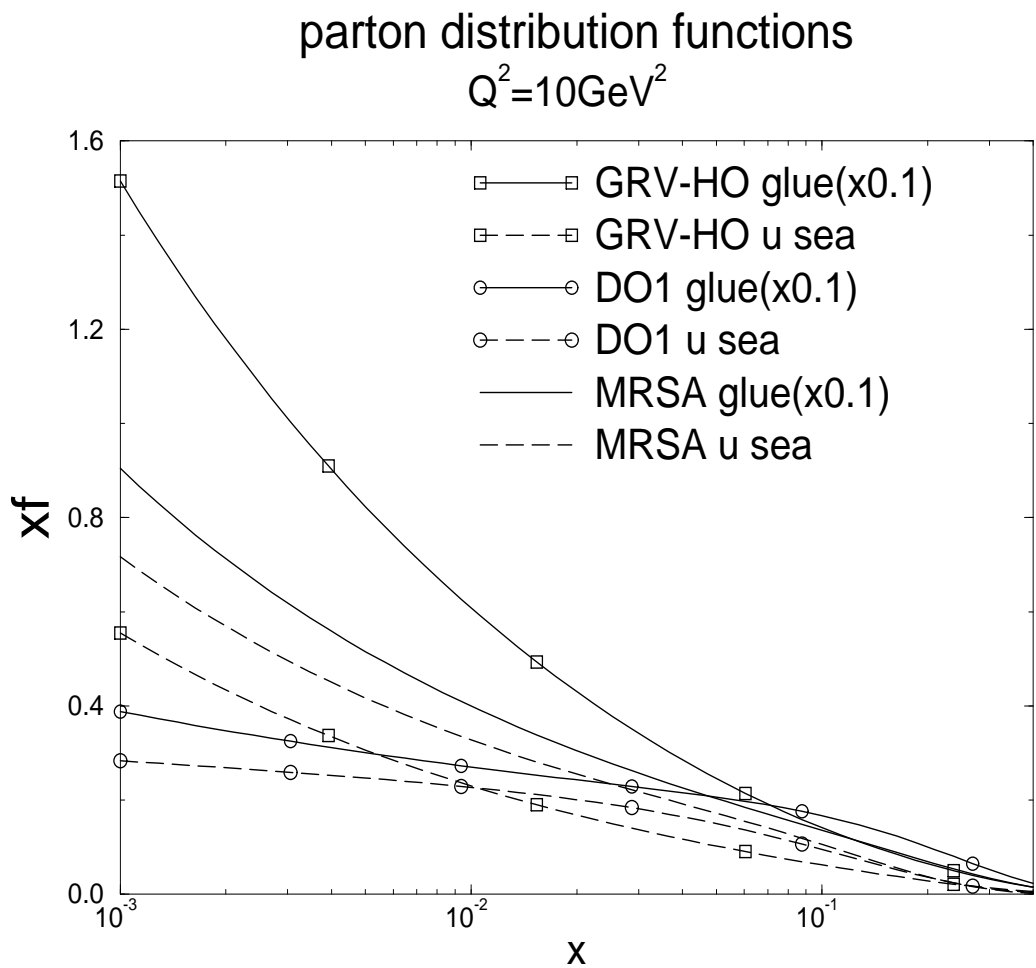,height=5.in,width=3.5in,angle=-90} 
\caption{
Parton momentum distribution functions for gluons and u sea-quarks as a
function of Bjorken $x$ for GRV-HO, DO1 and MRSA parametrizations.  
}
\label{fig:parton}
\end{figure}

For the production in $pp$ collisions, we use the light quark and gluon
structure functions from Gl\"uck et al. \cite{grv} and Duke-Owens \cite{do} for
comparison.
In Figure~\ref{fig:parton} we compare the parton structure functions from these
two parameterizations together with MRSA \cite{mrsa} parameterization.  The
MRSA set of parton distributions is very close to the latest parameterization
CTEQ4M  \cite{cteq}.

We consider the following two choices for the scale $Q^2$ in the coupling
constant $\alpha_s(Q^2)=12 \pi/ \left[ (33 - 2 n_f) \log(Q^2/\Lambda ^2)
\right]$.  The $Q^2$ scale in the parton distribution functions is taken to be
the same.

1. for $g g \rightarrow c\bar c, Q^2 = \hat s/2$;  for $q\bar q \rightarrow 
   c\bar c, Q^2 = \hat s$.  ($Q^2$ choice-1)

2. for both $g g \rightarrow c\bar c$ \& $q\bar q \rightarrow c\bar c, Q^2 =
   \hat s$.  ($Q^2$ choice-2)

We take  $n_f=4$ for charm quark production and $n_f=5$ for bottom quark
production. The QCD scale $\Lambda$ depends on the choice of parton
distribution functions and is given in table~\ref{tb:lambda}.

\begin{table}
\hspace{3cm}
\begin{tabular}{|l|c|} \hline
Parton distribution functions    & $\Lambda$(GeV) \\ \hline
GRV-LO set                   & 0.25 \\
GRV-HO set                   & 0.20 \\
Duke-Owens set 1(DO1)            & 0.20 \\
Duke-Owens set 2(DO2)            & 0.40 \\ \hline
\end{tabular}
\caption{Lambda scales in different parton distribution functions}
\label{tb:lambda}
\end{table}

To incorporate approximately the next-to-leading-order(NLO) corrections to the
above rates, we multiply the leading order results by a K-factor. In general,
the K-factor depends on the choice of parton distribution functions, the center
of mass energy of the collision, and the type of the projectile and target
particles. 
Calculations to order $O (\alpha_s^3)$ for the subprocesses were carried
out \cite{nason,beenakker}, and afterwards the calculations to order $O
(\alpha_s^3)$ for $p+p$ collisions were made \cite{berger,bm}.  For DO1,
$M_c=1.5 $ GeV, $Q^2=4M_c^2, P_{lab} = 100 - 1000$ GeV, the K-factor for $p+p$
collisions \cite{berger} was found to range from $2.85$ to $4.1$.  
There are other NLO calculations of charm production cross sections
\cite{sv,vogt_kfactor}, which reveal the dependence of the NLO K-factor on the
transverse momentum and rapidity of the initially produced charm. 

\hmysection{Comparison with Low-Energy Experimental $pp$ Data}

\begin{figure}[p]
\psfig{figure=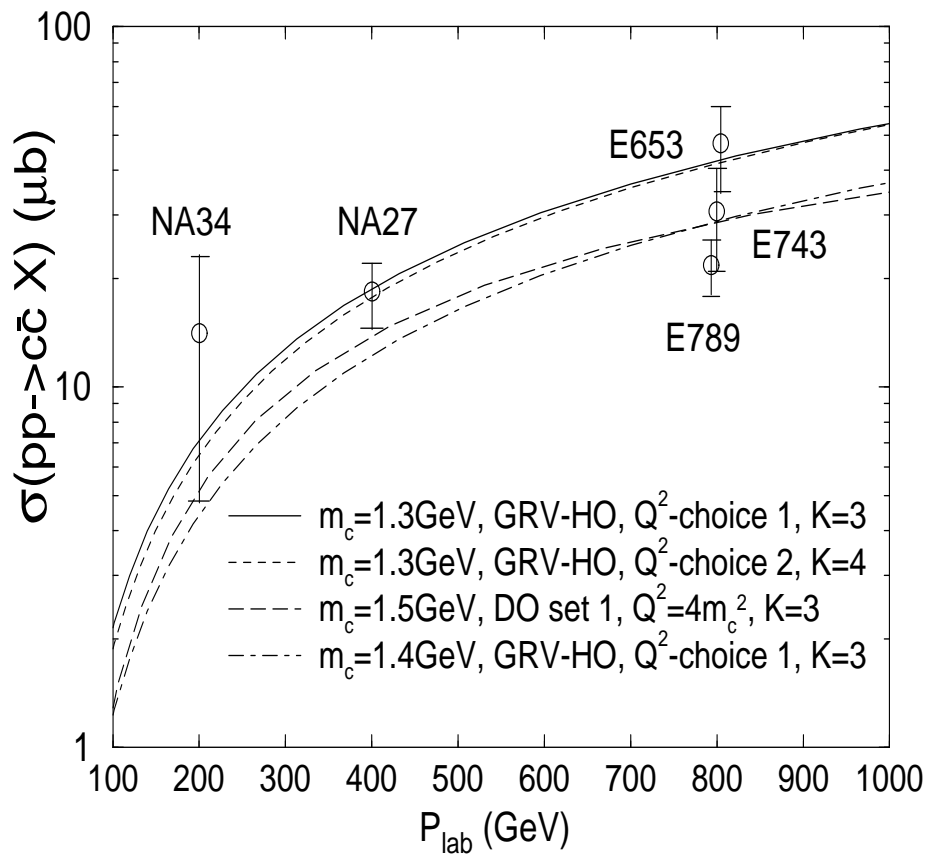,height=5.in,width=3.5in,angle=-90} 
\caption{
Cross sections for $pp \rightarrow c\bar c X$ as a function of $P_{lab}$.  
}
\label{fig:charm_exp}
\end{figure}

In Figure~\ref{fig:charm_exp} we compare the so-calculated charm cross section
to the limited data on inclusive $c\bar c$ production  in $pp$ collisions. 

The NA34 data for $\sigma _{charm}$ is taken directly from \cite{na34}.
The values for the other data lines are computed from D-meson cross sections 
according to the argument in \cite{goshaw} by using the published experiment 
results \cite{na27,e743,e653,e789}.  
By assuming that $D_s/\bar {D_s}$ production accounts for 8\% of the total
cross section, hidden charm accounts for 1\%, and $\Lambda_c \bar D/\bar
\Lambda_c D$ production is smaller than 6.1$\mu b$ at NA27, the conversion is
done as the following: 
\begin{eqnarray}
\sigma(pp\rightarrow c \bar cX) \sim \frac{\sigma(\bar
D/D)}{2(0.92\pm0.04)0.99} \left( 1+\frac{a(\leq 6.1)}{30.2\pm2.2} \right) 
\;\; .  
\end{eqnarray}
We then take $a=6.1 (0.)$ as the upper(lower) limit.  
Including the already-converted NA34/2 data, we have the following 
cross sections for $c\bar c$ in table~\ref{tb:ccbar}.

\begin{table}
\begin{tabular}{|r|c|c|c|} \hline 
Experiment & $P_{lab}(GeV)$ &published original data($\mu b$) &converted
$\sigma_{c\bar c}$($\mu b$) \\ \hline 
NA27 & 400 & $\Rightarrow \sigma(D/\bar D)=30.2 \pm 2.2 $ & $14.7-22.1$ \\
\hline 
E743 & 800 & $\Rightarrow \sigma(D/\bar D)=48^{+10}_{-8}$ & $21.0-40.5$ \\
\hline 
NA34/2 & 200 & & $14.1 \pm 9.3 $ \\ \hline
E653 & 800 & $\Rightarrow \sigma(D/\bar D)=76 \pm 10 $ & $34.7-60.1$ \\ \hline
E789 & 800 & $\Rightarrow \sigma(D/\bar D)\sim35.4 \pm 1.3 $ & $17.9-25.7$ \\
\hline 
\end{tabular}
\caption{$c\bar c$ cross sections from low energy charmed meson data}
\label{tb:ccbar}
\end{table}

Only statistical error is included in the above calculations.
Earlier experiment results \cite{tavernier} show larger uncertainties among
different experiments.     

In Figure~\ref{fig:charm_exp}, the solid line is our result with $M_c = 1.3$
GeV, $K=3$, $Q^2$ choice-1 and GRV-HO set.  
As a consistency check, we also plot the long-dashed curve using the same
parameters as in Figure~\ref{fig:charm_exp} of \cite{berger} (i.e. DO1,
$M_c=1.5 $GeV, $Q^2=4 M_c^2$) using constant $K=3$ for simplicity.  
Both the solid curve and the dashed curve fit the low energy data
reasonably well, so we use these two parametrizations for the following high
energy calculations in this chapter. 
Comparison of the solid and dot-dashed curves shows the strong dependence on
the assumed charm quark mass for the GRV-HO set. Comparing the solid and
dashed curve we see that different choices for the $Q^2$ scale can be
compensated for by shifts in the $K$ factor. These results, together with the
large uncertainty of data, emphasize the need to measure $pp$ and $pA$
reactions in order to control uncertainties in the initial charm production
rate, so that charm productions in $AA$ can be properly calculated. 

\begin{figure}[p]
\psfig{figure=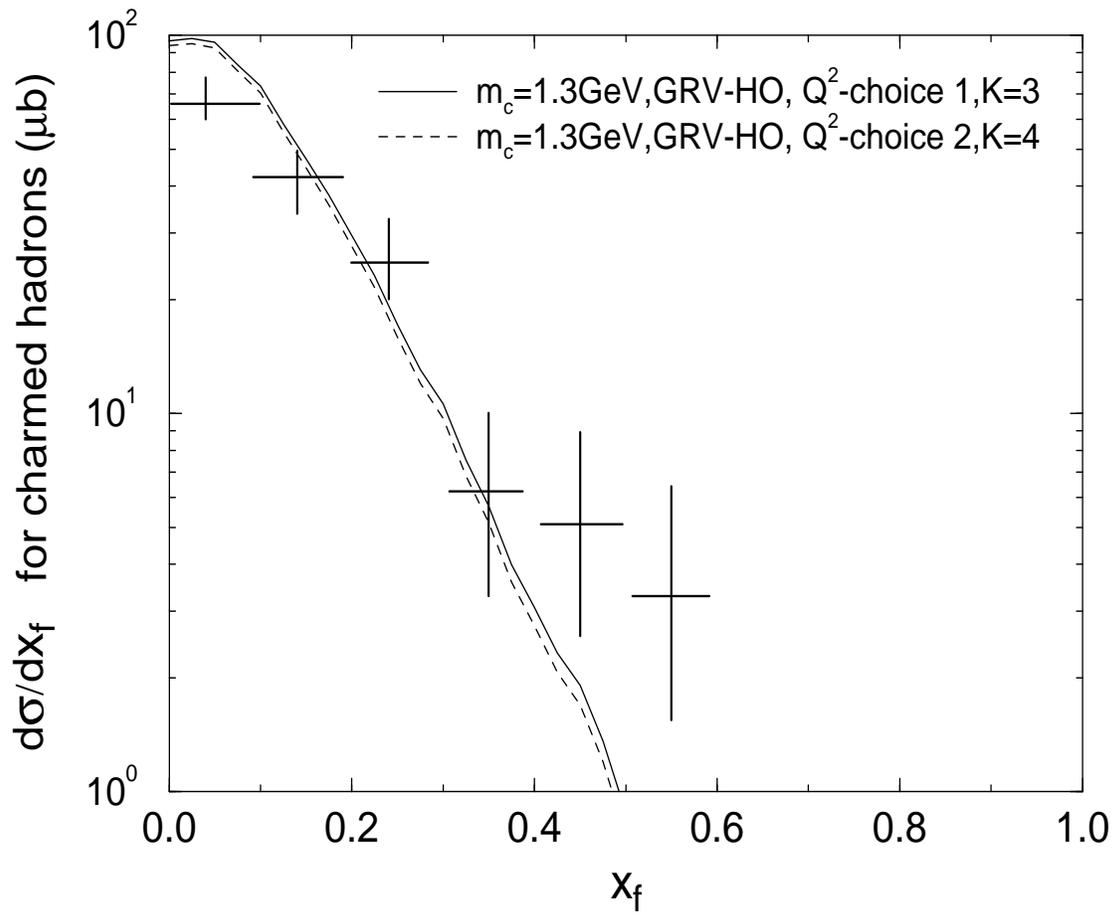,height=5.in,width=3.5in,angle=-90} 
\caption{
The production of charmed hadrons as a function of $x_f$ for $pp$ collisions
at $P_{lab}=400$ GeV.
}
\label{fig:charm_xf}
\end{figure}

\begin{figure}[p]
\psfig{figure=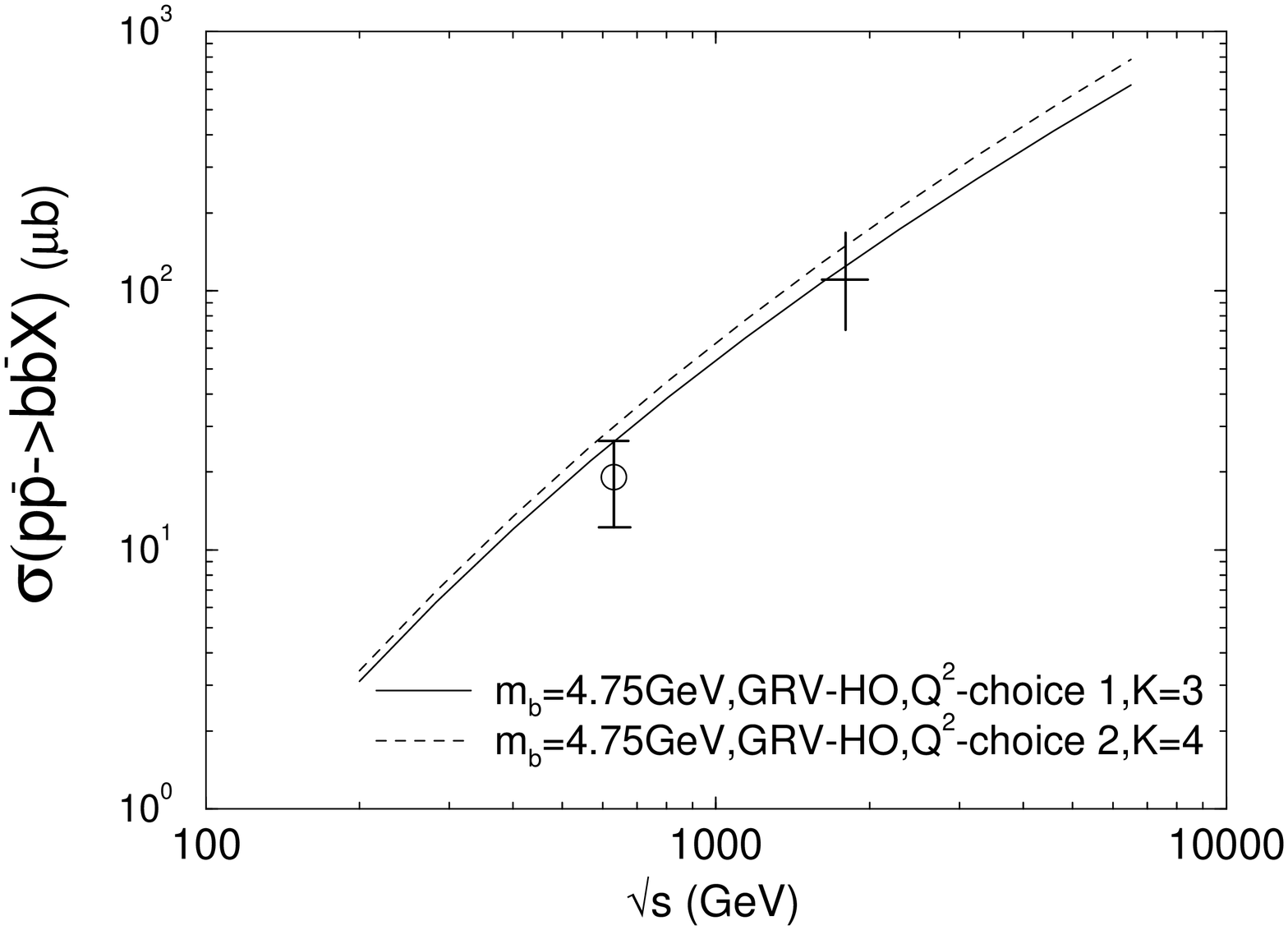,height=5.in,width=3.5in,angle=-90} 
\caption{
Cross sections for $p\bar p \rightarrow b\bar b+X$ vs $\protect \sqrt S$.
}
\label{fig:bottom}
\end{figure}

As a further check of the parameters we compare charmed hadron $x_f$ results in
Figure~\ref{fig:charm_xf} with 400 GeV $pp$ data \cite{na27}.
The solid curve represents the result $d\sigma/d x_f$ using the first
parameterization. The dashed curve represents the result using the second
parameterization.  Both curves assume a $\delta$ function charm fragmentation
function.  The realistic fragmentation function used in \cite{vogt_delta}
lowers the curves slightly and reveals the true high-$x_f$ intrinsic charm
component. 
In Figure~\ref{fig:bottom} we compare $b\bar{b}$ production.   
Here we take $M_b=4.75$ GeV as in \cite{bm}, with $K=3,n_f=5$.
The data point at $\sqrt S =630 $GeV is from \cite{ua1}: $\sigma(p\bar p
\rightarrow b + X)=19.3\pm7(exp.)\pm9(th.) \mu b$, and only the experimental
error is indicated in Figure~\ref{fig:bottom}.  
The dashed cross at $\sqrt S=1.8$ Tev is obtained indirectly from  \cite{bm},
and the error bar is only illustrative.   
At $\sqrt S =1.8 TeV$, our value is  $41.8 \mu b \times K = 125 \mu b$. This is
significantly larger than found in \cite{bm}.    


\hmysection{The A-scaling of Charm Productions}

The dependence of cross sections of charm productions on atomic number A in
$p+A$ collisions is usually parameterized as $\sigma^{pA}_D=\sigma^{pp}_D
A^\alpha$.  Earlier experiments gave $\alpha\simeq 0.76$ \cite{smallalpha}.
However, those experiments did not explicitly reconstruct $D$ mesons and had
lower statistics.  Later experiments \cite{alphaone,e789} reconstructed $D$
mesons and found $\alpha = 1.02 \pm 0.03 \pm 0.02$ \cite{e789}, which is
consistent with $\alpha=1.$ from the Glauber geometry.  
We note that the value $\alpha=0.76$ is taken from low energy experiments,
where energy conservation suppresses the contribution from multiple
collisions. At high energies, QCD factorization implies that $\alpha=1$ for
$p+A$ scaling is the appropriate scaling modulo small nuclear shadowing
effects. 
The nuclear shadowing effect on the A-scaling was estimated by Qiu
\cite{qiucharm}, and he found that the nuclear shadowing could only lead to a
small reduction in $\alpha$ ($\alpha \geq 0.96$) in low energy experiments
\cite{smallalpha}.  In the sections for initial charm productions, we neglect
the nuclear shadowing since it is a minor effect compared with effects in our
focus. 

\hmysection{Comparison with Results from Parton Cascade Model}

\begin{figure}[h]
\psfig{figure=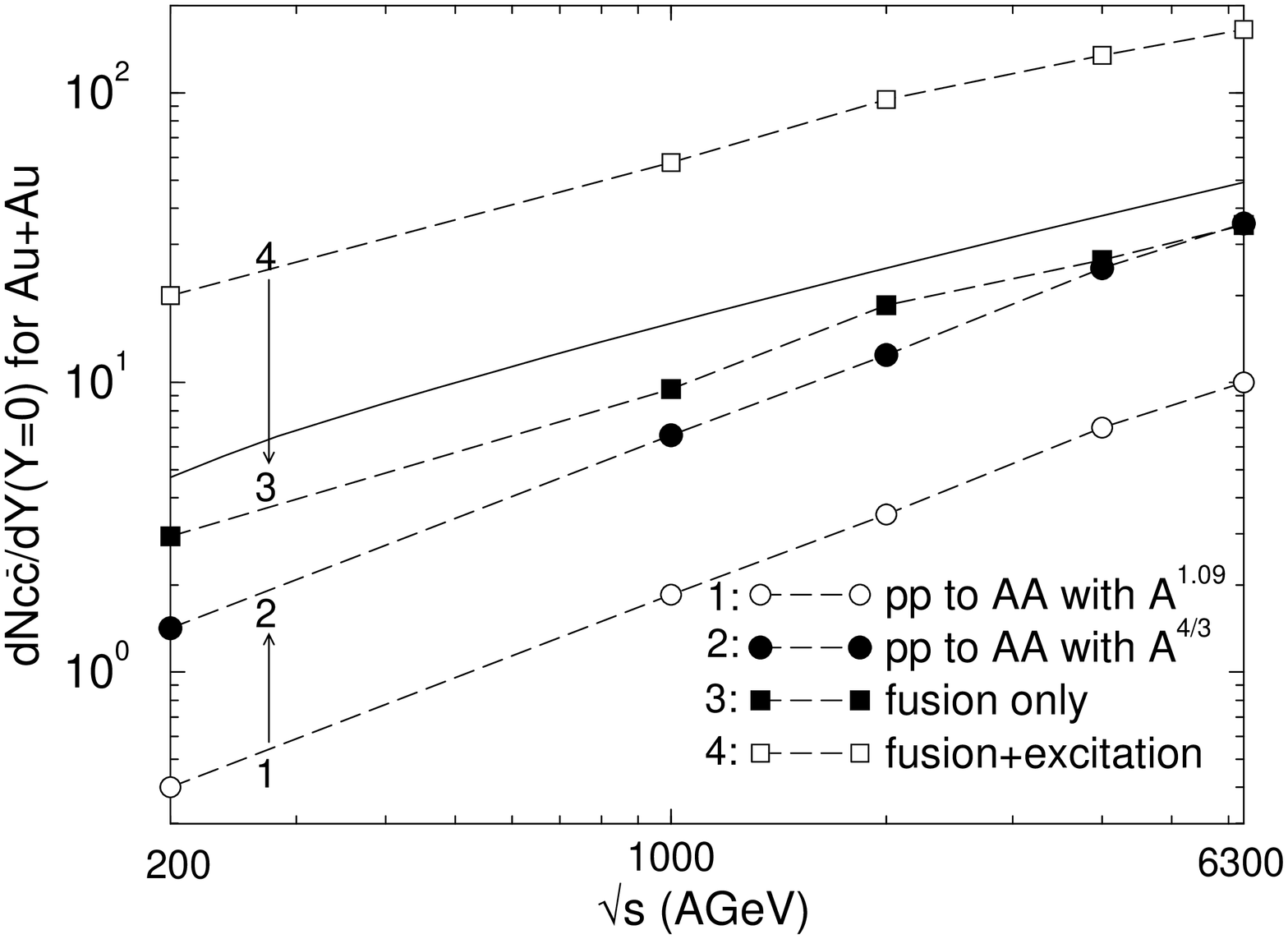,height=5.in,width=3.5in,angle=-90} 
\caption{
Comparison with Parton Cascade Model results on ${\left \protect( dN_{c\bar
c}/dY \right \protect)}_{Y=0}$: the rapidity density of charm and anti-charm
pairs for $Au+Au$ collisions as a function of $\protect \sqrt S/A$. 
}
\label{fig:klaus}
\end{figure}

Now we compare our results for the rapidity density of produced $c\bar{c}$
pairs at $Y=0$ with results of \cite{geiger}. Figure~\ref{fig:klaus} shows the
energy dependence in the range between RHIC and LHC (~$\sqrt s = 200 - 6300$
AGeV) for $Au+Au$ collisions.

The scaling from $pp$ results to $AA$ is 
\begin{eqnarray}
{\left( \frac{dN}{dY} \right)}^{AA}_{Y=0} = A^{\alpha +1/3}
{\left( \frac{d\sigma}{dY} \right)}^{pp}_{Y=0}/\sigma^{pp}_{inelastic} \;\; , 
\end{eqnarray}
where $\sigma^{pp}_{inelastic}$ is taken from \cite{goulianos}, which is
about 40.8$mb$ at RHIC energy and 73.5$mb$ at LHC energy.
Glauber geometry for central high $A+A$ collisions gives $\alpha=1$.  

In Figure~\ref{fig:klaus} the solid curve is the result we obtained using the
same parameters as for the solid curve in Figure~\ref{fig:charm_exp}.  The
parameterization for the dashed curve in Figure~\ref{fig:charm_exp} gives a
curve higher than the solid curve by 15\% to 30\%.  
The four long-dashed curves, curve1 to curve4, are all from PCM 
calculations \cite{geiger}. The top curve4 is the parton cascade model result
for the so-called QGP formation case, including both the fusion and the flavor
excitation processes.  That curve is much higher than our solid curve because
it includes the contribution from flavor excitation processes. Curve3, the
curve with solid squares, shows the contribution to curve4 from fusion
processes only (processes (1) and (2) in the notation of \cite{geiger}),
and curve3 is very close to our results. 
The bottom curve1 is the parton model result extrapolated to $AA$ from $pp$
using the $A^{0.76}$ scaling measured at much lower energies.  
It is lower than our solid curve by a factor of $12$ at the RHIC energy to $5$
at the LHC energy.  The main source of this difference is from the A-dependence
of $p+A$ cross sections.  The parton cascade model \cite{geiger} used an
$A^\alpha$ scaling with $\alpha=0.76$ \cite{smallalpha} instead of the value
$\alpha=1$ we use from Glauber geometry. 
To demonstrate the effect in different $A^\alpha$ scalings, we plot curve2,
which is the parton model result scaled from $pp$ to $AA$ by $A^{4/3}$, and it
is close to our results.   
As shown by the two arrows, curve4 becomes curve3 when the coherent
cancellation of flavor excitation processes is considered, and curve1 becomes 
curve2 when the high energy scaling is used. 
Therefore, the factor $\sim 50$ enhancement of charm production suggested in
 \cite{geiger}  comparing curve1 with curve4 for charm production at $RHIC$
is a consequence of the inclusion of incoherent flavor excitation processes and
the extrapolation from $pp$ to $AA$ via low energy scaling. 
The net dynamical enhancement in the PCM should be obtained by comparing curve2
with curve3.  In that case Figure~\ref{fig:klaus} leads to the expectation that
the pre-equilibrium charm production should be comparable to the initial fusion
rate.  This removes the bulk of the discrepancy between \cite{mw} and 
 \cite{geiger}.       

\hmysection{Summary}

We calculated initial charm production in nuclear collisions.  We demonstrated
the sensitive dependence on the choice of structure functions, the $Q^2$ scale,
and the K-factor.  The parameters were fixed by fitting the limited
available experimental data at lower energies.  We emphasized the need for new
measurements of $pp$ and $pA$ charm production to reduce the present large 
theoretical uncertainties.  We argued that the copious charm production 
predicted in \cite{geiger} was mainly due to the neglect of the coherent
suppression of  flavor excitation processes and the low energy $pp$ to $AA$
scaling.  Our calculated initial charm yields are close to those computed in
\cite{mw} and to the curve 2 in Figure~\ref{fig:klaus} from \cite{geiger}.   

%% file: sec5-pre.tex
\hmychapter{Pre-equilibrium Charm Production} 
\label{sec-pre}

In this chapter we consider the pre-equilibrium contribution to the charm yield
in $A+A$ collisions.
It is produced through final state interactions between partons in the dense
minijet plasma.  Here we only calculate the dominant contribution from minijet
gluon fusion. 

\hmysection{Spectrum of Minijets} 
\label{subsec-minijets}

\begin{figure}[p]
\psfig{figure=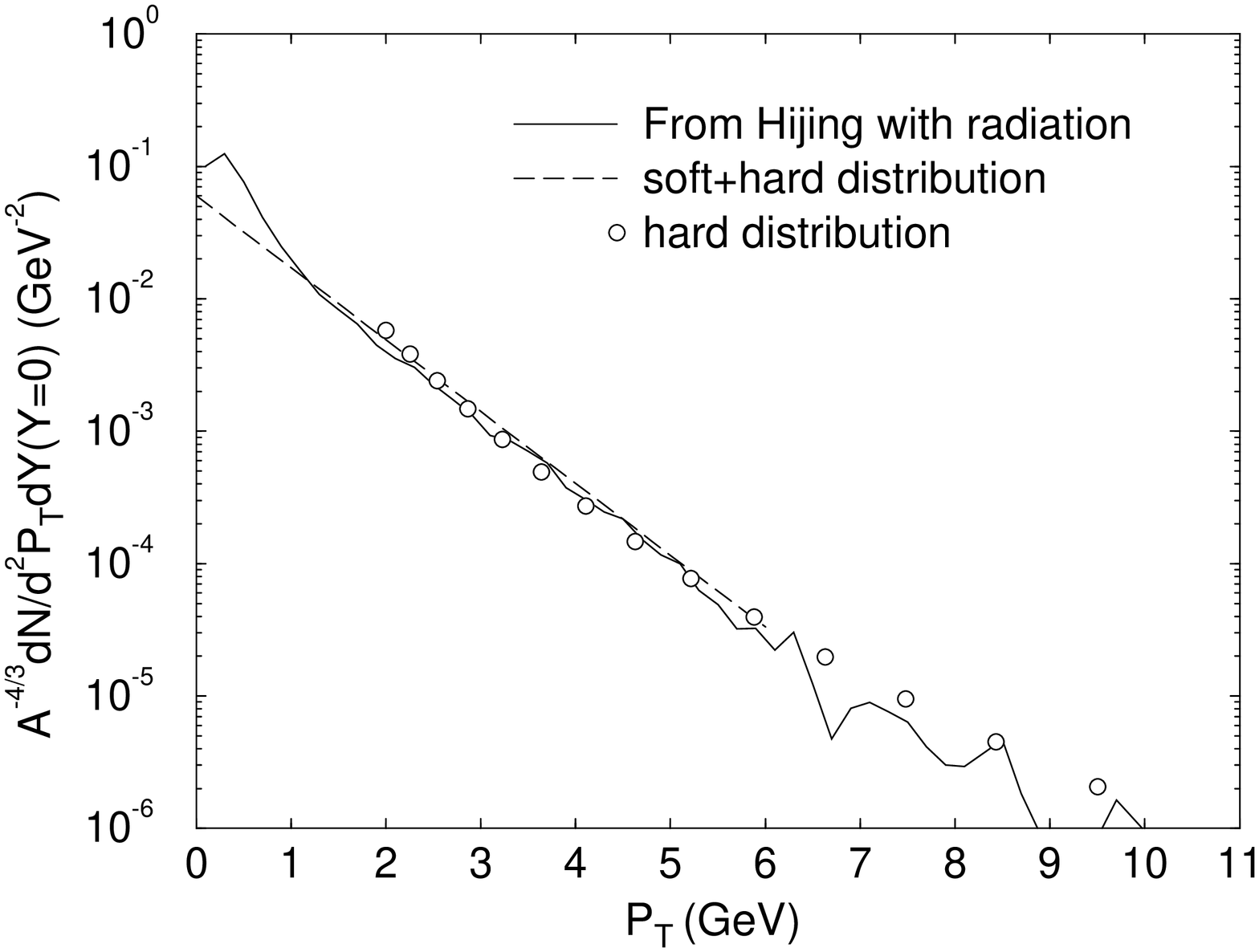,height=5.in,width=3.5in,angle=-90} 
\caption{
The minijet gluon $p_\perp$ distribution $A^{-4/3}\left\protect( dN/dy d
\vec{p_\perp} \right\protect)_{y=0}$. 
}
\label{fig:minijet}
\end{figure}

\begin{figure}[p]
\psfig{figure=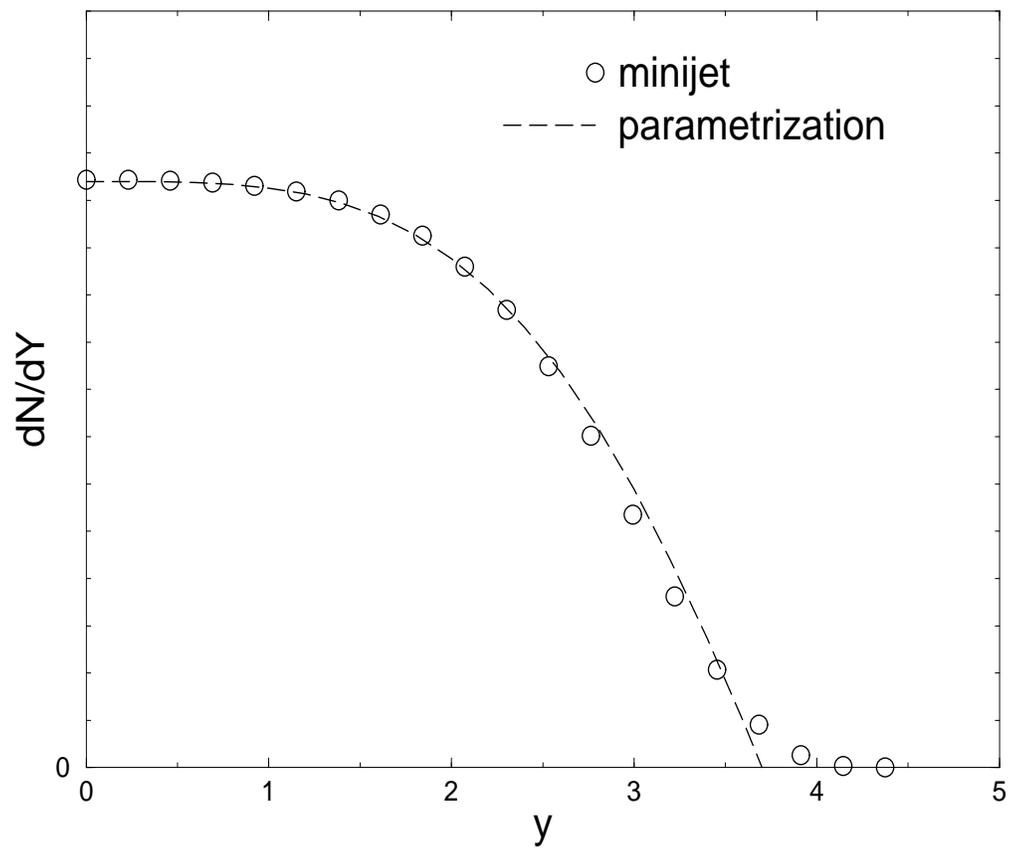,height=5.in,width=3.5in,angle=-90} 
\caption{
The minijet gluon rapidity distribution $dN/dy$.
}
\label{fig:minijety}
\end{figure}

The spectrum of minijet gluons in leading-order follows from \cite{ckr} 
\begin{eqnarray}
\frac{d \hat \sigma}{d \hat t}_{g g\rightarrow g g}
&=&\frac{9}{2} \frac{\pi \alpha_s^2}{\hat s^2} \left[3-\frac{\hat u\hat
t}{\hat s^2}-\frac{\hat u\hat s}{\hat t^2}-\frac{\hat s\hat t}{\hat u^2}
\right] \nonumber \;\; , \\[2ex]
\frac{d \hat \sigma}{d \hat t}_{g q\rightarrow g q}
&=&\frac{\pi \alpha_s^2}{\hat s^2} \left[ -\frac{4}{9}\frac{\hat u^2+\hat
s^2}{\hat u \hat s}+\frac{\hat u^2+\hat s^2}{\hat t^2} \right]  \;\; . 
\end{eqnarray}
The term minijets refers to unresolved jets at a scale
$p_\perp>{p_\perp}_{cut}$. The inclusive cross section to produce
minijets is given by   
\begin{eqnarray} 
\frac{d \sigma}{d y d p_\perp^2}
= \int d y_3 x_1 f_1 x_2 f_2 \frac{d \hat \sigma}{d \hat t}
(1+2\rightarrow 3+4) \;\; , 
\end{eqnarray}
where $f_1$ is the incident parton distribution evaluated at $x_1=p_\perp(e^y+
e^{y_3})/\sqrt s$ at a scale $Q^2={p_\perp}^2$.  The light-cone coordinates of
the initial and final partons are:
\begin{eqnarray}
p_1&=&\left[ 2x_1 p_0, 0, \vec 0 \right] \;\; , \nonumber \\
p_2&=&\left[ 0, 2x_2 p_0, \vec 0 \right] \;\; , \nonumber \\
p_3&=&\left[ m_\perp e^{y_3},m_\perp e^{-y_3},-\vec {p_\perp} \right] \;\; . 
\end{eqnarray}
The observed parton has momentum
\begin{eqnarray}
p=\left[ m_\perp e^{y},m_\perp e^{-y},\vec {p_\perp} \right] \;\; . 
\end{eqnarray}
The subprocess Mandelstam variables are $\hat s, \hat t$, and $\hat u$.  
For the calculation of minijet gluon fusion process in the following
Chapter~\ref{sec-correlation}, we choose $Q^2=\hat s$.  As in \cite{mw},
we use DO1 as the proton structure functions and $K=2$, $M_c=1.5$GeV for the
minijet production.  
The nuclear shadowing effect is taken from \cite{shadowing}. 
In Figure~\ref{fig:minijet} we show the resulting transverse momentum
distribution of mid-rapidity minijet gluons at $\sqrt s= 200$ AGeV in open
circles.  We call this distribution the hard distribution since it only
includes the hard gluons with ${p_\perp} > {p_\perp}_{cut}=2$GeV .  
The solid line in Figure~\ref{fig:minijet} is the output of the Monte Carlo
calculation via the HIJING model \cite{hijing} that includes initial and final
state radiation, and therefore fills the soft gluon region.  The dashed line is
our fit $0.06 e^{-1.25p_\perp}$, which is discussed below.  
In Figure~\ref{fig:minijety} we show the rapidity distribution of minijet
gluons. The circles represent our result from the initial production.  The
dashed line represents the fit $\cos\!{\left[ \frac{\pi
(\frac{y}{3.7})^{1.8}}{2} \right]}$, where $dN/dy$ scale is arbitrary.   


For convenience we have parameterized the minijet spectrum (both soft and hard
gluons) from the Monte Carlo calculation as the following:
\begin{eqnarray}
\frac{d N}{d y d {\vec {p_\perp}}}\equiv g(p_\perp)  \rho (y) A^{4/3} 
&=& 0.06 e^{-1.25 p_\perp} \cos\!{\left[ \frac{\pi (\frac{y}{3.7})^{1.8}}{2}
\right]} \; A^{4/3} \;\; , \\
\label{EQ:fit}
{\rm with }\; |y| &\leq& 3.7 \nonumber \;\; . 
\end{eqnarray}
In the following, we call this parameterized distribution the soft+hard
distribution. The soft+hard, hard, and Monte Carlo distributions are very close
to each other in the semi-hard  $p_\perp > 2$ GeV region at $y=0$, as seen in
Figure~\ref{fig:minijet}. 
The parameterized distribution falls underneath the Monte Carlo result in the
region $p_\perp< 1$ GeV. 
However, the soft component is strongly model-dependent as it requires the
furthest extrapolation from the pQCD hard domain. 
The Hijing yield in that region is due to initial and final state radiation.
Other contributions in this soft domain from coherent strings are possible
\cite{eskola1}. While most of the following results  are obtained with the
simple parameterization above, we will check the sensitivity to variations of
the soft component as well.  We also note that at bigger rapidity the $p_\perp$
spectrum falls more rapidly.  The above parameterization does not include that
property.  However, this property only lowers the high $p_\perp$ tail, and only
slightly changes the low $p_\perp$ part and the total number of the
pre-equilibrium charm.   

\hmysection{Two Extreme $\eta$ - $y$ Correlations}
\label{sec-correlation} 

In ideal Bjorken dynamics, the space-time rapidity
\begin{eqnarray}
\eta=1/2 \log[(t+z)/(t-z)] \;\; , 
\end{eqnarray}
and the true momentum rapidity
\begin{eqnarray}
y=1/2\log[(E+p_z)/(E-p_z)]
\end{eqnarray}
are assumed to be perfectly correlated. This is referred to as the
inside-outside picture, and the phase-space distribution function in Bjorken
correlation case has the form
\begin{eqnarray}
F(\vec x, \vec p, t)_{Bj}
&=&\frac{(2\pi)^3}{\tau \pi R_A^2 p_\perp} \frac{d N}{dy d {\vec {p_\perp}}}
\nonumber \\ 
&& \times \; \delta(\eta -y)  \Theta(\tau-\tau_i)
\Theta(\tau_f-\tau)   \;\; , 
\end{eqnarray}
where $\tau_i=0.1 fm/c$ is the minijet formation time.  $\tau_f\approx 1.7
fm/c$ as the proper time when the energy density of the pre-equilibrium
minijets falls by an order of magnitude to $\sim 2 GeV/fm^3$ due to rapid 
longitudinal expansion, and that is when we terminate the pre-equilibrium
stage.       

\begin{figure}[p]
\psfig{figure=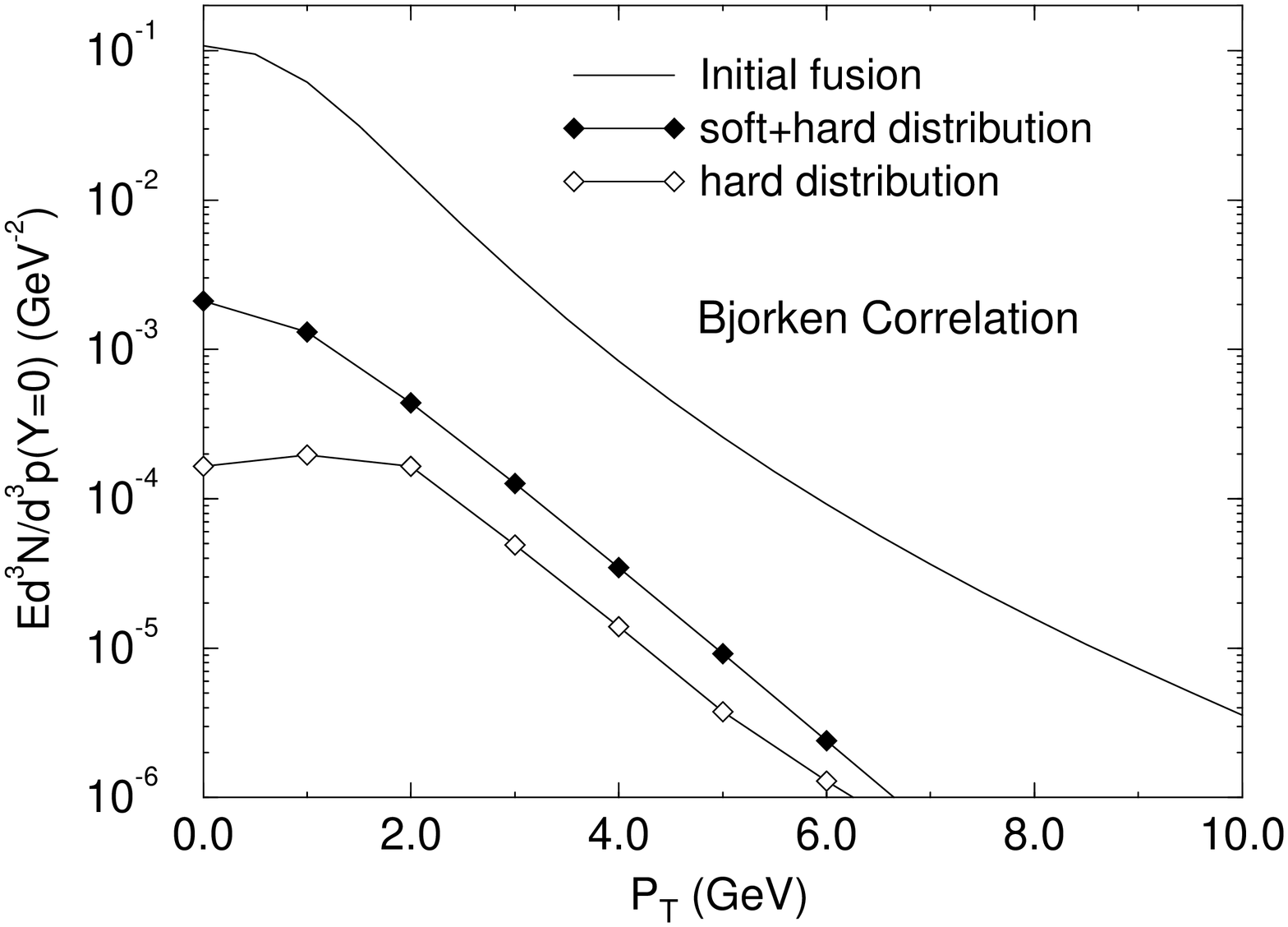,height=5.in,width=3.5in,angle=-90} 
\caption{
The distribution $\left \protect( E d^3N/d^3p \right\protect)_{y=0}$ of charm
quark production as a function of $p_\perp$ using $\delta
\protect(\eta-y\protect)$-correlation .  
}
\label{fig:bjorken}
\end{figure}

The phase space distribution is normalized such that
\begin{eqnarray}
\int \frac{F(\vec x, \vec p,t)_{Bj} d^3x}{(2\pi)^3}
=\frac{d^3N}{d^3p}=\frac{1}{E}\frac{d N}{dy d \vec{p_\perp}} \;\; , 
\end{eqnarray}
In this chapter we study the pre-equilibrium charm production at
$y=0$ \cite{mw}: 
\begin{eqnarray}
{\left( E\frac{d^3N}{d^3p} \right)}_{y=0}&=& \int{d^4 x} \int
\frac{1}{32(2\pi)^8} \frac{d^3{p_1}d^3{p_2}d^3{p^\prime}}{\omega_1  
\omega_2 E^\prime} \nonumber \\ 
&& \times \; F(\vec x,\vec p_1,t)~F(\vec x,\vec
p_2,t)~|M|^2~\delta^{(4)} \left( \sum P^{\mu} \right) \;\; . 
\label{EQ:edndp}
\end{eqnarray}
Denoting
\begin{eqnarray}
dN/dyd {\vec {p_\perp}} &\equiv& g(y,p_\perp) \;\; , \nonumber \\
\vec {{p_\perp}_1}&=&{p_\perp}_1 (\cos\!\phi_1,\sin\!\phi_1,0) \;\; , 
\end{eqnarray}
the ideal $\eta-y$ correlation leads to   
\begin{eqnarray}
{\left( E\frac{d^3N}{d^3p} \right)}_{y=0} &=& 
\int_{\tau_i}^{\tau_f} \frac{d \tau} {32(2\pi)^{2} \tau \pi R_A^2} \nonumber \\
&& \times \; \int d\eta d {p_\perp}_1 d {p_\perp}_2 d \phi_1 d \phi_2 \frac{
g(\eta,{p_\perp}_1) g(\eta,{p_\perp}_2) \delta\!(\sum E) |M|^2}
{E^\prime} \nonumber \\[2ex]     
&=& \frac {\ln (\tau_f / \tau_i)} {32(2\pi)^{\!2} \pi R_A^2} \int d\eta
d{p_\perp}_2 d \phi_1 d \phi_2 \nonumber \\ 
&& \times \; \frac {g(\eta,{p_\perp}_{1,0}) g(\eta,{p_\perp}_2) |M|^2 }
{{p_\perp}_2 \left[ 1-\cos\!(\phi_1-\phi_2) \right] -(E \cosh\eta-p\cos\phi_1)}
\;\; .  
\label{EQ:edndp.delta}
\end{eqnarray} 
In deriving the above, we have used kinematic relations
\begin{eqnarray}
E^\prime&=&({p_\perp}_1+{p_\perp}_2) \cosh\!\eta-E \;\; , \\[2ex]
\frac{\delta\! \left( \!\sum \!E\!\right)}{E^\prime}
&=&\frac{\delta({p_\perp}_1-{p_\perp}_{1,0})} {{p_\perp}_2
\left[ 1-\cos\!(\phi_1-\phi_2) \right]-(E \cosh\!\eta-p\cos\!\phi_1)} \;\; ,
\\[2ex]  
{p_\perp}_{1,0}&=&\frac{{p_\perp}_2 ( E \cosh\!\eta-p \cos\!\phi_2)}
{{p_\perp}_2 \left[ 1-\cos\!(\phi_1-\phi_2) \right]-(E
\cosh\!\eta-p\cos\!\phi_1)}   \;\; . 
\end{eqnarray} 
In Figure~\ref{fig:bjorken} we show the numerical results of  the above
integral in equation~(\ref{EQ:edndp.delta}).  

The solid line is the $p_\perp$-distribution for the initial charm production
from Chapter~\ref{sec-initial}.  
The curve labeled with solid diamonds represents the pre-equilibrium
contribution including both the  soft ($p_\perp<2$ GeV) and hard ($p_\perp> 2$
GeV) components of the minijet gluons.  The curve labeled with hollow diamonds
represents the pre-equilibrium contribution including only the hard component.
We see that the pre-equilibrium contribution in this strongly correlated case
is totally negligible.  This result is similar to the thermal charm production
contribution calculated in \cite{mw}, except that, in our case, the curve
extends to higher $p_\perp$ because of the broader initial minijet distribution
in $p_\perp$.    

\begin{figure}[p]
\psfig{figure=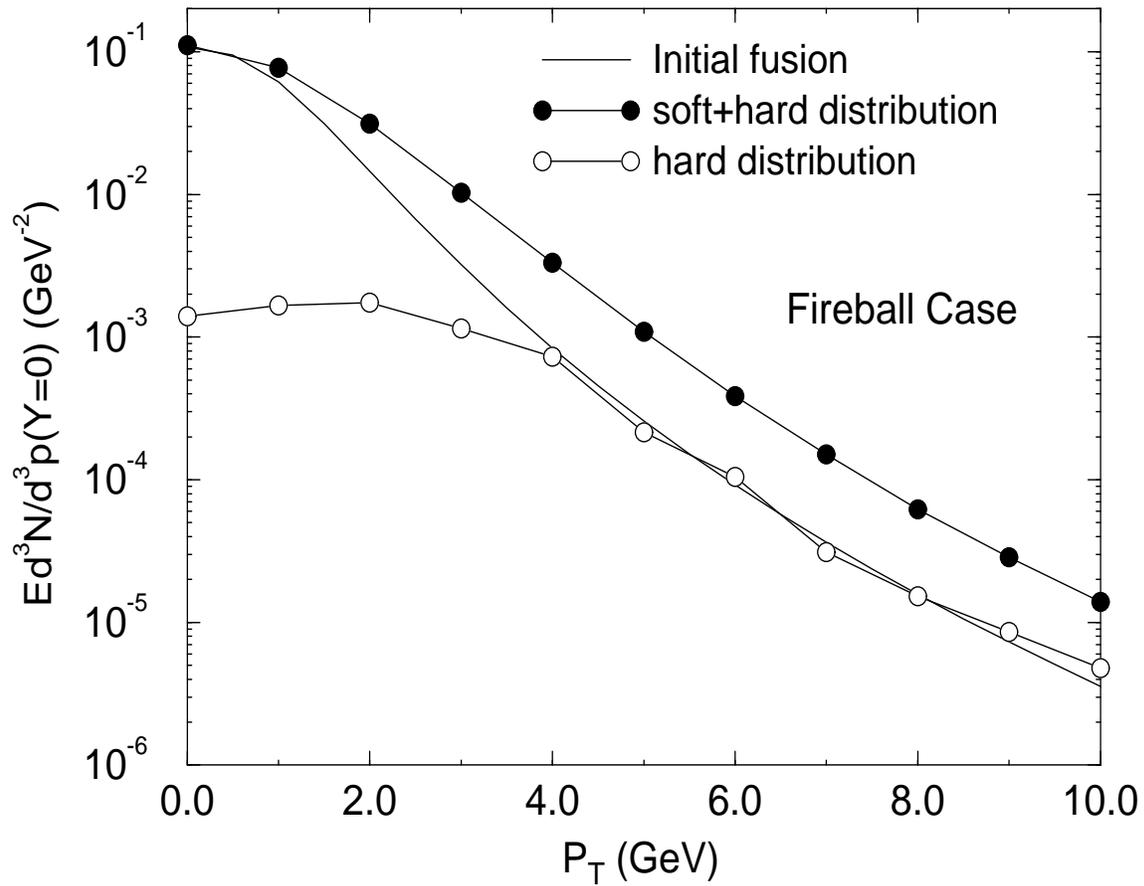,height=5.in,width=3.5in,angle=-90} 
\caption{
The distribution $\left\protect( E d^3N/d^3p \right\protect)_{y=0}$ of charm
quark production as a function of $p_\perp$ for the non-correlation case.  
}
\label{fig:fireball}
\end{figure}

Another extreme case, opposite to the ideal Bjorken picture, was considered in
 \cite{mw}.  In that uncorrelated $\eta-y$ case, the gluon distribution was
assumed to be completely uncorrelated as in an ideal thermal fireball.  This
assumption leads to  
\begin{eqnarray}
F(\vec x, \vec p, t)_{Fb}
=\frac {(2\pi)^3}{p} \frac{1}{V} \frac{d N}{dy d \vec{p_\perp}} \;\; . 
\end{eqnarray}
If one assumes a fixed volume $V=\tau_i \pi R_A^2$, then $\int dt \sim
\tau_f-\tau_i $, and 
\begin{eqnarray}
\int \frac{d^4x}{V^{\!2}} \sim \frac{1}{\pi R_A^2}
\frac{\tau_f}{\tau_i}~~~~\mbox { as in \cite{mw}. }
\end{eqnarray}
Then from equation~(\ref{EQ:edndp}), we have
\begin{eqnarray}
{\left( E\frac{d^3N}{d^3p} \right)}_{y=0}
=\frac{I(p)}{32(2\pi)^{\!2}}\int\!{\frac{d^4 x}{V^{\!2}}} 
=\frac{\tau_f/\tau_i}{32(2\pi)^{\!2}\pi R_A^2} I(p) \;\; , 
\label{EQ:edndp.un}
\end{eqnarray} 
where 
\begin{eqnarray}
I(p) &=& 
\int \frac{dy_1 dy_2 d{p_\perp}_2 d\phi_1 d\phi_2}{\cosh y_1 \cosh y_2}
\frac{ g(y_1,{p_\perp}_{1,0}) g(y_2,{p_\perp}_2) |M|^2} 
{D(y_1, y_2, {p_\perp}_2, \phi_1, \phi_2, p)}, \\[2ex]
\frac {\delta  \left( \sum E \right)}{E^\prime} 
&=&\frac {\delta({p_\perp}_1- {p_\perp}_{1,0})} 
{D(y_1, y_2, {p_\perp}_2, \phi_1, \phi_2, p)} \;\; , \\[2ex]
{p_\perp}_{1,0} &=& \frac{{p_\perp}_2 (E \cosh\! y_2 -p \cos\!\phi_2)}
{D(y_1, y_2, {p_\perp}_2, \phi_1, \phi_2, p)} \;\; , \\[2ex]
D(y_1, y_2, {p_\perp}_2, && \hspace{-0.9cm} \phi_1, \phi_2, p) 
= {p_\perp}_2 \left[ \cosh (y_1-y_2) -\cos(\phi_1-\phi_2) \right]\nonumber \\
&& \hspace{1cm} -(E \cosh y_1 -p \cos \phi_1) \;\; . 
\label{EQ:pt1.0} 
\end{eqnarray}
In Figure~\ref{fig:fireball} we show the pre-equilibrium charm production for
this uncorrelated case.  
The solid curve represents the initial charm production.  The curve labeled
with solid circles represents the pre-equilibrium contribution including both
the  soft ($p_\perp<2$ GeV) and hard ($p_\perp> 2$ GeV) components. 
It is much larger than the Bjorken-correlation case, and is comparable with the
initial charm yield, as shown in Figure~\ref{fig:fireball}.  This is similar to
the result in \cite{mw} in which the pre-equilibrium charm production has
almost the same magnitude and $p_\perp$-shape as the initial charm.     
The curve labeled with hollow circles represents the pre-equilibrium
contribution including only the hard component.

\hmysection{Minimally-correlated $\eta-y$} 

From the above results we conclude that the pre-equilibrium charm production is
very sensitive to the $\eta-y$ correlation of the minijets.
We consider here the simplest source of $\eta-y$ correlations resulting from
the minimal geometrical spread in initial production points required by the
uncertainty principle.  This type of correlations are included in the parton 
cascade model and discussed in \cite{geiger2}. The phase space
distribution function including such minimal correlation has the form  
\begin{eqnarray}
F(\vec x, \vec p, t)_{Min} &=&
{\cal N}\int\frac{d N}{dy d {\vec {p_\perp}}} 
\frac{\theta(\tau_{max}-\frac{t}{\cosh\! y})}
{1+(\frac{t_f(p)}{\Delta t})^2} \nonumber \\
&& \times \; \rho_0(\vec {x_0},t_0) \delta(\vec x-
\vec{x_0}-\vec{ v} \Delta t) d^3{x_0} dt_0 \;\; . 
\end{eqnarray}
The integration is over the space-time coordinates $(\vec{x}_0,t_0)$ of the
production points of the gluons.  These points are distributed according to a
normalized density $\rho_0(\vec{x}_0,t_0)$. The delta function arises to take
into account the free streaming of the partons from the production point, with
velocity $\vec{v}=\vec{p}/E$, where $E=p_\perp \cosh y$ and  $p_z=p_\perp \sinh
y$. The theta function defines what we mean by pre-equilibrium. The proper time
when the pre-equilibrium fusion is terminated is $\tau_{max}$, which is
determined below in Figure~\ref{fig:cutoff}. The theta function insures that
only those gluons with proper time less than $\tau_{max}$ contribute to
pre-equilibrium charm productions. 

The formation physics is included via the Lorentzian formation
factor \cite{gyuwang}
\begin{equation} 
[1+(t_f(p)/\Delta t)^2]^{-1} \;\; , 
\label{EQ:lorentzian}
\end{equation}
where $\Delta t=t-t_0$ is the elapsed time, and the formation time is given by 
\begin{equation}
t_f(p) \simeq \cosh\!y \frac{0.2 {\rm GeV}}{p_\perp} (fm) \;\; ,
\end{equation}
We note that, at the early time, the above formation factor more accurately
describes the interference phenomena suppressing production than the
conventionally assumed factor 
\begin{equation}
\theta[\Delta t - t_f(p)] \;\; ,
\label{EQ:theta}
\end{equation}

We assume
\begin{eqnarray}
\int{\rho_0(\vec {x_0},t_0)}d^3 x_0 dt_0=1 \;\; . 
\end{eqnarray}
In this case the normalization factor is 
\begin{eqnarray}
{\cal N}=\frac {(2 \pi)^3} {E} \;\; , 
\end{eqnarray}
so that   
\begin{equation}
\lim_{t\rightarrow \infty} 
\frac {1} {(2 \pi)^3} \int F(\vec x, \vec p, t)_{Min} d^3 x=
\frac {d^3 N} {d^3 p} \;\; . 
\end{equation}
As discussed in \cite{geiger2}, the production points are spread along the
beam axis according to the uncertainty principle by an amount $\delta z\equiv d
\sim \hbar /p_\perp$, since the dominant parton interaction leading to a $y=0$
parton with final $p_\perp$ has an initial longitudinal momentum $x P_0 \sim
p_\perp$. We take as a particular model  
\begin{equation}
d=\frac{0.2}{p_\perp}(fm) \;\; . 
\end{equation}
Clearly this is only a rough guess, but it allows us at least to investigate
the sensitivity of the results to a particular $\eta-y$ correlation that
results from this spatial spreading of the production points. We emphasize that
it is precisely the uncertainty of the initial space-time formation physics
that leads us to investigate the possibility of open charm production as an
experimental probe of this physics.  

Given the above assumption we take
\begin{eqnarray}
\rho_0(\vec{x_0},t_0)
=\frac{1} {\pi R_A^2} \delta (t_0) \frac{e^{-z_0^2/(2 d^2)}}{\sqrt {2\pi} d}
\;\; ,  
\end{eqnarray}
where $d$ is the mean spread for gluons depending on $p_\perp$ from above.
This distribution only spreads out the production points along the beam axis. 
A more realistic treatment would also include a spread in the time coordinate.

\begin{figure}[p]
\psfig{figure=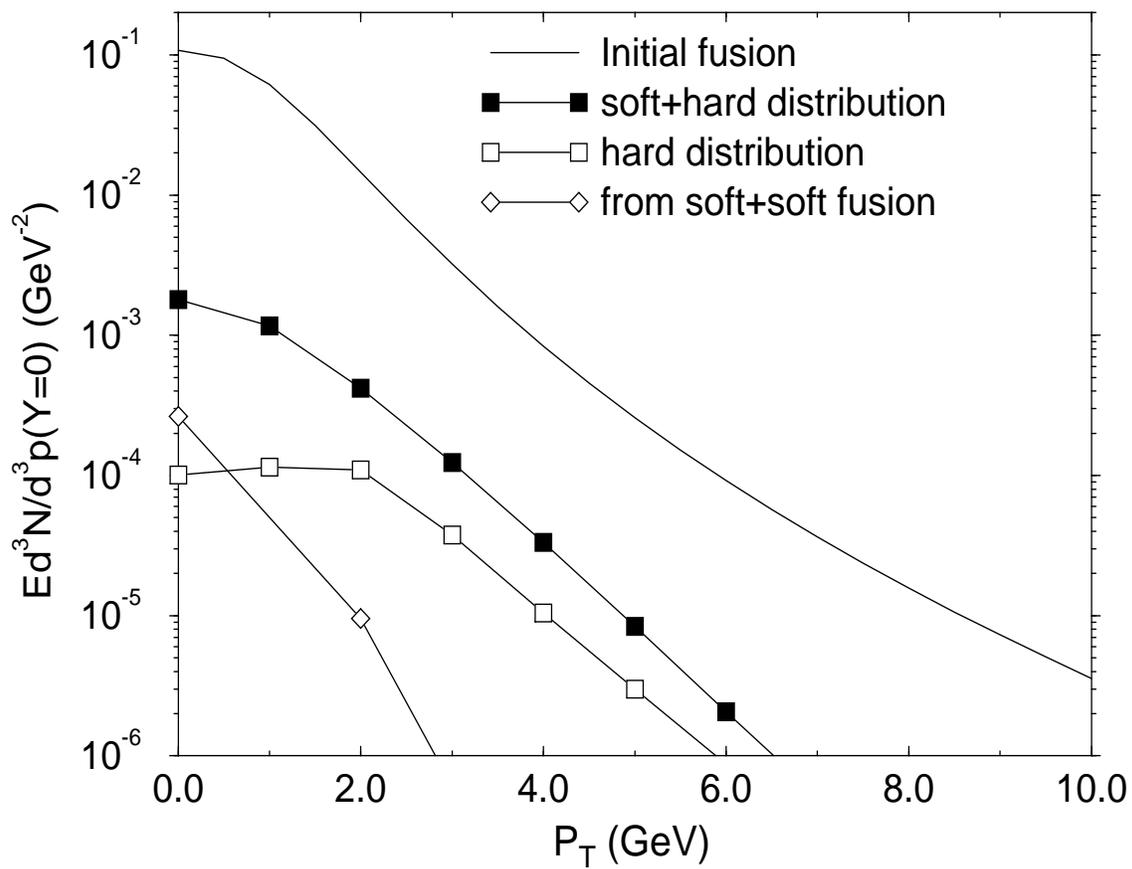,height=5.in,width=3.5in,angle=-90} 
\caption{
The distribution $\left\protect( Ed^3N/d^3p \right\protect)_{y=0}$ of charm
quark production as a function of $p_\perp$ using minimal $\eta-y$
correlation. 
}
\label{fig:minimal}
\end{figure}

\begin{figure}[p]
\psfig{figure=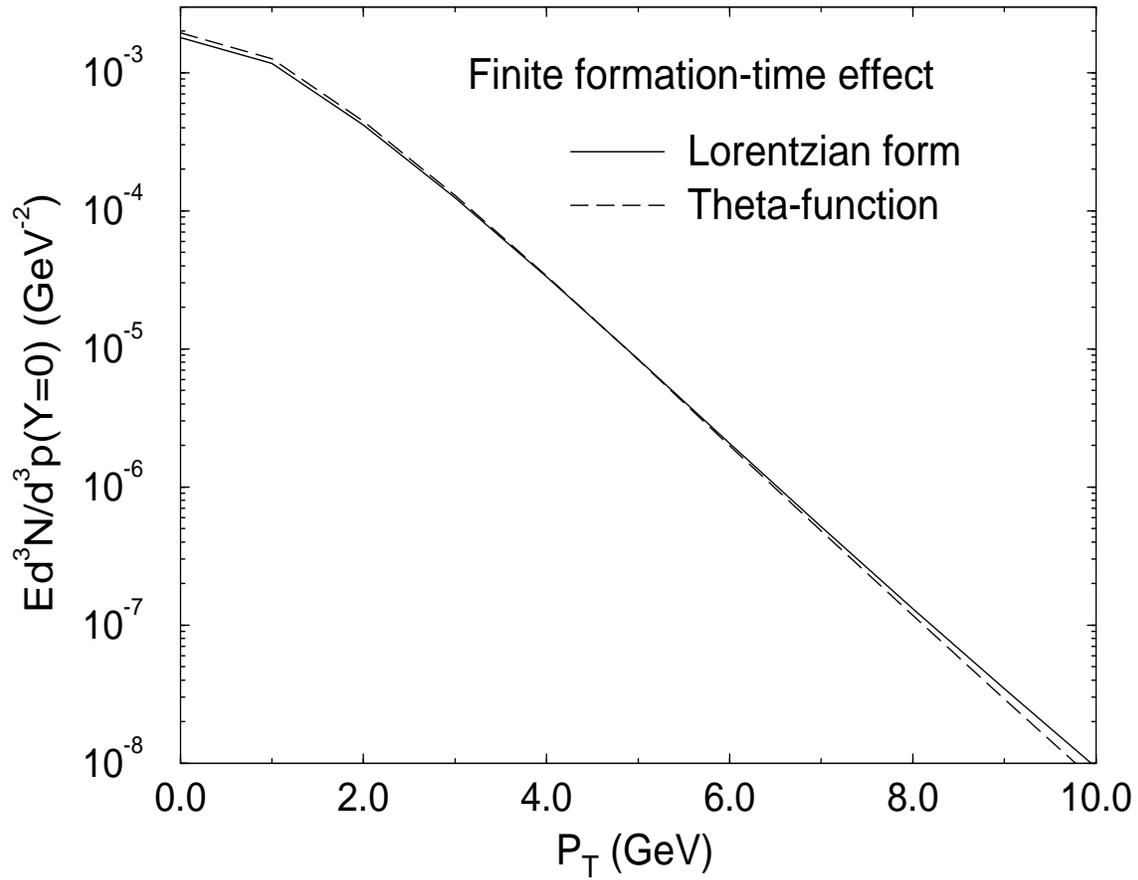,height=5.in,width=3.5in,angle=-90} 
\caption{
The distribution $\left\protect( Ed^3N/d^3p \right\protect)_{y=0}$ of charm
quark production using different formation-time probability distributions. 
} 
\label{fig:formation}
\end{figure}

Neglecting transverse expansion, we obtain finally 
\begin{eqnarray}
F(\vec x, \vec p, t)_{Min} &=& 
\frac{(2\pi)^3}{\sqrt {2\pi} \pi R_A^2} \frac{p_\perp}{0.2}
e^{-(z-\tanh\!y\:t)^{2} (\frac{p_\perp}{0.2})^{\!2}/2} \nonumber \\
&& \times \; \frac{1}{p} \frac{d N}{dy~d {\vec {p_\perp}}}
\frac{\theta(\tau_{max}-\frac{t}{\cosh\! y})} {1+(\frac{0.2 \cosh\!y}{p_\perp  
t})^2} \;\; .  
\label{EQ:Fmin}
\end{eqnarray}
Denote
\begin{eqnarray}
a_1=\tanh y_1 \;\; , \nonumber \\
a_2=\tanh y_2 \;\; , \nonumber \\ 
b_1=(\frac{{p_\perp}_{1,0}}{0.2})^2/2 \;\; , \nonumber \\
b_2=(\frac{{p_\perp}_2}{0.2})^2/2 \;\; , 
\end{eqnarray}
then after integration over $z$, we have the final expression as the following:
\begin{eqnarray}
{\left( E\frac{d^3N}{d^3p} \right)}_{y=0} &=& 
\frac{\sqrt {\pi}}{16 (2\pi)^4 R_A^2} \int dy_2 d y_1 d{p_\perp}_2
d \phi_2 d \phi_1 \nonumber \\[1ex]   
&& \hspace{-1.in} \times \; \frac{{p_\perp}_{1,0} \frac{{p_\perp}_{1,0}}{0.2}
\frac{{p_\perp}_2}{0.2} |M|^2} {{p_\perp}_2 \cosh\!y_1\cosh\!y_2
(E\cosh\!y_2~-p\cos\!\phi_2)} \frac{g(y_1, {p_\perp}_{1,0}) g(y_2,
{p_\perp}_2)} {\sqrt {b_1+b_2}} \nonumber \\[1ex]   
&& \hspace{-1.in} \times \; \int_0^{t_f} dt \frac{e^{\frac{-(a_1-a_2)^2
t^2}{1/b_1+1/b_2}}} { \left[ 1+(\frac{0.2 \cosh\!{y_1}} {{p_\perp}_{1,0} t})^2
\right] \left[ 1+(\frac{0.2 \cosh\!{y_2}}{{p_\perp}_2 t})^2 \right] }
\label{EQ:edndp.min} \;\; . 
\end{eqnarray} 
In the above
\begin{eqnarray}
t_f=\tau_{max} \min\,(\cosh {y_1}, \cosh {y_2}) \;\; , 
\end{eqnarray}
and ${p_\perp}_{1,0}$ is the same as in equation~(\ref{EQ:pt1.0}).  Note that
by using the unit $GeV$ for momentum and unit $fm$ for time, the expression
${\left( E\frac{d^3N}{d^3p} \right)}_{y=0}$ in
equations~(\ref{EQ:edndp.delta}), (\ref{EQ:edndp.un}), and (\ref{EQ:edndp.min})
has the dimension $GeV^{-4}fm^{-2}$.  We require a factor $(\hbar c)^2 \sim
(0.2GeV fm)^2$ to convert it to the dimension $GeV^{-2}$, which we have used in
Figure~\ref{fig:bjorken}, Figure~\ref{fig:fireball},Figure~\ref{fig:minimal}
and Figure~\ref{fig:formation}. 

The numerical results of the above equation~\ref{EQ:edndp.min} are shown in
Figure~\ref{fig:minimal}.  The curve  labeled with solid squares includes
both components while that labeled with hollow squares includes only the
fusion of hard gluons. The curve labeled with diamonds represents the
contribution from fusion of soft gluons both with  $p_\perp<2$GeV. This shows
that the pre-equilibrium contribution comes mainly from the fusion of soft and
hard gluons.  

In Figure~\ref{fig:cutoff} the energy density at $z=0$ as a
function of proper time is shown assuming minimal correlation and Lorentzian
formation probability.  The solid curve includes both soft and hard components
while the dashed curve is calculated using the hard distribution and includes
only the hard component.  The energy density increases and reaches maximum at
the time about $0.1 fm/c$, then the energy density decreases linearly to $\sim
2  GeV/fm^3$ at $\sim 1.7 fm/c$ for soft+hard distribution. We choose the above
time as the cutoff $\tau_{max}$.

\begin{figure}[p]
\psfig{figure=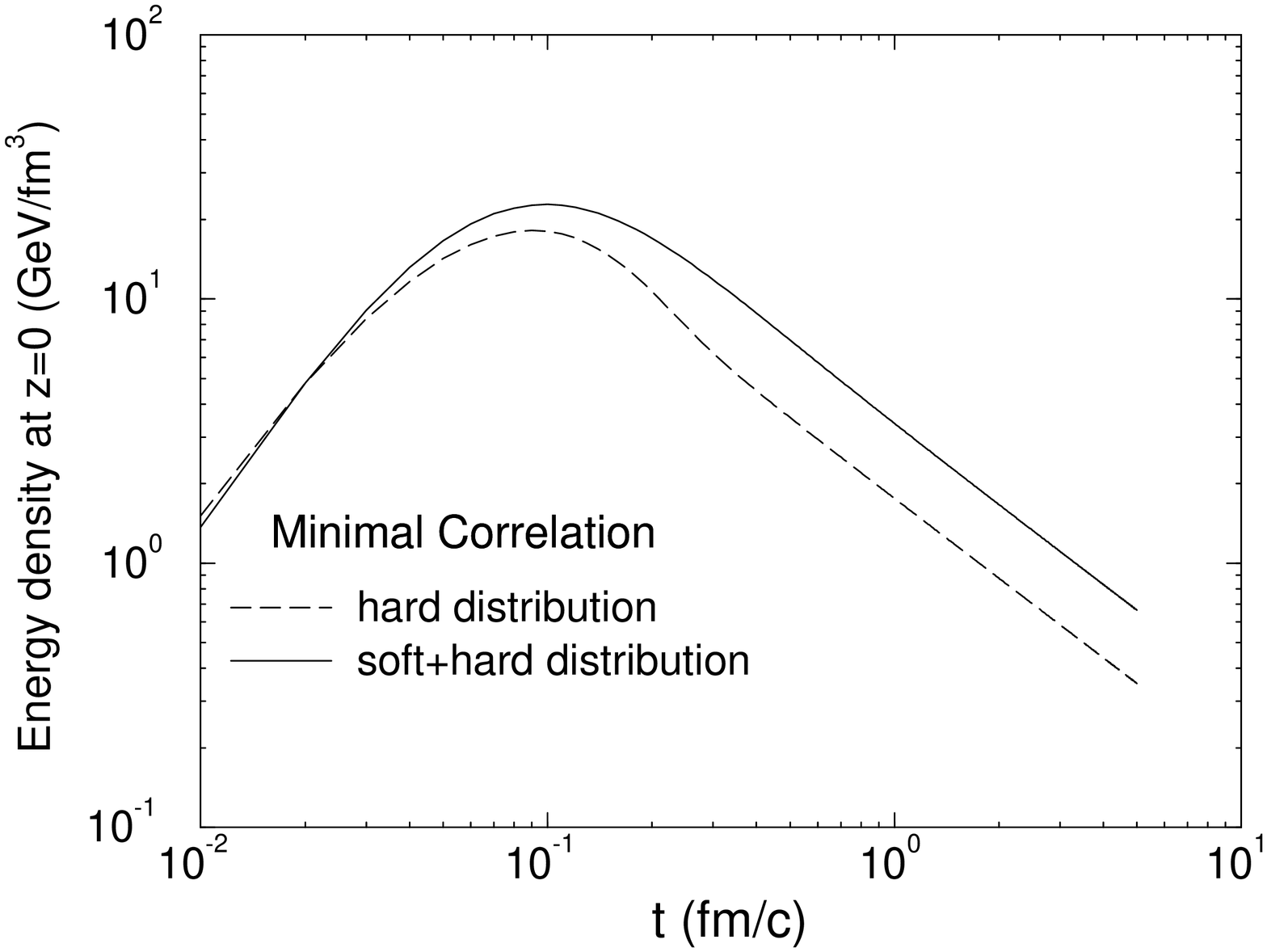,height=5.in,width=3.5in,angle=-90} 
\caption{
The energy density at $z=0$ as a function of proper time.
}
\label{fig:cutoff}
\end{figure}

\hmysection{Why Is the Pre-equilibrium Charm Yield So Small?: An Estimate}

To understand why the pre-equilibrium charm yield 
is so small compared to the initial yield as found through tedious
numerical calculations in the previous chapter,
we consider here the calculation of the total
number of  pre-equilibrium charm pairs. 
The expression for that number is given by
\begin{eqnarray}
N=\frac{(\hbar c)^2}{4 (2 \pi)^6} \int d^4x \int \frac{d^3 p_1}{\omega_1}
\frac{d^3 p_2}{\omega_2} F(\vec x, \vec p_1, t) F(\vec x, \vec p_2, t) \hat s
\hat \sigma(\hat s) \;\; , 
\label{EQ:nccbar}
\end{eqnarray}
where $\hat \sigma(\hat s)$ is the integrated cross section for the process $gg
\rightarrow c \bar c$, see equation~(\ref{EQ:ggfusion}).
Our main strategy  is to estimate the mean difference between the two gluon
rapidities, then, from the kinematical constraint on charm production ($\hat s
\geq 4 M_c^2$), estimate the effective lower cutoff  for $p_\perp$ of the
minijet gluons.  Thus we separate the $p_\perp$ integrals from the rapidity
integrals, and have a rough estimate for the total number of charm pairs. 

For the fireball case, 
\begin{eqnarray}
F(\vec x, \vec p, t)_{Fb}
&=&\frac{(2\pi)^3}{p}\frac{1}{V} \frac{d N}{dy d \vec{p_\perp}}  \;\; , \\[2ex]
\hat s&=&2 {p_\perp}_1 {p_\perp}_2 \left[ \cosh(y_1-y_2) -
\cos(\phi_1-\phi_2) \right] \;\; . 
\label{EQ:hats}
\end{eqnarray}
For the Bjorken case, 
\begin{eqnarray}
F(\vec x, \vec p, t)_{Bj}
&=&\frac{(2\pi)^3}{p_\perp}\frac{d N}{dy d \vec{p_\perp}} \frac{\delta(\eta
-y)}{\tau \pi R_A^2} \Theta(\tau-\tau_i) \Theta(\tau_f-\tau) \;\; , \\[2ex]
\hat s &=& 2 {p_\perp}_1 {p_\perp}_2 \left[1 - \cos(\phi_1-\phi_2) \right] 
\;\; .  
\end{eqnarray}
For all the cases, we use the  fit to the
gluon distribution given by  equation~(\ref{EQ:fit}),
where
\begin{eqnarray}
g(p_\perp) \equiv a e^{-b p_\perp} = 0.06 e^{-1.25 p_\perp} \;\; . 
\end{eqnarray}
Therefore,
\begin{eqnarray}
N_{Fb}&=& 
\frac {(\hbar c)\!^2 \frac{\tau_f}{\tau_i} A^{8/3}}{4 \pi R_A^2}
\int dy_1 \frac {\rho (y_1)}{\cosh {y_1}} 
\int dy_2 \frac {\rho (y_2)}{\cosh {y_2}} \nonumber \\
&& \times \; \int d{p_\perp}_1 g({p_\perp}_1) \int d{p_\perp}_2 g({p_\perp}_2) 
\int d\phi_1 \int \! d\phi_2 \hat s \hat \sigma (\hat s)
\nonumber \;\; , \\[2ex] 
N_{Bj}&=& \frac {(\hbar c)\!^2 \ln \!\frac {\tau_f}{\tau_i} A^{8/3}}{4 \pi
R_A^2}  \int d \eta \left[ \rho (\eta) \right]^2  \nonumber \\
&& \times \; \int d{p_\perp}_1 g({p_\perp}_1) \int d{p_\perp}_2 g({p_\perp}_2) 
\int d\phi_1 \int d\phi_2 \hat s \hat \sigma (\hat s) \;\; . 
\end{eqnarray}
The dominant contribution is coming from the vicinity of the production
threshold where $\hat s=4 M_c^2=9 GeV^2$ \cite{combridge}.  From this we make
the following rough estimates:  
\begin{eqnarray}
\hat s \hat \sigma (\hat s) &\sim& \alpha ^2(\hat s)  \sim 0.06\;
\;\; , \nonumber\\[2ex] 
\int d{p_\perp} g(p_\perp) &\sim& \int_{p_c}^\infty d{p_\perp}
g(p_\perp)  \sim \frac {a}{b e^{b p_c}} \;\; , 
\end{eqnarray}
where $p_c$ is the effective cutoff value
for ${p_\perp}_1$ and ${p_\perp}_2$ from the requirement $\hat s \geq 4 M_c^2$.

For the fireball case, 
\begin{eqnarray}
\langle {\cosh (y_1-y_2)} \rangle&\equiv & 
\frac {\int \! dy_1 dy_2 \frac {\rho (y_1)}{\cosh {y_1}} \frac {\rho
(y_2)}{\cosh {y_2}} \cosh (y_1-y_2)}{\int \! dy_1 dy_2 \frac {\rho (y_1)}{\cosh
{y_1}} \frac {\rho (y_2)}{\cosh {y_2}} } 
\sim 4.0 \;\; . 
\label{EQ:rapidity}
\end{eqnarray}
Since the minijet $p_\perp$ spectrum is dropping almost exponentially, the
production heavily favors the smaller cutoff $p_c$, so the mean value of $\cos
(\phi_1-\phi_2)$ is most likely to be negative.  We take 
\begin{eqnarray}
\langle {\cos (\phi_1-\phi_2)} \rangle \sim -0.5 \;\; . 
\end{eqnarray}
Then
\begin{eqnarray}
\hat s \sim 9 p_c^2 \;\; , 
\end{eqnarray}
so the effective cutoff for the fireball case is 
\begin{eqnarray}
p_c \sim 1.0 {\rm GeV} \;\; . 
\end{eqnarray}
On the other hand, for the Bjorken case, 
\begin{eqnarray}
y_1=y_2=\eta \;\; , \Rightarrow \hat s \sim 3 p_c^2 \;\; , \nonumber \\
\Rightarrow p_c \sim 1.73 {\rm GeV} \;\; . 
\end{eqnarray} 
Using the same values as in Chapter~\ref{sec-correlation}: $\tau_i=0.1 fm$, 
$\tau_f=1.0 fm$ for fireball case, $\tau_f^{\prime}=1.7 fm$ for Bjorken case,
and  
\begin{eqnarray}
\int  dy_1 \frac {\rho (y_1)}{\cosh {y_1}} \sim 2.8 \;\; , \nonumber \\
\int  d \eta \left[ \rho (\eta) \right] ^2 \sim 4.9 \;\; . 
\end{eqnarray}
We then have the estimate for the total number of the pre-equilibrium charm:
\begin{eqnarray}
N_{Fb}&\sim&\frac {(\hbar c)\!^2 \frac {\tau_f}{\tau_i} A^{8/3}}{4 \pi R_A^2} 
\;2.8^2 (\frac{a}{b e^{1.0b}})^2 (2\pi)^2 \left[ \hat s \hat \sigma (\hat s)
\right] \sim 3.5 \;\; , \nonumber\\ [2ex]
N_{Bj}&\sim&\frac {(\hbar c)\!^2 \ln \!\frac{\tau_f^{\prime}}{\tau_i} A^{8/3}}
{4 \pi R_A^2} \;4.9 (\frac{a}{b e^{1.73b}})^2 (2\pi)^2 \left[ \hat s \hat
\sigma (\hat s) \right] \sim 0.098 \;\; . 
\end{eqnarray}
Therefore we estimate $N_{Fb}/N_{Bj} \sim 35$, in rough agreement with the
detailed numerics.  We see that the large increase going from the Bjorken case
to the fireball case results mainly from the different $p_\perp$ cutoff.  In
the uncorrelated fireball case, one allows particles with different rapidities
to interact with each other (see equation~(\ref{EQ:rapidity})), thus more low
$p_\perp$ gluons can take part in the interaction.  Since the minijet $p_\perp$
spectrum drops almost exponentially, the fireball case produces much more
pre-equilibrium charm than the Bjorken case (a factor of 6 increase from the
smaller $p_\perp$ cutoff).  Although the questionable linear proper time
dependence in the fireball case also gives an considerable increase (about a
factor of 3.5), it is not as important as the correlation effect.   

For the minimal correlation case, the estimate is unfortunately not as
straightforward.    The phase space distribution function is  
\begin{eqnarray}
F(\vec x, \vec p, t)_{Min} &=& 
\frac{(2\pi)^3}{\sqrt {2\pi} \pi R_A^2} \frac{e^{-(z-\tanh\!y\:t)^{\!2}
(\frac{p_\perp}{\hbar c})^{\!2}\!/2}}{\hbar c \cosh y}  \nonumber \\
&& \times \frac{d N}{dy~d {\vec {p_\perp}}}
\theta(\tau_{max}-\frac{t}{\cosh\! y}) \theta(\frac{\hbar c \cosh\!
y}{p_\perp}-\tau_{max}) \;\; ,   
\end{eqnarray}
and $\hat s$ is the same as in equation~(\ref{EQ:hats}).  In the above
distribution function we choose to use the $\theta$-function for the
formation-time effect.   

Using equation~(\ref{EQ:nccbar}) and after the integration over $z$, we have
\begin{eqnarray}
N_{Min}&=&\frac {(\hbar c)^2 A^{8/3}} {4 \pi R_A^2 \sqrt \pi} 
\int \! d{p_\perp}_1 g({p_\perp}_1) \int \! d{p_\perp}_2
g({p_\perp}_2) 
\int \! d\phi_1 \int \! d\phi_2 \nonumber\\
&& \hspace{-1.cm} \times \; \int dy_1 \frac {\rho (y_1)}{\cosh {y_1}} 
\int \! dy_2 \frac {\rho (y_2)}{\cosh {y_2}}  \hat s \hat \sigma (\hat s)
\int_{t_{min}}^{t_{max}} \! dt\; \frac {e^{\frac{-(a_1-a_2)^2
t^2}{1/b_1+1/b_2}}} {\sqrt {1/b_1+1/b_2}}  \;\; , 
\label{EQ:NMin}
\end{eqnarray}
where 
\begin{eqnarray}
t_{min}=\hbar c \max \left( \frac{\cosh y_1}{{p_\perp}_1},\frac{\cosh
y_2}{{p_\perp}_2} \right) \nonumber \;\; , \\[2ex]
t_{max}=\tau_f \min (\cosh y_1, \cosh y_2) \;\; , 
\end{eqnarray}
and $a_1,a_2,b_1,b_2$ are defined the same as in equation~(\ref{EQ:Fmin}). 

We estimate that for the dominant part of the integral
\begin{eqnarray}
\frac{1}{b_1}+\frac{1}{b_2} \sim \left( \frac {2 \hbar c}{p_c} \right)^2
\nonumber \;\; , \\[2ex]  
t_{min} \sim \frac{\hbar c \cosh \bar y}{p_c} \;\; , \nonumber \\[2ex]  
t_{max} \sim \tau_f \cosh \bar y \;\; , 
\end{eqnarray} 
where $\bar y=(|y_1|+|y_2|)/2$.  
Now let $u=t \;p_c/(\hbar c \cosh \bar y)$, then the last 3-dimensional
integral in equation~(\ref{EQ:NMin}) without the factor $\hat s \hat \sigma
(\hat s)$ is  
\begin{eqnarray}
J &\sim&
\int \! dy_1 \frac {\rho (y_1)}{\cosh {y_1}} 
\int \! dy_2 \frac {\rho (y_2)}{\cosh {y_2}} \frac {1}{2} \cosh \bar y
\nonumber \\
&& \times \; \int_{1}^{\frac{\tau_f p_c}{\hbar c}} du\; e^{-[ \frac{\sinh
(y_1-y_2) \cosh \bar y }{2 \cosh y_1 \cosh y_2} ] ^2 u^2} \;\; . 
\end{eqnarray}
The $u$-integral gives a dependence on $\tau_f$ which is similar to the
logarithmic dependence in the Bjorken case, and the exponential form in the
integrand forces the spread $y_1-y_2$ to be small. Numerically, by taking
$\tau_f \sim 1.7 fm/c$ , $p_c \sim 2.0$ GeV (as the first-step value) in the
$u$-integral, the above 3-dimensional integral is $J \sim 19.1$.  When the
integrand is weighed by $\cosh (y_1-y_2)$, the integral is $\sim 23.3$.  So  
\begin{eqnarray}
&&\langle {\cosh (y_1-y_2)} \rangle \sim 23.3/19.1 \sim 1.22 \;\; , \nonumber
\\[2ex] 
&&\Rightarrow \hat s \sim 3.44 p_c^2 \;\; , \Rightarrow p_c \sim 1.62 {\rm GeV}
\;\; .  
\end{eqnarray}
Note that the above determined value of $p_c$ is insensitive to the
first-step $p_c$ value we tried in the $u$-integral.   

Therefore for the total pre-equilibrium charm number,
\begin{eqnarray}
N_{Min} &\sim& \frac {(\hbar c)^2 A^{8/3}} {4 \pi R_A^2 \sqrt \pi} 
\int \! d{p_\perp}_1 g({p_\perp}_1) \int \! d{p_\perp}_2
g({p_\perp}_2) \int \! d\phi_1 \int \! d\phi_2 
\;J\;\hat s \hat \sigma (\hat s)
\nonumber \\[2ex]
&\sim& \frac {(\hbar c)^2 A^{8/3}} {4 \pi R_A^2} 
\frac {J}{\sqrt \pi} (\frac{a}{b e^{1.62b}})^2 (2\pi)^2 \left[ \hat s \hat
\sigma (\hat s) \right]  \sim 0.10 \;\; . 
\end{eqnarray}
Therefore $N_{Min}/N_{Bj} \sim 1$.
From the above estimate we can see that although the minimally-correlated case
allows particles with different rapidities to interact, the dominant
contribution still comes from the region where the two gluons have almost the
same rapidity, thus there is no sizeable enhancement in the pre-equilibrium
charm yield.  The Minimally-Correlated case is very much like the Bjorken case
in that the dominant contribution comes from $y_1 \simeq y_2$ region. 

As a comparison to the above rough estimates in this chapter, the numerical
integration gives $N_{Fb}=3.8$, $N_{Bj}=0.093$, and $N_{Min}=0.078$, so
$N_{Min}/N_{Bj} \sim 80\%$. 

\hmysection{Discussion}
\label{sec-discussion} 

In order to see the formation-time effect, in Figure~\ref{fig:formation} we
compare the results from two different formation-time assumptions.  The solid
curve is obtained using the Lorentzian form in equation~(\ref{EQ:lorentzian}),
and the dashed curve using the theta function form in
equation~(\ref{EQ:theta}).    
The result from this $\theta$-function is about 10\% higher at $p_\perp=0$GeV,
and 10\% lower at $p_\perp=9$GeV.
The lack of sensitivity to the formation-time physics is due to the relative
large $p_\perp$ for the gluon minijets in the charm production process.  There
would be more sensitivity had the production been dominated by low $p_\perp$
components.     

We also see that for the soft+hard distribution, soft gluons significantly
increase the pre-equilibrium charm production in both low-$p_\perp$ and
high-$p_\perp$ region, with the largest increase in low-$p_\perp$ region.  It
is interesting to identify where the enhancement comes from.  
In Figure~\ref{fig:minimal}, the curve with diamonds shows the contribution
from the fusion of soft gluons both with $p_\perp<2$ GeV, and the curve with
hollow squares shows the contribution from the fusion of hard gluons both with
$p_\perp>2$ GeV.  These two curves are both very low compared with the curve
calculated from the soft+hard distribution.  Therefore the enhancement going
from the hard distribution to the soft+hard distribution is mainly due to the
fusion of hard and soft minijet gluons.   

We have noted before that our fit for the minijet gluon spectrum falls below
the Monte Carlo result from HIJING calculation.  We can fit the soft gluons
from HIJING better by using $0.265 e^{-2.6 p_\perp}$ for $p_\perp \in (0, 1.1)$
GeV, and use the old fit $0.06 e^{-1.25p_\perp}$ for higher-$p_\perp$ gluons.
This new fit gives us more very soft gluons.  Using the new fit, we have done
the calculation for minimally-correlated case, and the result is different
only by less than 10\%.  We therefore conclude that super-soft gluons are
not very important for the pre-equilibrium charm production.

\hmysection{Summary}
\label{sec-summary} 

For the pre-equilibrium charm production, we studied the
effect of correlations between the rapidity $y$ and space-time rapidity $\eta$
of minijet gluons. 
In the ideal Bjorken-correlated case, where $\eta=y_1=y_2$, the pre-equilibrium
charm production is negligible compared with the yield due to initial gluon
fusion.  
For the opposite extreme fireball case, corresponding to uncorrelated $y$ and
$\eta$,  the pre-equilibrium charm production is almost a factor of 50 larger
than in the Bjorken-correlated case and is comparable with the initial charm
yield \cite{mw}.  With the estimates of the total pre-equilibrium charm number,
we found that the difference comes mainly from the $\eta-y$
correlation. Therefore, the pre-equilibrium charm production is very sensitive
to the $(\eta-y)$ correlations in the initial state.     

In order to investigate the effect of more realistic correlations that may
exist in the initial minijet plasma, we introduced a minimal correlation model
taking into account  the uncertainty principle along the lines of
\cite{geiger2}.  This minimal correlation is similar to the ideal Bjorken
correlation case and produces negligible pre-equilibrium charm compared with
the initial charm yield.  

%% file: sec6-ratio.tex
\hmychapter{Gluon Shadowing In Lepton Pair Spectra From Open Charm Decay}
\label{sec-ratio}

Dileptons from charm decay are probably dominant relative to the conventional
signals from the plasma at RHIC such as thermal dileptons 
 \cite{vogt_charmdilepton}.  
Here we offer a different point of view on the dilepton measurement
 \cite{lm_dilepton}. 
Instead of considering it merely as a background, we propose to use
opposite-sign lepton pairs ($ee$,$e\mu$ and $\mu\mu$) from open charm decay in
$p+A$ collisions as a measure of nuclear shadowing effects.  
We focus here on $p+A$ collision rather than $A+A$ to minimize the
combinatorial $\pi,K$ decay backgrounds and other final state interaction
effects.  

\hmysection{Introduction}

The spectrum of dileptons produced in heavy ion collisions at high energies has
been proposed in the past to provide  information about the dynamical evolution
of quark-gluon plasmas (see, e.g., the review by Ruuskannen in
\cite{dilepton_review}).  Unlike hadronic probes, this spectrum is sensitive
to the earliest moments in the evolution, when the energy densities are an
order of magnitude above the QCD confinement scale (1 GeV/fm$^3$).  However,
the small production cross sections for pairs with invariant mass above 1 GeV 
together with the large combinatorial background from decaying hadrons
necessitate elaborate procedures to uncover the signal from the noise.  The
PHENIX detector \cite{cdr}, now under construction at the Relativistic Heavy
ion Collider (RHIC), is designed to measure $ee,e\mu$, and $\mu\mu$ pairs to
carry out this task.  One of the important background sources in the few
GeV mass range arises from semileptonic decay of charmed hadrons ($D \bar D$).
As shown recently in \cite{vogt_charmdilepton}, the expected thermal signals in
that mass range may only reach 10\% of pairs from open charm
decay.  It is much more difficult to precisely measure these signals
than to measure dileptons from charm decay, and special kinematical cuts and
precise $e\mu$ measurements are necessary to uncover  the thermal signals
 \cite{cdr}.  

There have been several attempts \cite{geiger,mw} to take advantage of the
large open charm background as a probe of the evolving gluon density.  Most
mid-rapidity charm pairs are produced via gluon fusion \cite{brodsky}.
Therefore, the dileptons from charm decay carry information about the
distribution of primordial gluons before hadronization.  In fact, the
inside-outside cascade nature of such reactions greatly suppresses
 \cite{lm_charm,lmw} all sources of charm production except  the initial
perturbative QCD source. Therefore, the open charm background is dominated by
the {\em initial} gluon fusion ($gg\rightarrow c\bar{c}$) rate.  

The initial $c\bar c$ rate depends strongly on the  nuclear gluon structure
function \cite{vogt_shadowing}, $g_A(x,Q^2)$. The quantity of fundamental
interest \cite{muellerq,eskolaq} is the  gluon shadowing function 
\begin{equation}
R_{g/A}(x,Q^2)=\frac {g_A(x,Q^2)} {Ag_N(x,Q^2)} \;\; . 
\label{EQ:shad} 
\end{equation}
The point of this study is to demonstrate that the $A$ dependence of continuum
dilepton pairs in the few GeV mass region provides a novel  probe of that
unknown gluon structure.  We also show that the required measurements of
$ee$,$e\mu$, and $\mu\mu$ pair yields in  $p+A\rightarrow \ell^+ \ell^- X$ at
$\sqrt s=200$ AGeV are not only experimentally feasible at RHIC but also that
the open charm signal can be easily extracted via the proposed PHENIX detector.

A  key advantage of the continuum dilepton pairs from open charm decay over
those from  $J/\psi$ decay is that it is possible to test the applicability of
the underlying QCD dynamics at a given fixed $\sqrt s$  by  checking  for a
particular scaling property discussed below.  In  quarkonium  production the
mass is fixed and the required scaling can be checked only by varying the beam
energy.  In $p+A\rightarrow J/\psi$ production  the required scaling was
unfortunately found to be violated in the energy range  $20< \sqrt{s} < 40$
AGeV \cite{moss}, thus precluding a determination of $R_A$.  Possible
explanations for the breakdown of QCD scaling for $J/\psi$ production and its
anomalous  negative $x_F$ behavior \cite{vogt_shadowing,moss} include
interaction of next-to-leading order quarkonium Fock state with nuclear matter
 \cite{satz_octet}, nuclear and co-mover $J/\psi$ dissociation \cite{huf}, and
parton energy loss mechanisms \cite{gavin}.  It remains an open question
whether the required scaling will set in at RHIC energies $60 < \sqrt s < 200$
AGeV.  

We propose to use $ee$,$e\mu$ and $\mu\mu$ pairs from open charm decays to
probe the shadowing effects in nuclei.    
These three measurements cover three different rapidity regions as a result of
the PHENIX detector geometry.
In Chapter~\ref{sec-ratio} we calculate the lepton pair spectrum from open
charm decay in $p+Au$ at 200 GeV/A using two commonly used shadowing scenarios:
the parameterization which does not depend on scale or parton component
\cite{hijing}, and the scale-dependent, component-dependent shadowing
\cite{shadowing}.      
The difference between these two shadowing scenarios is significant.
The ratio curve of dilepton $dN/dMdy$ spectra from $p+A$ to those from $pp$ has
a similar shape to the shadowing curve.  
We then show that there is an approximate scaling among the ratio curves at
different dilepton pair-mass $M$, which enables us to bring them together to
the shadowing function curve. 
In Chapter~\ref{sec-bg} we first estimate the original backgrounds of the
lepton pair spectra.  
The large number of electrons from Dalitz decay and the muons from
random decay of pions or kaons results in large backgrounds for $e\mu$ and
dilepton experiments.  
Without suppression and kinematical cuts, it is much higher than the signal
from charm decay.  In chapter~\ref{sec-phenix}, we consider the PHENIX
detector geometry, a specific kinematical cut and the suppression factors of
the background leptons.  We calculate the lepton pair signal and backgrounds
entering the detector, and the signal-to-background ratio is very promising.
For $ee$ and $e\mu$ spectra, the signal is bigger than the background.  For
$\mu\mu$ spectra, the signal is smaller than the background, 
but with the help of like-sign subtraction, it is possible to also observe the
$\mu\mu$ signal from open charm decay.
In chapter~\ref{sec-single}, we show the signal and background for the single
inclusive electron and muon spectrum.  
The signal from single inclusive leptons reflects the shadowing effects less
effectively, and the signal-to-background ratio there is much smaller than that
for the lepton pair case. 
Finally in chapter~\ref{sec-sum} we give the discussion and summary.   

\hmysection{A Recent Attempt to Extract Nuclear Gluon Shadowing}

Recently there appeared an attempt \cite{pirner} to extract nuclear gluon
shadowing from the high statistics $F_2^A$ data on $S_n$ and $C$ taken by NMC
\cite{mucklich}.

Parton structure functions depend on Bjorken $x$, the momentum fraction of
the parton in the frame where the nucleon (or nucleus) is moving very fast, and
$Q^2$, the momentum scale at which the parton structure function is being
resolved.
The $Q^2$ evolution equations for the gluon number density, $G(x)$, and the
quark number densities, $q_i(x)$ ($i$ from $1$ to $2n_f$), are given by the
Altarelli-Parisi equations \cite{ap} as 
\begin{eqnarray}
\frac {\partial G (x)} {\partial t} &=&
\frac {\alpha}{2 \pi} \int_x^1 \frac{dy}{y} \left [
\sum_j P_{G q_j} (z) q_j(y) + P_{G G} (z) G(y) \right ]  \;\; , \nonumber
\\[2ex] 
\frac {\partial q_i (x)} {\partial t} &=&
\frac {\alpha}{2 \pi} \int_x^1 \frac{dy}{y} \left [
\sum_j P_{q_i q_j} (z) q_j(y) + P_{q_i G} (z) G(y) \right ] \;\; , 
\label{EQ:ap}
\end{eqnarray}
where $z \equiv x/y$, and $t \equiv \ln (Q^2/Q_0^2)$.  
There are four splitting functions, $P_{G q_j}(z),P_{G G}(z),P_{qq}(z),P_{q_i
q_j}(z)$ and $P_{q_i G}(z)$, and a function $P_{AB}(z)$ is basically
proportional to the probability that a parton $B$ splits to a parton $A$ with
fraction $z$ of the momentum of the parent parton $B$.

Based on the Altarelli-Parisi equation (\ref{EQ:ap}) for quarks, 
the $Q^2$ evolution equation for the quark distribution 
\begin{eqnarray}
F_2(x,Q^2) = x \sum_i e_i^2 q_i(x,Q^2)
\end{eqnarray}
is therefore given as
\begin{eqnarray}
\frac {\partial F_2(x)} {\partial t} &=&
\frac {\alpha}{2 \pi} x \int_x^1 \frac{dy}{y} \left [
\sum_{ij} e_i^2 P_{q_i q_j} (z) q_j(y) + \sum_i e_i^2 P_{q_i G} (z) G(y)
\right ] \\[2ex]
&=&
\frac {\alpha}{2 \pi} x \int_x^1 \frac{dz}{z} \left [
\sum_i e_i^2 P_{qq} (z) q_i(\frac {x}{z}) + \sum_i e_i^2 
P_{qG} (z) G(\frac {x}{z}) \right ] \\[2ex]
&\equiv &
\frac {\alpha}{2 \pi} \left (
P_{qq} \ast F_2 + x \sum_i e_i^2 P_{qG} \ast G \right ) \;\; .
\label{EQ:p_qg} 
\end{eqnarray}

The above two convolutions are defined as
\begin{eqnarray}
P_{qG} \ast G &=&
\int_x^1 \frac{dz}{z} P_{qG} (z) G(\frac {x}{z})
= \int_x^1 \frac{dz}{2z} [z^2+(1-z)^2] G(\frac {x}{z}) \\[2ex]
P_{qq} \ast F_2 &=&
\int_x^1 dz P_{qq} (z) F_2(\frac {x}{z}) \;\; , \nonumber \\
&=&
\int_x^1 dz \frac {N_c^2-1}{2N_c} \left [
\frac {1+z^2}{(1-z)_+}+\frac {3}{2} \delta (1-z) \right ] F_2(\frac {x}{z})
\nonumber \\
&=&
\frac {4}{3} \! \int_x^1 \! dz \! \left [
\frac {(1+z^2)F_2(x/z)-2F_2(x)}{(1-z)}+ \! \frac {3}{2} \! \delta (1-z)
F_2(\frac {x}{z}) \right ] \! ,
\end{eqnarray}
with the divergence at $z=1$ being regularized by the definition
\begin{eqnarray}
\int_0^1 dz \frac {f(z)}{(1-z)_+} \equiv \int_0^1 dz \frac {f(z)-f(1)}{(1-z)}
\label{EQ:plus}
\;\; .
\end{eqnarray}

Using the approximation
\begin{eqnarray}
\int_x^1 dz [z^2+(1-z)^2] \frac{x}{z} G(\frac {x}{z})
&\simeq & \frac{4x}{3} G(2x) \;\; ,
\end{eqnarray}
one can relate the second term in eq.(\ref{EQ:p_qg}), $P_{qG} \ast G$, directly
to the gluon density function $G(x)$.  Note that the first term in
eq.(\ref{EQ:p_qg}), $P_{qq} \ast F_2$, is only related to $F_2(x)$, which can
be computed from data. 

Thus from the $Q^2$ dependence of the quark distribution $F_2$, one can
extract some information on the gluon density $G$.  
In the above, one neglects the effects of higher order corrections and the
non-linear recombination terms \cite{muellerq}.  
The non-linear term probably does not have a large magnitude unless the parton
momentum fraction $x$ is extremely small \cite{muellerq,shadowing,kumano}.  

\begin{figure}[p]
\psfig{figure=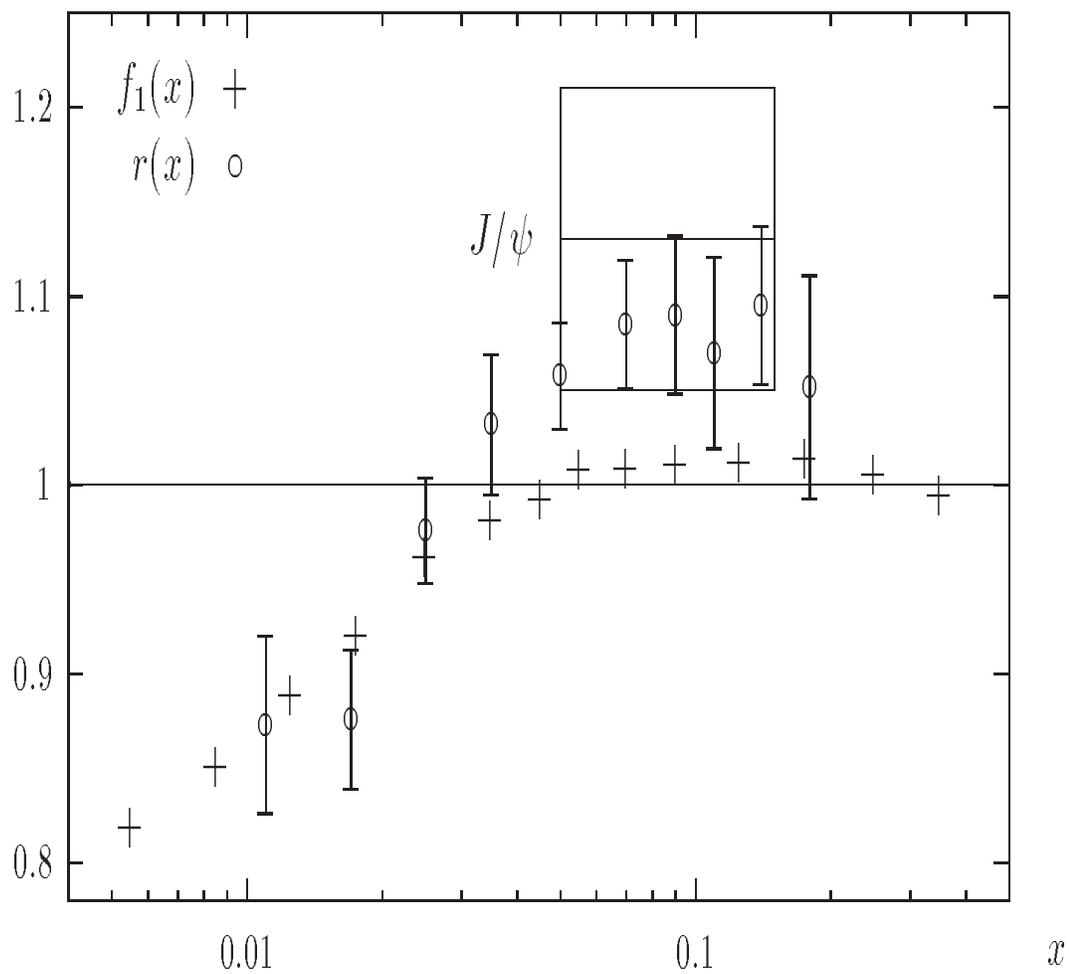,height=5.in,width=5.5in,angle=0} 
\caption{
The ratio $r(x)=G^{Sn}(x)/G^C(x)$ of tin to carbon gluon density, together with
the ratio of structure function, $f_1(x)=F^{Sn}_2(x)/F^C_2(x)$, as a function
of $x$, from \protect{\cite{pirner}}.
}
\label{fig:ratio_g}
\end{figure}

In Figure~\ref{fig:ratio_g} \cite{pirner}, the so-extracted ratio of gluon
density of tin to carbon $r(x,Q^2)=G^{S_n} (x,Q^2)/G^C (x,Q^2)$ is plotted
against the gluon momentum fraction $x$ in small $x$ range $0.01 < x < 0.1$,
together with the $F_2(x,Q^2)$ data from which the ratio $r(x,Q^2)$ is
extracted. The ratio of gluon density shows both the shadowing effect in the
lower range of $x$ and the anti-shadowing effect in the higher range of $x$,
similar to the case of quark shadowing. 

\hmysection{The $Q^2$ Evolution of Nuclear Parton Distributions}

There are mainly two steps to determine nuclear shadowing effects.  First we 
need to know the parton distribution functions in nuclei at some initial
momentum scale $Q_0$.  Given the input nuclear parton distributions at the
scale $Q_0$, we then calculate the parton structure functions at any values of
scale $Q$ according to evolution equations.  As long as $Q$ is within the
perturbative QCD domain, $Q^2$ evolution equations are known from pQCD
\cite{ap,muellerq}. 

The input parton structure functions, $G(x,Q^2_0)$ and $q_i(x,Q^2_0)$, contain
all non-perturbative nuclear effects at the scale $Q_0$.  They are thus, in
principle, not calculable within pQCD.  However, it is not hard to imagine that
nuclear shadowing effects occur to nuclear parton distributions \cite{qiu}.
We can imagine that shadowing effects occur when partons from different
nucleons in nuclei begin to overlap spatially.  For a nucleus with momentum $p$
per nucleon, the longitudinal size of the parton with momentum $xp$ is
$\sim 1/(xp)$.  The longitudinal separation of different nucleons is $\sim
2r_0m/p$, where $m$ is the nucleon rest mass, and $r_0$ is the nucleon radius.
Thus shadowing effects occur when $x < 1/(2r_0m)$.  Shadowing effects grow as
partons overlap more, until the partons are overlapping completely.  This
saturation of shadowing effects happens when the longitudinal size of the
parton is about equal to the longitudinal size of the nucleus, i.e. $x \sim
1/(2R_Am)$, where $R_A$ is the radius of the nucleus.  

$Q^2$ evolution equations of parton density functions are usually given by
Altarelli-Parisi equations (AP) \cite{ap}, as expressed in eqs.~(\ref{EQ:ap}). 
They can be rewritten separately for the gluon momentum distribution $xG$, the
flavor singlet sea quark part $xS$, and the flavor nonsinglet valence quark
part $xV$: 
\begin{eqnarray}
Q^2 \frac {\partial }{\partial Q^2} xG(x,Q^2) &=&
\frac {\alpha (Q^2)}{2\pi} \int_x^1 \frac {dy}{y} 
\left [ z P_{Gq}(z) y \left ( V(y,Q^2) + S(y,Q^2) \right ) \right .\nonumber \\
&& \hspace{1cm} \left . + z P_{GG}(z) y G(y,Q^2) \right ]
\label{EQ:ap_g2} \;\; , \\[2ex]
Q^2 \frac {\partial }{\partial Q^2} xS(x,Q^2) &=&
\frac {\alpha (Q^2)}{2\pi} \int_x^1 \frac {dy}{y} 
\left [ z P_{qq}(z) y S(y,Q^2) \right . \nonumber \\
&& \hspace{1cm} \left . + z P_{qG}(z) y G(y,Q^2) \right ]
\label{EQ:ap_s2} \;\; , \\[2ex]
Q^2 \frac {\partial }{\partial Q^2} xV(x,Q^2) &=&
\frac {\alpha (Q^2)}{2\pi} \int_x^1 \frac {dy}{y} z P_{qq}(z) y V(y,Q^2) \;\; ,
\end{eqnarray}
where the sea quark momentum distribution $xS=xu_s, xd_s$ or $xs_s$. 

The four splitting functions are given as \cite{ap}
\begin{eqnarray}
P_{qG}(z) &=& \frac {1}{2} \left ( z^2 + (1-z)^2 \right ) \;\; , \\[2ex]
P_{GG}(z) &=& 6\left ( \frac {1-z}{z} + \frac {z}{(1-z)_+} + z(1-z) \right )
\nonumber \\
&& + \left ( \frac {11}{2}-\frac {n_f}{3} \right ) \delta (1-z) \;\; , \\[2ex]
P_{qq}(z) &=& \frac {4}{3} \frac {1+z^2}{(1-z)_+} + 2 \delta (1-z) \;\; ,
\\[2ex] 
P_{Gq}(z) &=& \frac {4}{3} \frac {1+(1-z)^2}{z} \;\; ,
\end{eqnarray}
where $(1-z)^{-1}_+$ is defined in eq.~(\ref{EQ:plus}).

The AP-evolution means that as $Q^2$ increases, we can see more small-$x$
partons and less high-$x$ ones, because high-momentum partons lose energy by
radiation.  Thus the parton density can become very large at sufficiently small
$x$.  However, as the parton density becomes huge at small $x$, the
AP-evolution equations have to be modified \cite{muellerq} because gluons then
have more probability to recombine to reduce the numbers of partons in the
nuclei.  Mueller and Qiu \cite{muellerq,qiu} modified the AP-equations by
calculating the recombination probabilities for gluons to go into gluons or
into quarks.  The AP-evolution equation(\ref{EQ:ap_g2}) for gluons and
equation(\ref{EQ:ap_s2}) for sea quarks are modified to the following
AP+MQ-equations: 
\begin{eqnarray}
Q^2 \frac {\partial }{\partial Q^2} xG(x,Q^2) &=&
\frac {\alpha (Q^2)}{2\pi} \int_x^1 \frac {dy}{y} 
\left [ z P_{Gq}(z) y \left ( V(y,Q^2) + S(y,Q^2) \right ) \right .\nonumber \\
&& \hspace{1cm} \left . + z P_{GG}(z) y G(y,Q^2) \right ]
\nonumber \\ 
&& \hspace{1cm} -\frac {n}{Q^2} R_{gg\rightarrow g}(x,Q^2)
\label{EQ:mq_g} \;\; , \\[2ex] 
Q^2 \frac {\partial }{\partial Q^2} xS(x,Q^2) &=&
\frac {\alpha (Q^2)}{2\pi} \int_x^1 \frac {dy}{y} 
\left [ z P_{qq}(z) y S(y,Q^2) \right . \nonumber \\
&& \hspace{1cm} \left . + z P_{qG}(z) y G(y,Q^2) \right ]
\nonumber \\ 
&& \hspace{1cm} -\frac {n}{Q^2} \left [ R^{(1)}_{gg\rightarrow
q}(x,Q^2)+ R^{(2)}_{gg\rightarrow q}(x,Q^2) \right ] \;\; , 
\end{eqnarray}
where the nuclear number density $n \sim 1/(\pi R^2)$, and the contributions to
the evolution from recombinations of gluons into gluons or into quarks are 
\begin{eqnarray}
R_{gg\rightarrow g}(x,Q^2)&=&\frac {4\pi^3}{N^2-1} \left [\frac {\alpha(Q^2)
C_A}{\pi} \right ]^2 \int_x^1 \frac {dy}{y} \left [yG(y,Q^2) \right ]^2 \;\; ,
\\[2ex] 
R^{(1)}_{gg\rightarrow q}(x,Q^2)&=&\frac {4\alpha^2(Q^2)T_f\pi}{N(N^2-1)} 
\left ( \frac {4N^2}{15}-\frac {3}{5} \right ) \left [xG(x,Q^2) \right ]^2
\;\; , \\[2ex] 
R^{(2)}_{gg\rightarrow q}(x,Q^2)&=& -\frac {2\alpha(Q^2)T_f}{\pi} 
\int_x^1 \frac {dy}{y} z \bar \gamma_{FG}(z) y G_{HT}(y,Q^2) \;\; .
\end{eqnarray}
In the above $T_f=1/2$, $C_A=N=3$, $\bar \gamma_{FG}(z)=-2x+15x^2-30x^3+10x^4$.
\begin{eqnarray}
Q^2 \frac {\partial }{\partial Q^2} yG_{HT}(y,Q^2) &=&
-R_{gg\rightarrow g}(y,Q^2)
\end{eqnarray}
gives the evolution for the higher-dimensional gluon distribution $G_{HT}$.
This distribution function is basically not calculable \cite{qiu}.  Since it
comes from two-gluon interaction, it has the form, at some scale $Q_0$, as  
\begin{eqnarray} 
yG_{HT}(y,Q_0^2) = K_{HT} \left [ yG(y,Q_0^2) \right ]^2 \;\; , 
\end{eqnarray}
where $K_{HT}$ is a constant.

From the above AP+MQ-equations, the modification terms from two-gluon
recombinations to the AP-equations are basically of the form
\begin{eqnarray}
\alpha^2(yG)^2/(R^2 Q^2) \;\; .
\end{eqnarray}
We take gluons for
example, and {\it roughly} represent the modified evolution
equation(\ref{EQ:mq_g}) as  
\begin{eqnarray}
Q^2 \frac {\partial xG(x,Q^2)}{\partial Q^2} \;\sim \;
\alpha \frac {6}{2\pi} xG(x,Q^2) - \alpha^2 \frac {4\pi n}{Q^2} \left [
xG(x,Q^2) \right ]^2 \;\; . 
\label{EQ:symbol}
\end{eqnarray}
The modification term, the second on the right-hand side in
eq.(\ref{EQ:symbol}), therefore slows down the $Q^2$ evolution of the gluon
density function.  To make a rough estimate of the onset of non-linear
recombination effects, we equal the two terms in eq.(\ref{EQ:symbol}), the
usual AP-evolution term and the gluon recombination term, and get the condition
as 
\begin{eqnarray}
xG(x,Q^2) \sim \frac {3Q^2}{4\pi^2 n \alpha} \sim \frac {Q^2 R^2}{\pi \alpha} 
\;\; .
\end{eqnarray}
The modification will be significant only when $x$ is so small that the above
condition is reached.  If we take $R=1fm$, $Q \simeq 2$GeV, the above condition
will be reached only for extremely small $x$ where $xG(x,Q^2) \sim 100$.  
Thus the recombination is likely to be a very small correction to the usual
$Q^2$ evolution equations of parton structure functions. 

Modified evolution equations (AP+MQ) was first numerically solved by Qiu
\cite{qiu}, and then by Eskola \cite{shadowing}. 
Eskola assumes a shadowing ansatz at an initial scale $Q_0=2$GeV for heavy
nuclei with $A \simeq 200$, then evolves shadowed Duke-Owens parton
distributions \cite{do} according to both AP equations and modified evolution
equations (AP+MQ). 
The MQ-modification slows down the increase of the parton densities with $Q^2$,
and therefore shadowing effects go away slower with the increase of $Q^2$ than
in the AP-evolution case.  However, the difference between the AP-evolution and
the AP+MQ-evolution with this particular choice of input parton distributions
is found to be only on the order of 10\% \cite{shadowing}, thus the mainline of
the evolution of parton structure functions is given by the usual AP-equations.

\hmysection{Nuclear Shadowing in Open Charm Production}

We use two common shadowing scenarios: the parameterization which does not
depend on scale or component (HIJING's scaling) \cite{hijing}, and the
scale-dependent, component-dependent shadowing (Eskola's shadowing)
\cite{shadowing}.  
The production for $c\bar c$ in one $A+B$ collision at a fixed impact-parameter
$\vec b$ is given by \cite{mw} 
\begin{eqnarray}
\frac {dN^{AB}}{dp_\perp^2 dy_1 dy_2}
= K \int d^2r \sum _{a,b} x_a \Gamma_{a/A}(x_a, Q^2, \vec r) 
x_b \Gamma_{b/B}(x_b, Q^2, \vec b-\vec r) \frac {d \sigma_{ab}}{d \hat t} 
\;\; ,  
\label{EQ:dncc} 
\end{eqnarray}
where 
\begin{eqnarray}
\Gamma_{a/A}(x, Q^2,\vec r)=T_A(\vec r) f_{a/N}(x,Q^2) R_{a/A}(x,Q^2, \vec r)
\end{eqnarray}
is the nuclear parton density function in terms of the known nucleon parton
structure functions $f_{a/N}(x,Q^2)$, the  nuclear thickness function $T_A(\vec
r) = \int dz n_A(\sqrt {z^2+\vec r ^2})$, and the unknown
impact-parameter-dependent shadowing function $R_{a/A}(x,Q^2, \vec r)$.   
The conventional kinematic variables $x_{a,b}$ are the incoming parton light
cone momentum fractions, $\hat t=-(p_b-p_1)^2$, and the final $c,\bar{c}$ have
rapidities $y_1,y_2$ and transverse momenta $\vec{p}_{1\perp}=-\vec{p}_{2\perp}
\equiv \vec{p}_\perp$.  Next-to-leading order corrections to the lowest order
parton cross sections ${d \hat \sigma_{ab}}/{d \hat t}$ are approximately taken
into account by a constant $K$ factor \cite{sv,vogt_kfactor}.   We use the
recent MRSA \cite{mrsa}  parton structure functions.  From a fit to low energy
open charm production data in $pp$ collisions, we fix $m_c=1.4$ GeV, $K=3$,
$Q^2 = \hat s/2$ for $g g \rightarrow c\bar c $ and $\hat s$ for $q\bar q
\rightarrow c \bar c$.  This choice of parameters leads to a charm pair cross
section of $340 \mu b$ for $pp$ collisions at RHIC.  For
impact-parameter-averaged collisions, the integral over the transverse vector 
in eq.(\ref{EQ:dncc}) leads to a factor:  
\begin{eqnarray}
BA R_{b/B}(x_b,Q^2) R_{a/A}(x_a,Q^2)/\sigma ^{BA}_{in} \;\; , 
\end{eqnarray}
where $\sigma ^{BA}_{in}$ is the inelastic $B+A$ cross section and
$R_{a/A}(x_a,Q^2)$ is the impact-parameter-averaged shadowing function. 

Therefore for $c\bar c$ production in impact-parameter-averaged $p+Au$
collisions, we have 
\begin{eqnarray}
\frac {d\sigma^{p+Au}}{dp_\perp^2 dy_1 dy_2}
= K A \sum _{a,b} x_a f_{a/N}(x_a, Q^2)  x_b f_{b/N}(x_b, Q^2) 
R^{p+Au}_{b/Au}(x_b,Q^2)\frac {d \sigma_{ab}}{d \hat t}, 
\end{eqnarray}
where $R^{p+Au}_{b/Au}(x_b,Q^2)$ is the impact-parameter-averaged shadowing
function for the parton $b$ in $Au$ nucleus.  

\begin{figure}[p]
\psfig{figure=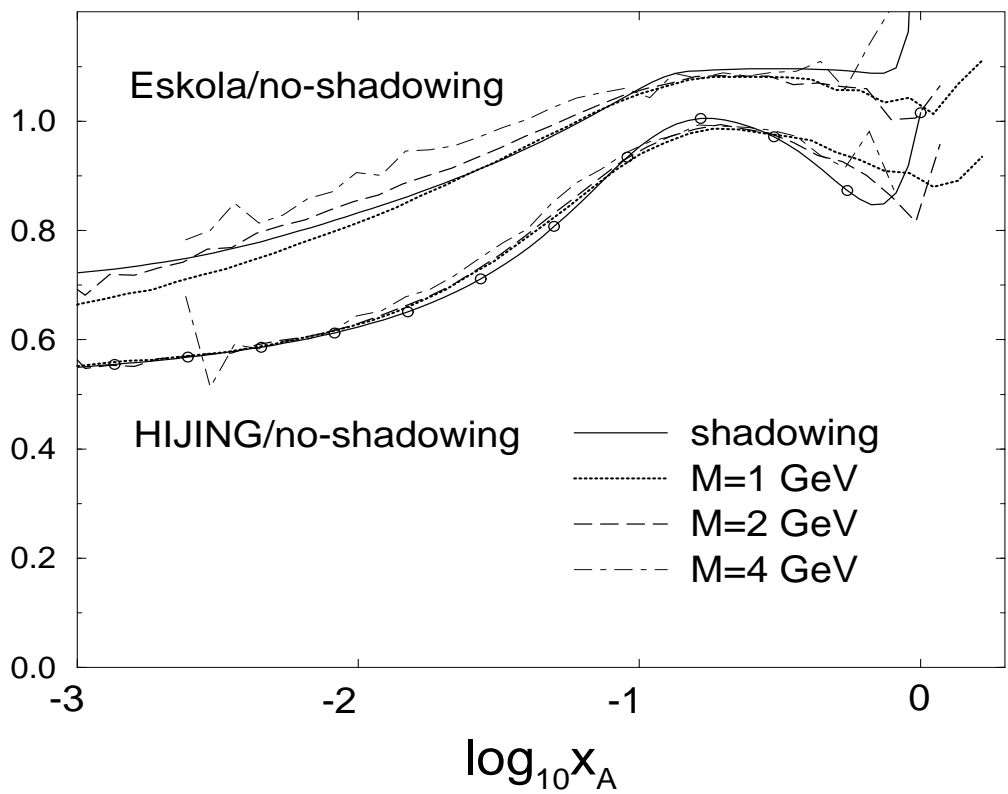,height=5.in,width=5.in,angle=-90} 
\caption{
Dilepton $dN/dMdy$ ratio curves (shadowed over unshadowed) 
as a function of the scaling variable $\log _{10} x_A$ at different masses, and
the comparison with shadowing curves.   
}
\label{fig:scaling}
\end{figure}

The two different shadowing functions are plotted in Figure~\ref{fig:scaling}.
It is clear that HIJING's scenario has much stronger shadowing than Eskola's
scenario in almost all regions of interest, and Eskola's scenario has
anti-shadowing in $x \geq 0.1$ region.  These features are reflected in the
final $e\mu$ spectrum, as we will show in the following chapter.    

\hmysection{$e\mu$ Spectrum from $D/\bar D$ Decay}
\label{sec-ddbar}

Since the mass difference of $e$ and $\mu$ has little effect on the
semileptonic decay of charmed meson, the spectra of dilepton pair $ee$ and
$\mu\mu$ from charm decay have shapes almost identical to the $e\mu$ spectrum.
Since we include both $e^+\mu ^-$ and $e^-\mu ^+$ pairs, and assume the average
branching ratio for $D \rightarrow \mu+X$ is the same as for $D \rightarrow
e+X$, the $e\mu$ signal is twice as large as the $ee$ and $\mu\mu$ signal.  

To compute the $D\bar D$ pair distribution, a hadronization scheme for
$c \rightarrow D$ must be adopted.  In $e^+e^-\rightarrow c\bar{c}\rightarrow
D\bar{D}X$ hadronization can be modelled via string fragmentation or fitted,
for example, via the Peterson fragmentation function (see \cite{vogt_delta}).
The final $D$ carries typically only a fraction $\sim 0.7$ of the original $c$
momentum. However, in $pp$ collisions, charm hadronization is complicated by
the high density of partons produced during beam jet fragmentation.  In this
system recombination or coalescence of the heavy quark with co-moving partons
provides another mechanism which in fact  seems to be dominant at least at
present energies.  The inclusive $pp \rightarrow DX$ data is best reproduced
\cite{vogt_delta,e769} with the delta function fragmentation $D(z)=\delta
(1-z)$ in all observed  $x_f$ and moderate $p_T$ regions.  This observation can
be understood in terms of a coalescence model if the coalescence radius is $P_c
\sim 400$ MeV.  We assume that hard fragmentation continues to be the dominant
mechanism at RHIC energies, and therefore no additional A-dependent effects
from hadronization will arise.  However, this assumption must be tested
experimentally.  Therefore our main dynamical assumption is that the
$pp\rightarrow D$ transverse momentum distributions can be accurately
reproduced from the QCD level rates assuming hard fragmentation, as at present
energies \cite{e769}.  At RHIC this can be checked either via single inclusive
leptons \cite{cdr} or directly via  $K\pi$ \cite{star}. 

With this assumption, the impact-parameter-averaged $D\bar D$ pair distribution
in $p+A$ is given by 
\begin{eqnarray}
\frac {dN^{pA}}{dp_\perp^2 dy_3 dy_4}
&=& \frac {KA}{\sigma ^{pA}_{in}} \frac {E_3 E_4}{E_1 E_2} \nonumber \\
&& \hspace{-1cm} \times \sum _{b,a} x_b f_{b/N}(x_b, Q^2)  x_a
f_{a/N}(x_a, Q^2) R_{a/A}(x_a,Q^2)\frac {d \hat \sigma_{ab}}{d \hat t} \;\; .
\end{eqnarray}
Note that there is no $p_\perp$ kick here, and $D$ and $\bar D$ in a pair have
opposite azimuthal directions.  

After the above spectrum $dN/dp_\perp^2 dy_3 dy_4$ for the charmed mesons is
calculated, we use Monte Carlo programs to simulate their semileptonic decays
 \cite{jetset}.
We generate $D/\bar D$ events according to the above spectrum.  
For each event, we decay one charmed meson in its rest frame to an electron,
and decay the other charmed meson in its rest frame to a muon isotropically,
according to the parameterized semileptonic energy spectrum.  We then boost the
electron and muon back according to the momentum of its meson parent.  We then
calculate the invariant mass 
\begin{eqnarray}
M=\sqrt {(p_e+p_\mu)^2}
\end{eqnarray}
and the rapidity 
\begin{eqnarray}
y=\tanh ^{-1} [(p_e^{\parallel} +p_{\mu}^{\parallel})/ (E_e +E_\mu)]
\end{eqnarray}
for the pair and have the final $e\mu$ pair spectrum $dN/dMdy$.    

\begin{figure}[p]
\psfig{figure=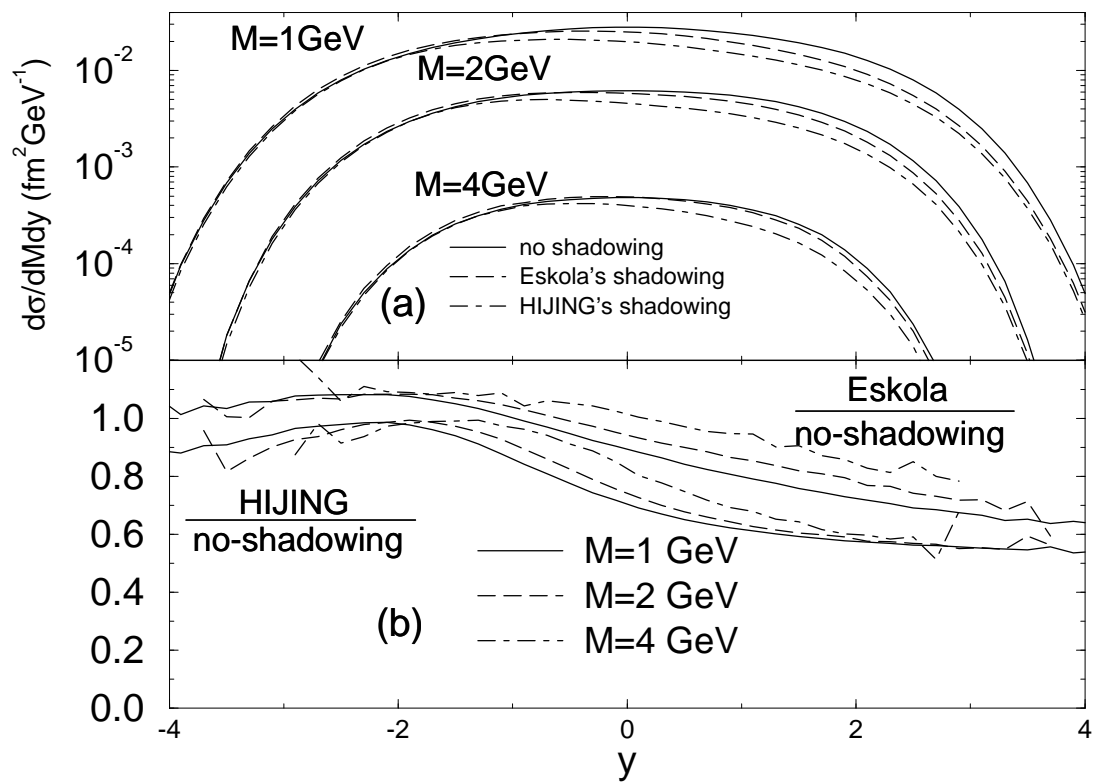,height=5.in,width=3.5in,angle=-90} 
\caption{
(a) The production of opposite-sign $e\mu$ pairs as a function of pair-rapidity
$y$ at pair-masses $M=1, 2, 4$ GeV;  
(b) The ratio of shadowing curves over non-shadowing curves in (a).
}
\label{fig:ccbar}
\end{figure}

In Figure~\ref{fig:ccbar}(a) we plot the absolute rate $d\sigma^{e\mu}/dMdy$
from the above Monte Carlo calculation, and it includes both $e^+ \mu^-$ and
$e^- \mu^+$ signals.  The solid, dashed and dot-dashed curves represent the 
non-shadowing, Eskola's shadowing, and HIJING's shadowing case, respectively.
From top to bottom, the three sets of solid, dashed and dot-dashed curves
correspond to different pair-masses $M=1, 2, 4$ GeV.
In Figure~\ref{fig:ccbar}(b), the ratio of the curves in (a) from two shadowing
cases to that from the non-shadowing case is plotted at pair-masses $M=1$
(solid), $2$ (dashed), $4$ (dot-dashed) GeV.  The upper(lower) set of three
curves corresponds to Eskola(HIJING)'s shadowing.  
The curves in Figure~\ref{fig:ccbar}(b) resemble the shapes of shadowing
curves in Figure~\ref{fig:scaling}.    
Note that in Figure~\ref{fig:scaling} the abscissa is in logarithmic scale and
in Figure~\ref{fig:ccbar} the abscissa is in linear scale.    
We see that the shadowing effect is much stronger in positive pair-rapidity
region.  This is expected since in $p+A$ collisions positive pair-rapidity
corresponds to smaller $x$ from nucleus $A$ and thus deeper shadowing.  For
Eskola's scenario some anti-shadowing effect shows up in the negative
pair-rapidity region.  One may notice that in Figure~\ref{fig:ccbar}(b) we have
a narrower rapidity region at higher invariant masses. 
That is because at the edges of the rapidity region we have fewer Monte Carlo
events, and we discard the edges to restrain big statistical fluctuations.  

\hmysection{Scaling}
\label{sec-scaling}

Since gluon fusion dominates, there is an approximate scaling in terms of the
light cone fraction $x_A$ that one of the gluon carries from nucleus $A$:
\begin{eqnarray}
\ln x_A= -y_{c\bar{c}}+ \ln( M_{c\bar{c}}/\sqrt s) \;\; . 
\end{eqnarray}

\begin{figure}[p]
\psfig{figure=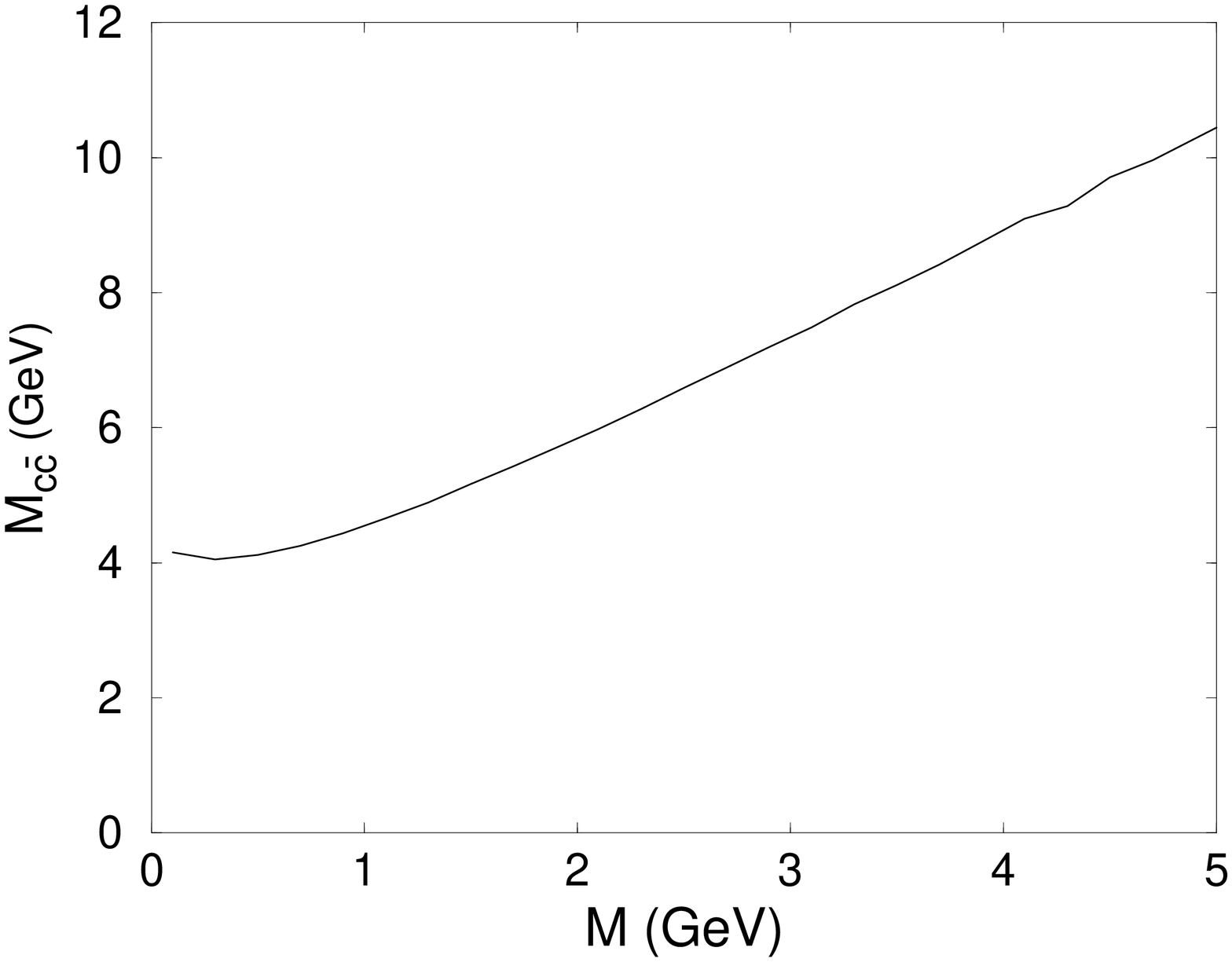,height=5.in,width=3.5in,angle=-90} 
\caption{
The average mass $M_{c\bar c}$ as a function of dilepton mass
$M$.  
}
\label{fig:mass}
\end{figure}

While the lepton pair-mass $M$ and rapidity $y$ fluctuate around a mean value
for any fixed $M_{c\bar{c}}$ and $y_{c\bar{c}}$, on the average 
\begin{eqnarray}
\langle y \rangle \approx \langle y_{c\bar{c}} \rangle \;\; , 
\end{eqnarray}
and $\langle M_{c\bar{c}} \rangle$ can be well approximated by a linear
relation   
\begin{eqnarray}
\langle M_{c\bar{c}} \rangle \approx \beta M+ M_0
\end{eqnarray}
over the lepton pair-mass range $1 \le M \le 4$GeV.  For the delta function
fragmentation case, $\beta\approx 1.5$ and $M_0 \approx 3.0$ GeV are determined
by the D-meson decay kinematics. In Figure~\ref{fig:mass} we plot the average
$M_{c\bar{c}}$ as a function of $M$, the mass of the correlated lepton pair
from the $c\bar c$ decay with the delta function fragmentation.
The above relation between the average masses is determined from the following
histogram plot generated from Monte Carlo calculation, which showed the event
number density on the plane of charm pair-mass $M_{c\bar c}$ and the mass $M$
of the dielectron from the charm semileptonic decays (the unit of mass is GeV
in the figure). 
\begin{figure}[p]
\psfig{figure=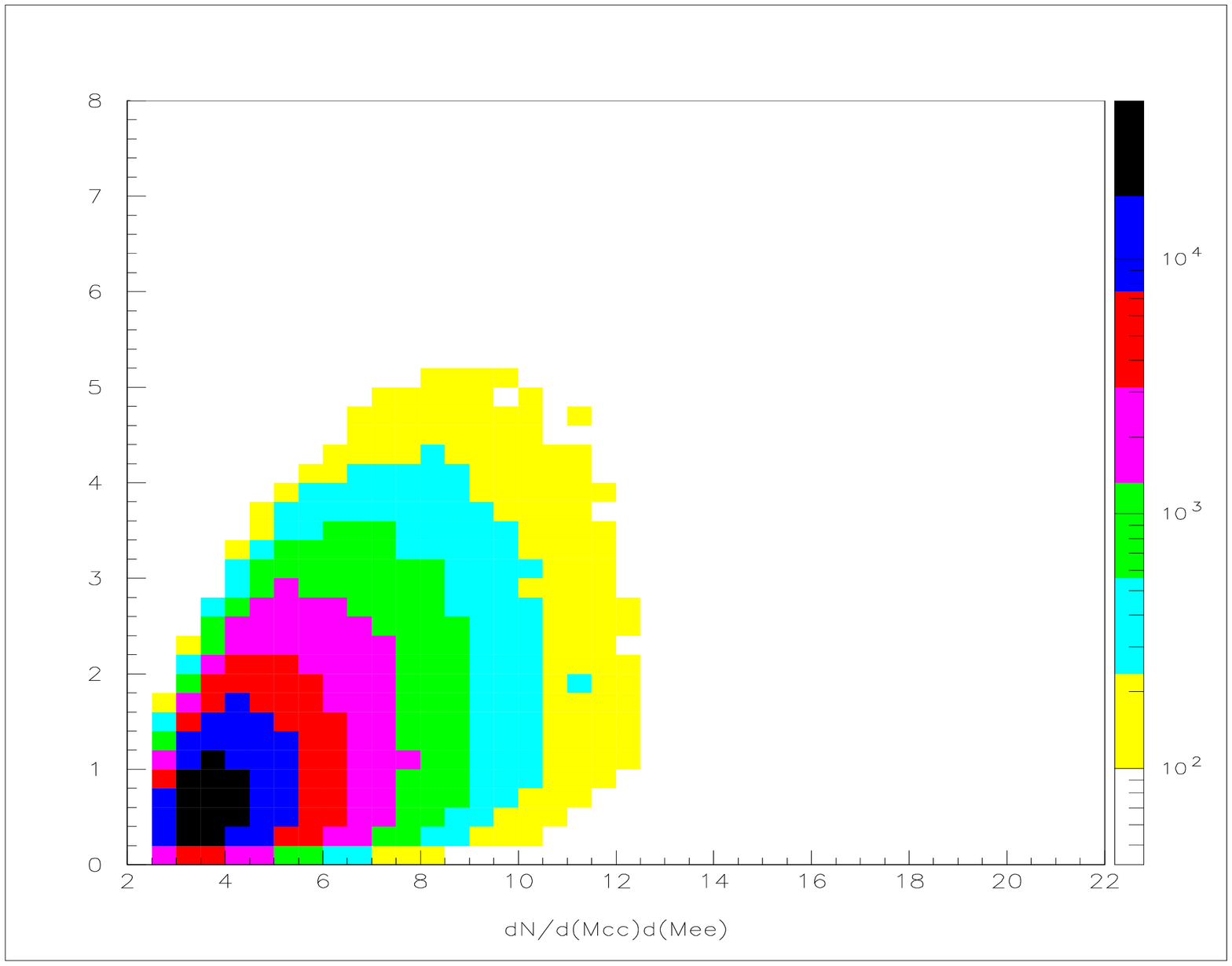,height=4.5in,width=5.in,angle=0} 
\caption{
Probability distribution of $M$, the dilepton mass from charm pair decay, as a
function of $M_{c\bar c}$ in $p+Au$ collision.
}
\label{fig:mass_relation}
\end{figure}

Therefore,
\begin{eqnarray}
\ln x_A \approx -y+ \ln [(\beta M + M_0) /\sqrt s] \;\; . 
\end{eqnarray}
The ratio of the dilepton $dN/dMdy$ spectra in $pA$ to scaled $pp$ for
different pair-masses is thus expected to scale approximately as
\begin{eqnarray}
&&R^{pA}_{e\mu} ( M,y=-\ln x_A + \ln [ (\beta M+M_0)/
\sqrt s ] \;) \nonumber \\ [2ex]
&&\equiv \frac {1} {\nu} \frac {dN^{pA}_{e\mu}} {dN^{pp}_{e\mu}}
\approx R_{g/A}(x_A,Q^2\!\approx\! (\beta M\!+\!M_0)^2\!/2\;) \;\; , 
\label{EQ:scaling}
\end{eqnarray}
where $\nu \equiv A {\sigma ^{pp}_{in}}/{\sigma ^{pA}_{in}} \sim A^ {1/3}$.  

In order to test gluon dominance and the accuracy of the above approximate
scaling, we compare in Figure~~\ref{fig:scaling} the gluon shadowing function
to the above dilepton ratio from the Monte Carlo calculation.  
The upper solid curve is Eskola's gluon shadowing \cite{shadowing} for $Q^2=10$
GeV$^2$, and the lower solid curve is shadowing from HIJING \cite{hijing}. 
The other six curves are ratios of dilepton $dN/dMdy$ spectra of shadowed
$p+Au$ over those from unshadowed $p+Au$, as given by eq.(\ref{EQ:scaling}) as
a function of the scaling variable $x_A$ for three different dilepton masses.
In Figure~\ref{fig:scaling} we first plot all the ratio curves in terms of
reversed pair-rapidity $-y$, then we shift the ratio curves at $M=1,2,4$ GeV to
the left by $3.79, 3.51$, and $3.10$ respectively.

Overall, with an approximate scaling the ratio of spectra from $p+A$ to
those from $pp$ reflects the nuclear shadowing function well. 
We conclude that this ratio can therefore serve to map out gluon shadowing in
nuclei.  Note that in the case of Eskola's shadowing, even the $Q^2$ dependence
of the shadowing function is visible through the rise of the ratio curves with
$M$ in the small $x$ region.

%% file: sec7-bg.tex
\hmysection{Original Backgrounds}
\label{sec-bg}
 
It is important to estimate the dilepton background to see if the proposed
signal is experimentally feasible.
In this chapter we estimate some dominant backgrounds without taking any
suppression or kinematical cuts.  
In chapter~\ref{sec-phenix} we calculate the signal and background dileptons
for the PHENIX detector \cite{cdr}, taking into account the detector geometry
and specific kinematical cuts. 

The electron background comes from Dalitz decay of $\pi^0$ mainly, and also 
from Dalitz decay of higher mass particles and resonances.  Also we have
electron background from photon conversion. 
For dielectron measurements, $\pi^0$ Dalitz decay background
($\pi^0~\rightarrow~e^+e^-\gamma$) is quite problematic. 
The correlated pair from Dalitz decay has an invariant mass smaller than the
pion mass, so ideally we can eliminate all these electrons and positrons by
invariant mass constraint.   
However, electrons with small momenta may be easily lost in the strong
magnetic field in the detector, in addition, we will not have $4\pi$ solid
angle coverage. 
Although we can reject many of these leptons by using techniques like
small-angle cuts for dielectron pairs, we still have background from $\pi^0$
Dalitz decay, and this is also a problem for $e\mu$ measurements.  
The main muon background comes from the random decays of charged pions and
kaons.  They can decay in the free space before the hadron absorber, and they
can also decay in the absorber or leak through. 
According to the momentum resolution in the PHENIX muon detector ($\sigma
(\theta_p)=0.4^\circ, \sigma (\phi_p)=1.6^\circ$) \cite{cdr}, it is very
difficult to differentiate these muons from muons coming directly from charm
decay.

Let us estimate the dielectron numbers in a central $p+Au$ collision at 200
GeV/A at RHIC. 
Given $N$ as the number of $\pi^0$s, $P$ as the branching ratio of $\pi^0
\rightarrow e^+ e^- \gamma$, the average number of combinatoric (uncorrelated)
dielectron pairs is then $N(N-1) P^2$.  
The HIJING Monte Carlo calculation \cite{hijing} gives about 38 $\pi^0$s, and
the branching ratio of $\pi^0$ Dalitz decay is about $1.2\%$.  Therefore the
average number of dielectrons originating from the $\pi^0$ background is 0.2.
Compared to the next largest source (dileptons from the $D/\bar D$ decay), 
this background is enormous.
Using HIJING's shadowing, we have 0.05 $D/\bar D$ pairs.
With the average branching ratio of $D \rightarrow e+X$ at about $12\%$, 
the dielectron yield from charm decay is about 0.0007, a factor of 300 below
the original background from $\pi^0$ Dalitz decay. 
For dimuon numbers in a central $p+Au$ collision at RHIC, HIJING
gives about 34 $\pi^+$s and $\pi^-$s each, and they almost all decay to muons.
Thus, the original dimuon background from charged pions is $1.1\times 10^3$,
which is enormous compared with the dimuon yield from charm decay (also about
0.0007). 
For opposite-sign $e\mu$ numbers, the original background originating from
$\pi^0$ and $\pi^+(\pi^-)$ is 31., compared to $e\mu$ from charm decay (about
0.0014).  

The above provides us with a rough idea of the opposite-sign lepton pair
backgrounds before we suppress them.  We must suppress them in order to observe
signals from charm decay and thus probe nuclear shadowing. 

\hmysection{Backgrounds Entering the Detector}
\label{sec-phenix}

In this chapter we study the backgrounds entering the detector by including
detector-related effects: the suppression factor for leptons, proposed PHENIX
detector geometry, and a specific kinematical cut for the events.  These are
based on the RHIC report \cite{cdr}. 
Then we compare the backgrounds to the lepton pairs signal from charm decay.  

The electrons from $\pi^0$ Dalitz decay could be suppressed by the small-angle
cut for dielectron pairs since they have an invariant mass less than the pion
mass. 
We suppose we could reject 90\% of the $\pi^0$ Dalitz electrons \cite{zhang}.
We also have electron background from photon conversions, which has a softer
spectrum. 
We suppose we have the same rejection rate of 90\% for the conversion
electrons.
Each of the two photons decayed from $\pi^0$ has a 0.5\% probability of
conversion to electron/positron pair, and the $\pi^0$ Dalitz decay branching
ratio is 1.2\%; as an upper limit we will simply double the electron background
from $\pi^0$ Dalitz decay to include the contribution from photon conversion.  
From HIJING program, in terms of number of electrons from $\pi^0$ Dalitz decay,
there is approximately 40\% more electrons from $K^-$ decay. 
We also assume a rejection rate of 90\% for these electrons based on the
estimates of free space decay probability and angular resolution for the
vertex.  Thus, the number of background electrons after rejection is 24\% of
the number of $\pi^0$ Dalitz electrons, which means we have a suppression
factor of about 4 for the original background electron number from $\pi^0$
Dalitz decay. 
We generate the background electrons from the electron spectrum calculated from
HIJING program (using proper weights for $\pi^0$ Dalitz and $K^-$ Dalitz
events), since electrons from $K^-$ have a harder spectrum, and thus a greater
probability of passing the kinematical cut. 
 
The background muons can come from decays in the free space before the hadron 
absorber, which is about 32 centimeters($L$) from the nominal vertex.  
For a charged pion with mass $m$, energy $E$, and proper decay time $\tau$, the
probability of the decay in free space is about $Lm/ \tau E$.  We would use
an energy cut of $2$ GeV for muons, therefore this probability is less than
0.3\%. 
There is also a background from particles that have leaked through the
absorber, which decay to muons in flight inside the muon arm.  From the RHIC 
report, for charged hadrons above $1$ GeV in the muon arm for a central $AuAu$
collision at $200$ GeV, the leakage number is 4.8, which is about 0.1\% of the
total number of charged  pions.  
Their decay to muons is also suppressed by the long proper lifetime $\tau$
and Lorentz $\gamma$ factor as above, and those which have not decayed will
mostly be differentiated by the muon identifier.  So the probability of finding
a muon above 1 GeV from the decays of leakage particles is about 0.003\%.
Considering the factor from the geometry and energy cuts explained below, the
probability from leakage before these cuts is also 0.3\%.  The charged hadrons
could also decay in the absorber during the shower.   
After all, we take 1\% of the muon number from charged pion decay as the
estimate of the muon background passing the hadron absorber and muon
identifier, which means we have a suppression factor of 100 for the original 
background muon number.  
We generate background muons from the muon spectrum calculated from HIJING
program, which is mainly from $\pi^+$, $\pi^-$ two-body decay, and, in very
small amounts, from decay of heavier particles such as $K^+$ and $K^-$.   

For the PHENIX detector geometry, the electron arm barrel covers electron
pseudo-rapidity range 
\begin{eqnarray}
-0.35 < \eta _e < 0.35 \;\; , 
\end{eqnarray}
and azimuthal angle range
\begin{eqnarray}
\phi_e \in \pm (22.5^\circ ,112.5^\circ ) \;\; . 
\end{eqnarray}
The muon arm endcap covers the polar angle range
$10^ \circ < \theta < 35^\circ$ and almost the full azimuthal angle.  Thus, we 
only require muon pseudo-rapidity range to be 
\begin{eqnarray}
1.15 < \eta _\mu < 2.44 \;\; . 
\end{eqnarray}
For the kinematical cut used to improve the signal-to-background ratio, we take
\begin{eqnarray}
E_e > 1 {\rm GeV} \;\; , \\
E_\mu > 2 {\rm GeV} \;\; . 
\end{eqnarray}
We also require the relative azimuthal angle
of the lepton pair to be $\phi _{l^+l^-} > 90^\circ $, and this improves the
ratio by a factor of 2.  
We find the following percentages for the numbers of leptons which pass the
above energy and pseudo-rapidity requirements: 1.1\% for electrons from charm
decay, 2.7\% for muons from charm decay, 0.02\% for background electrons, and
1.0\% for background muons.  The small percentage for background electrons is
due to the small mass of the particles which make Dalitz decays, and this
greatly lowers the electron background.  Recently, a second muon arm at the
opposite end is proposed to increase the muon rapidity coverage.

\begin{figure}[p]
\psfig{figure=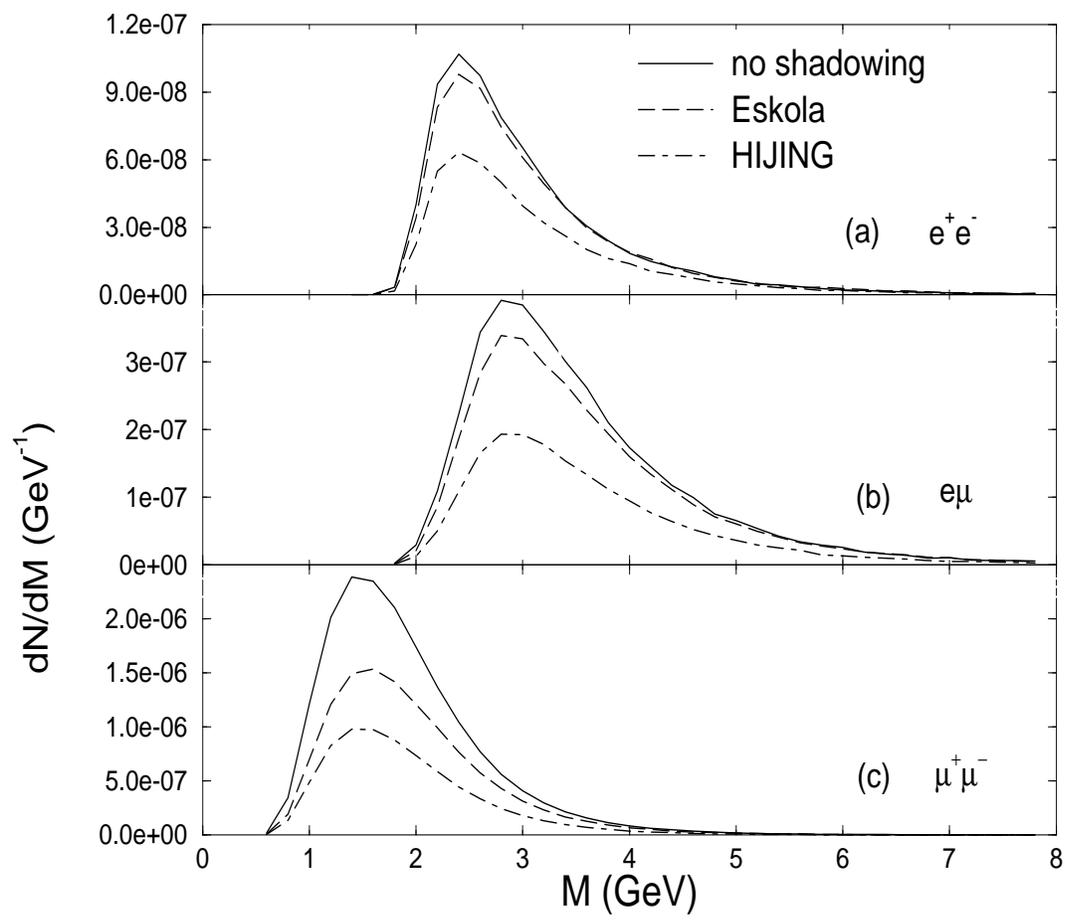,height=5.in,width=3.5in,angle=-90} 
\caption{
$dN/dM$ spectra for lepton pairs entering the PHENIX detector for different
shadowing cases. 
}
\label{fig:dndm_sg}
\end{figure}

\begin{figure}[p]
\psfig{figure=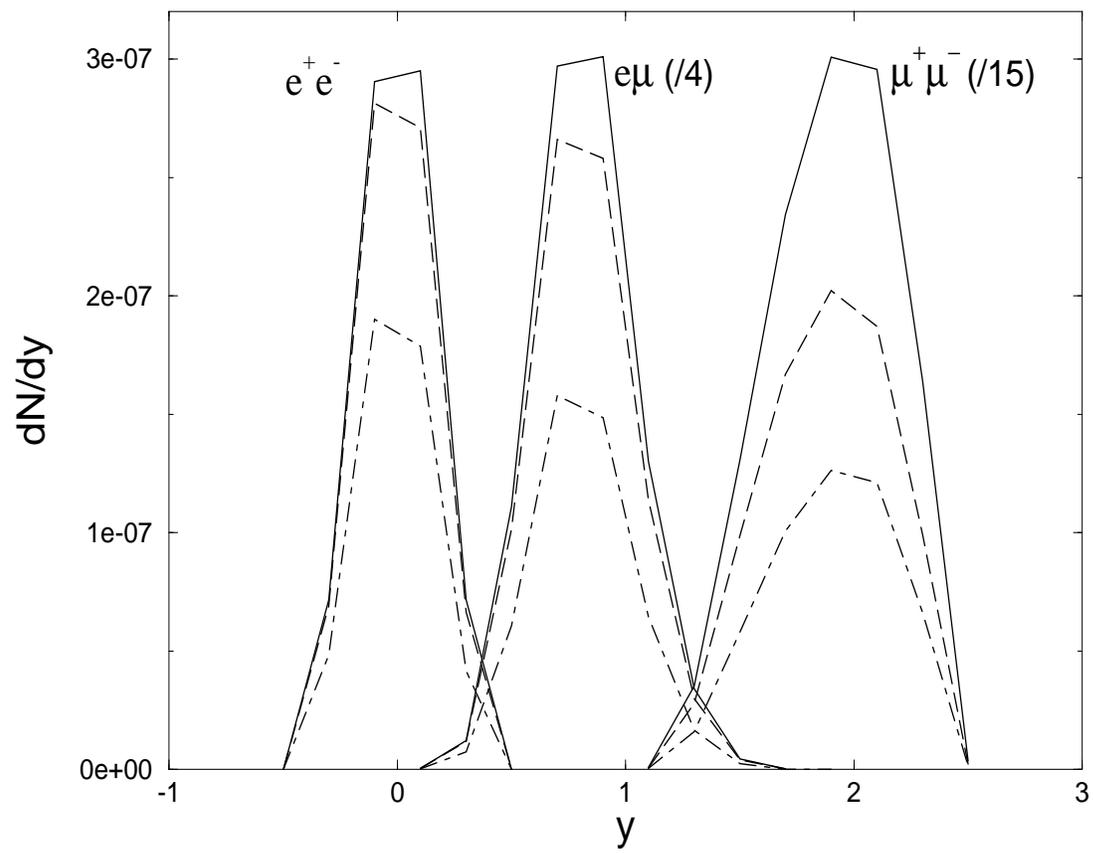,height=5.in,width=3.5in,angle=-90} 
\caption{
$dN/dy$ spectra for lepton pairs entering the detector for different shadowing
cases. 
}
\label{fig:dndy_sg}
\end{figure}

Considering the above suppression factors resulting from the geometry and the
kinematical cuts, we calculate the signal and the backgrounds entering the
detector.  
First, the $ee$, $e\mu$ and $\mu\mu$ signals entering the detector from charm
decay are calculated in three different shadowing cases.   
The $dN/dM$ plots of different shadowings are shown in
Figure~\ref{fig:dndm_sg},  and the $dN/dy$ plots are shown in
Figure~\ref{fig:dndy_sg}.  
In Figure~\ref{fig:dndm_sg}, results from non-shadowing case (solid), Eskola's
shadowing (dashed), and HIJING's shadowing (dot-dashed) are plotted.  Figure~
(a), (b), (c) are for opposite-sign $ee$, $e\mu$ and $\mu\mu$, respectively.
In Figure~\ref{fig:dndy_sg}, the left, middle and right set of three curves
correspond to opposite-sign $ee$, $e\mu$ and $\mu\mu$ signals. 
Note that the $e\mu$ and $\mu\mu$ yields are scaled down by a factor of 4
and 15, respectively.
Due to detector geometry and kinematical cuts, $ee$, $e\mu$ and $\mu\mu$
spectra cover pair-rapidity regions centered at about 0, 1 and 2, respectively 
(the second PHENIX muon arm can reach pair-rapidity region centered around $-1$
and $-2$).  
As shown in Figure~\ref{fig:ccbar}(b) for the signal without cuts, the
shadowing effect is stronger in larger pair-rapidity region.


\begin{figure}[p]
\psfig{figure=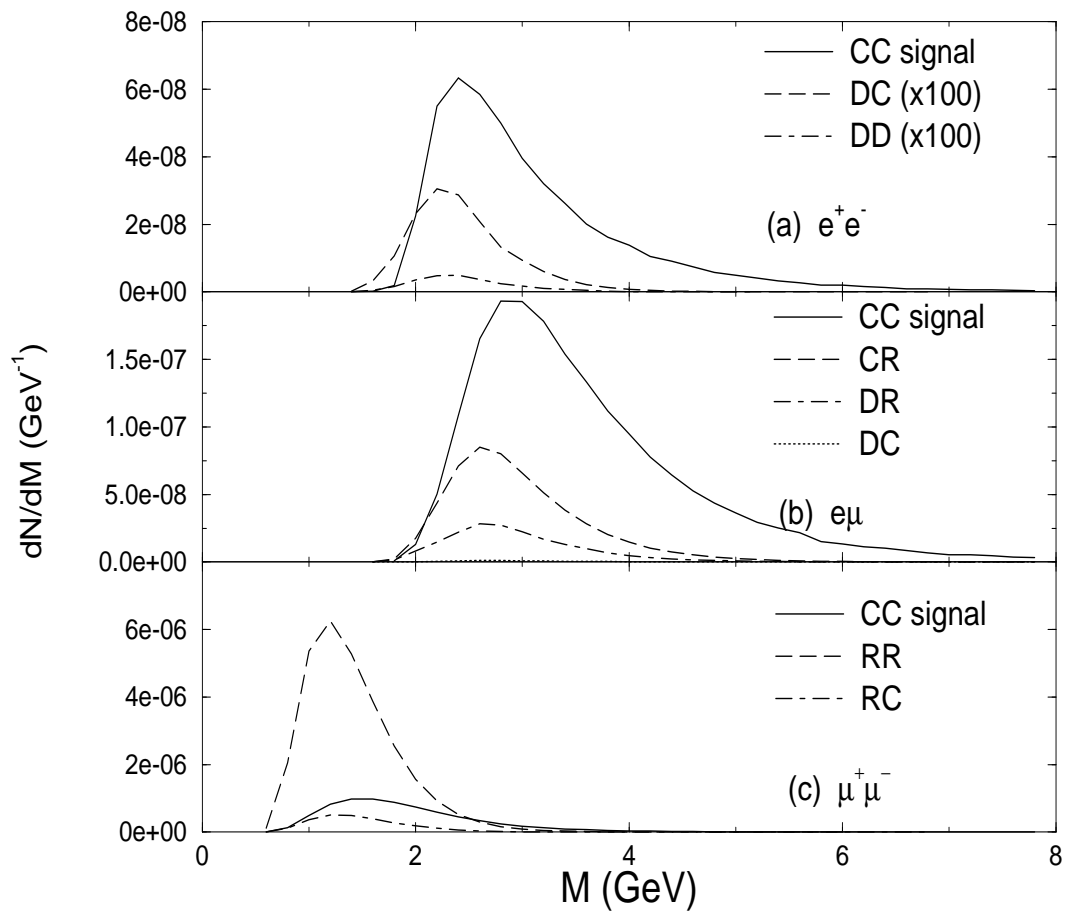,height=5.in,width=3.5in,angle=-90} 
\caption{
Charm signal and backgrounds entering the detector as a function of the lepton
pair-mass $M$.   
}
\label{fig:dndm_bg}
\end{figure}

\begin{figure}[p]
\psfig{figure=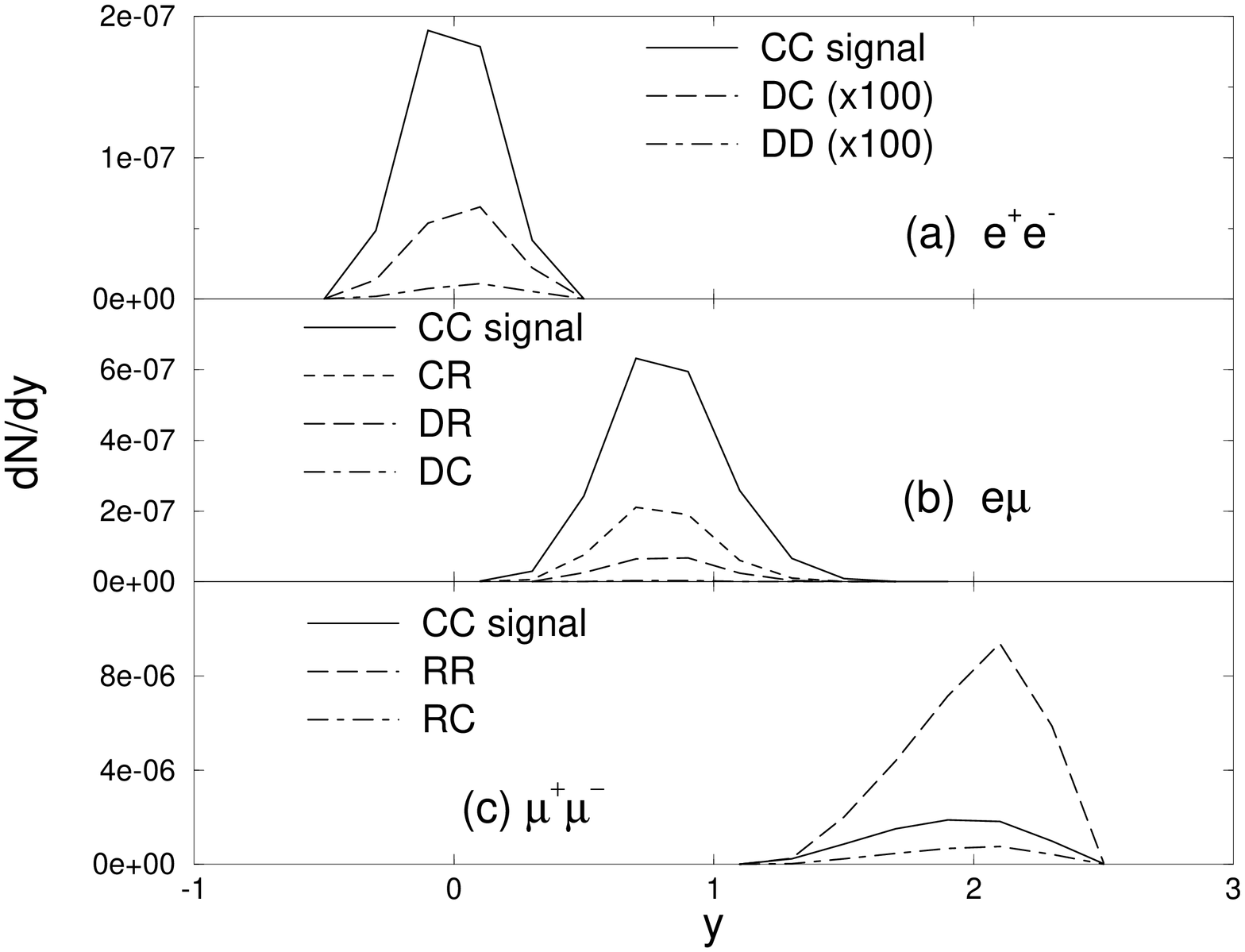,height=5.in,width=3.5in,angle=-90} 
\caption{
Charm signal and backgrounds entering the detector as a function of the lepton
pair-rapidity $y$. 
} 
\label{fig:dndy_bg}
\end{figure}

We then generate background electrons and muons based on HIJING calculation,
and calculate the $ee$, $e\mu$ and $\mu\mu$ backgrounds entering the detector.
We compare the backgrounds with the calculated signal from the HIJING shadowing
case.  The $dN/dM$ plots of signal and backgrounds are shown in
Figure~\ref{fig:dndm_bg}, and the $dN/dy$ plots are shown in
Figure~\ref{fig:dndy_bg}.  
In Figure~\ref{fig:dndm_bg}, the solid curves represent the signals from open
Charm decay --- labeled CC signal.  The label C refers to the lepton
originating from open Charm decay, label D refers to the electron originating 
from Dalitz and photon conversion, and label R refers to the muon originating
from Random decay of pions and kaons. In (a), the dielectron backgrounds are
represented by the dashed curve DC and the dot-dashed curve DD; and both are
scaled up by a factor of 100. 
In (b), the opposite-sign $e\mu$ backgrounds are represented by the dotted
curve CR, the dashed curve DR, and the dot-dashed curve DC.   
In (c), the dimuon backgrounds are represented by the dashed curve RR and the
dot-dashed curve RC.
The notations in Figure~\ref{fig:dndy_bg} are the same as in Figure~\ref{fig:dndm_bg}.  


We find that the signal-to-background ratio for $ee$ is very large, thus the
dielectron signal from open charm decay is the easiest to extract.  That ratio
falls to about 2.5 for $e\mu$, and about 1/4 for $\mu\mu$.  
Note that we have not included the backgrounds from Drell-Yan and thermal
production that will most change the above signal-to-background ratio for $ee$.

One could further suppress the lepton pair backgrounds by like-sign
subtraction, especially in the $\mu\mu$ channel.   
Since the particle production has a fairly good charge symmetry, the background
opposite-sign lepton pairs are almost the same as the like-sign pairs.  The
signals from open charm decay are almost all opposite-sign pairs, so we could
subtract like-sign pairs from opposite-sign pairs to suppress the backgrounds.
The efficiency of the like-sign subtraction depends on the charge and isospin
asymmetry of the collision.  With the help of like-sign subtraction, the $e\mu$
and $\mu\mu$ signal could be observed. 

\hmysection{Single Inclusive Lepton Spectrum}
\label{sec-single}

\begin{figure}[p]
\psfig{figure=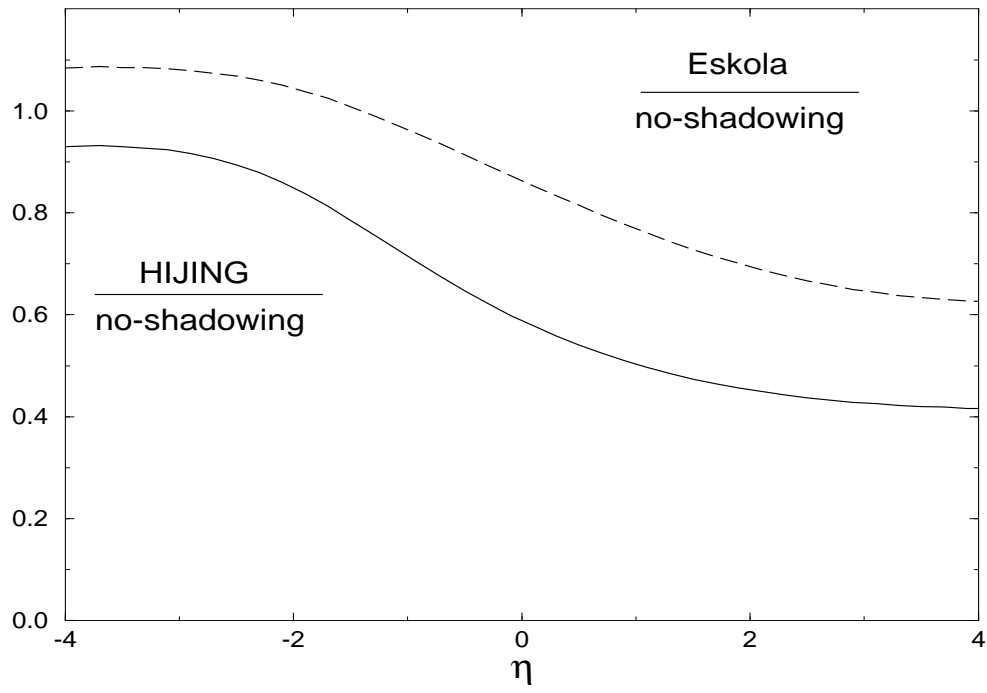,height=5.in,width=3.5in,angle=-90} 
\caption{
Ratios of single electron signals from the two shadowing cases over those
from the non-shadowing case as a function of the pseudo-rapidity $\eta$. 
}
\label{fig:single}
\end{figure}

The kinematics here are considerably less restricted, so the relation between
the kinematic variable $E, \eta$ for the single lepton and the gluon Bjorken
variable $x_A$ are more indirect.  Thus, the single inclusive lepton spectrum
reflects the nuclear shadowing effects less effectively.  For the signal from
open charm decay, the ratios of electron pseudo-rapidity ($\eta=1/2
\ln[(1+\cos \theta)/(1-\cos \theta)]$) distributions $dN/d\eta$ from the two
shadowing cases over those from the non-shadowing case are shown in
Figure~\ref{fig:single}.  The solid(dashed) curve represents HIJING(Eskola)'s
shadowing.   

Also the signal-to-background ratio in the detector is not as good as that for
the lepton pair case.  The estimated ratio is approximately 3 for single
electrons and 1/22 for single muons (with HIJING's shadowing), compared with
about 260 for $ee$, 2.5 for $e\mu$ and 1/4 for $\mu\mu$.  
Given the above small signal-to-background ratio, it is difficult to observe
single muons.  Since the single electron and muon spectrum cover pair-rapidity
ranges around 0 and $0, 2$ respectively, we can only observe mid-rapidity
single electrons.  While for the the three pair spectra, they cover rapidities
around 0, 1  and 2, and have reasonable signal-to-background ratio.  So we
conclude that although the single electron spectrum signal could be seen as an
easy initial measure of shadowing effects, the shadowing effects will be seen
much more clearly in the lepton pair spectrum than in the single spectrum.

\hmysection{Cronin Effect and Energy Loss}
\label{sec-cronin}

Cronin effect and energy loss will weaken the back-to-back type of correlation
between the two charm mesons in the pair, and thus weaken the scaling behavior
among the dilepton $dN/dMdy$ ratio curves showed in Figure~\ref{fig:scaling}.
However, in the following they are estimated to lead to distortions of up to
$10\%$ in $p+A$ collisions at RHIC energies.  

From studies of the nuclear dependence of transverse momenta of $J/\psi$
production, the typical increment, $\delta p_\perp^2(A)$, due to multiple
collisions is limited to $\sim 0.34$ GeV$^2$  even for  the heaviest nuclei
\cite{gavinmg}.  This momentum spread is shared between the $c$ and $\bar{c}$,
and distributed approximately as  
\begin{eqnarray}
g(\delta p_\perp)=e^{-\delta p_{\perp}^2/\Delta ^2}/\pi \Delta ^2 \;\; , 
\end{eqnarray}
where $\Delta^2 \sim 0.17$GeV$^2$.  
For energy loss, we assume  that the charm quark loses an energy $\delta E$ in
the lab frame, and the charm quark $p_\perp$ is reduced by $(1-\epsilon)$,
where  
\begin{eqnarray}
\epsilon=\delta E/[m_\perp \sinh (y+y_0)] \;\; . 
\end{eqnarray}
Combining these two effects, we may write for the charm quark that the final
transverse momentum is 
\begin{eqnarray}
\vec p_\perp=(\vec {p_{\perp}^{ini}} + \vec \delta p_\perp) (1-\epsilon) \;\; .
\end{eqnarray}
Assuming that the D-meson spectrum without these two effects can be expressed
as
\begin{eqnarray}
F(p_\perp) \equiv d^2 N/d m_{\perp}^2 \;\; , 
\end{eqnarray}
then it becomes
\begin{eqnarray}
F^{\prime}(p_\perp)=\int F({p_{\perp}^{ini}}) g(\delta p_\perp) d\vec \delta
p_\perp \;\; . 
\end{eqnarray}
with 
\begin{eqnarray}
F(p_\perp) \propto e^{-\alpha p_\perp}\;\; ; \alpha \simeq 1.3 \;{\rm GeV}^{-1}
\end{eqnarray}
from HIJING.  Expanding the convolution to lowest order,  the relative change
of the D-meson spectrum is given by 
\begin{eqnarray}
F^{\prime}(p_\perp)/F(p_\perp) \simeq 1-\alpha p_\perp \epsilon +\alpha^2
\Delta^2 /4 \;\; , 
\end{eqnarray}
This is also the relative change of the $c\bar c$ pair spectrum when $M_{c\bar
c}=2 m_\perp$ in case of $y_1=y_2$, $p_{\perp 1}=p_{\perp 2}$.  Take
\begin{eqnarray}
m_\perp \sim 3 {\rm GeV} \;\; , \\
\delta E=10 {\rm GeV} \;\; , \\
\cosh y_0 \simeq 100 \;\; , 
\end{eqnarray}
then the relative change in the pair spectrum is estimated to be 
\begin{eqnarray}
F^{\prime}(p_\perp)/F(p_\perp) \simeq 1\;-0.1( e^{-y}- 1) \;\; . 
\end{eqnarray}

\hmysection{Effect of Energy Loss in $A+A$ Collisions}
\label{sec-loss}

In heavy ion experiments, dileptons from a few GeV region were suggested to be
possible signals of the quark-gluon plasma; however, they were later claimed
 \cite{vogt_charmdilepton} to be dominated by the dileptons from open charm
decay.  Nevertheless, it has been suggested recently  \cite{loss} that energy
loss is an important factor which could soften greatly the dilepton spectrum
from open charm decays.   

In $A+A$ collisions, open charm must pass through the dense parton medium
formed after the initial collision; it will therefore lose a certain amount of
energy, typically $2$GeV/fm. 
Moreover, the two charm mesons in a pair tend to have opposite azimuthal
angles, so the probability that both of them can escape without substantial
energy loss is very small. 
Considering that we are interested in dileptons with energies of a few GeV,
the dilepton spectrum from correlated open charm (also open bottom) decay will
be greatly softened compared to the conclusion in \cite{vogt_charmdilepton},
and is possibly comparable to the signal from the quark-gluon plasma.

In order to get a better understanding of the magnitude of the softening of the
charm dilepton signal, we can take into account the following facts:  
in a finite size nucleus, partons lose energy in finite number of collisions,
not by the simple linear energy loss formula.   Therefore the energy loss has
more fluctuations than in the linear assumption case.  Secondly, the dilepton
$p_\perp$ spectrum depends crucially on the angular distribution, thus it is
important to calculate the open charm production to the next-to-leading order
in order to take into account the fact that the open charm pair are not always
back-to-back azimuthally.

In the mid-rapidity region in $p+A$ collisions, there are much fewer partons
than in the $A+A$ case, and the most energy loss occurs in the target rapidity
region where $y<<0$.  This energy loss effect was estimated in the above
chapter \ref{sec-cronin} and shown to be very small.      

\hmysection{Discussion and Summary}
\label{sec-sum}

We use Delta function as the charm quark fragmentation function to D meson.
The above analysis depends on the validity of this basic assumption regarding
the hard fragmentation of charm quarks.  Although this assumption is being
supported by the low energy pp data \cite{vogt_delta,e769}, it should be
checked explicitly via the single inclusive measurements of $D$ production at
RHIC.  

In this study, we included only the geometrical part of the experimental
acceptance in our calculations.  In order to get the detailed background
spectra for photon conversions, leakage muons, etc, the full detector
simulation should be carried out. 
Since the geometry is asymmetric for $p+Au$ collisions, energy loss will shift
the rapidity of the particles toward the target rapidity region, which is an
additional shift to the larger rapidity shift due to shadowing effects. 
Jet quenching could also change the rapidity distribution slightly if the
nuclear density is high, particularly for the high $p_\perp$ part.  
We must separate contributions due to other effects such as energy loss
in order to clearly observe the interesting shadowing curves shown in
Figure~\ref{fig:scaling}. 

We choose $p+Au$ since it has much smaller combinatoric background.  It also
demonstrates the shadowing effects more clearly in the $dN/dy(M)$ spectrum for
the lepton pairs as shown in Figure~\ref{fig:scaling}.

In summary, we calculated the lepton pair spectra from open charm decay in two
different shadowing scenarios.  By scaling the ratios for different mass ranges
according to eq.(\ref{EQ:scaling}), we showed that the dilepton rapidity
dependence of those ratios on $x_A$ reproduces well the underlying gluon
shadowing function defined in eq.(\ref{EQ:shad}).  Finally we showed that the
measurements required to extract the gluon shadowing are experimentally
feasible at RHIC.  The signal can be seen in $p+Au$ collisions, some even
before using like-sign subtraction.     

We conclude by emphasizing the importance of determining gluon shadowing in
$p+A$ to fix theoretically the initial conditions in $A+A$.  In $A+A$ the open
charm decay is regarded as an annoyingly large background that must be
subtracted to uncover the thermal signal.  In $p+A$ that background becomes the
signal needed to determine the incident gluon flux in $A+A$.  The continuum
charm dileptons in $p+A$ at RHIC are likely to provide a unique source of
information on the low $x_A$ nuclear gluon structure at least until HERA is
capable of accelerating heavy nuclei.

%% file: outlook_8.tex
\hmychapter{Outlook}
\label{sec-outlook}

In this thesis work, we have investigated the open charm production in nuclear
collisions and whether it can serve as a probe of the quark-gluon plasma phase.
We found that the additional open charm production from the pre-equilibrium
stage is negligible if we allow strong correlations between the space-time
coordinate and the momentum of minijet partons.  Contrary to an earlier result,
which claimed an enormous enhancement in the open charm production given the
formation of the quark-gluon plasma, we showed that open charm is not likely to
be a good probe of the plasma phase formed in relativistic heavy ion
collisions.   

We then proposed to use the open charm as a probe of the initial parton
structure function in nuclei, and thus turned our attention from probing the
quark-gluon plasma to probing the nuclear shadowing effect.
The dominance of the open charm in lepton-pair spectra with a few GeV energy
makes it possible to observe the open charm via high-mass lepton-pairs in $p+A$
collisions.  To make a better connection with ongoing experimental projects, we
then estimated the signal-to-background ratio for the proposed lepton-pair
measurements for the PHENIX detector at RHIC, and showed that these
measurements are feasible.   

In our study, we find that the pre-equilibrium charm production depends
strongly on the space-time and momentum correlation of minijet partons.  
Our model is a minimal modification of the Bjorken correlation by taking into
account the uncertainty principle.  The exact strength of the correlation of
minijet partons is yet an open question, and needs to be further investigated.

The assumption on the hard fragmentation function of charm quarks needs to be
tested in high energy hadroproductions.  
The latest Fermilab data support this assumption.
However, we need to further study the dependence of the hardness of
fragmentations on both the beam atomic number, $A$, and the beam energy.

Charmonium productions provide us with very interesting experimental data.  The
uncertainty from the fudge factors needs to be addressed in order to calculate
the absolute yield of charmonium states.

Another open problem that needs future investigation is the effect of the
energy loss.  It has been suggested that, due to the energy loss of charm
quarks, the large initial open charm production could be kinematically shifted
to the lower-mass region in dilepton spectra in $A+A$ collisions.  To evaluate
this possibility, parton cascade simulations will be needed.  Such models are
currently being formulated, and it will soon be possible to carry out detailed
numerical simulations.  The energy-loss effect could have interesting
consequences on the PHENIX search for direct thermal dileptons.  In addition,
a detailed analysis on dilepton spectra from open charm decays in $A+A$
collisions could provide information on the still-controversial energy-loss
mechanism in the QGP.

%% file: appendix_9.tex
\addcontentsline{toc}{chapter}{Appendix}
\hmysection*{Appendix}

\setcounter{equation}{0}
\def\theequation{A.\arabic{equation}}

In the appendix, we shall consider the impact-parameter-dependence of nuclear
shadowing effects. 

For the shadowing used in HIJING model \cite{hijing}, the density function is: 
\begin{eqnarray}
\Gamma_{a/A}(x, Q^2,\vec r)=T_A(\vec r)  f_{a/N}(x,Q^2) S_A(x, \vec r) \;\; , 
\end{eqnarray}
where $f_{a/N}(x,Q^2)$ is the parton structure function in a nucleon, $\vec r$
is the transverse vector of parton $a$ inside nucleus $A$, $T_A(\vec r)$ is the
transverse density function of the nucleus, and for a uniform-sphere nucleus
\begin{eqnarray}
T_A(\vec r)=(3A/2\pi R_A^2)\sqrt {1-r^2/R_A^2} \;\; . 
\end{eqnarray}
$S_A(x,\vec r)$ describes shadowing effects with the following
parameterization, which is independent of the parton species and the $Q^2$ 
scale:
\begin{eqnarray}
S_A(x,\vec r) &\equiv& \frac {f_{a/A}(x,\vec r)}{Af_{a/N}(x)} \nonumber \\[2ex]
&=&1+1.19\ln^{1/6}A[x^3-1.5(x_0+x_L)x^2+3x_0x_Lx] \nonumber \\[2ex]
&&-\left[\alpha_A(\vec r)-\frac{1.08(A^{1/3}-1)}{\ln(A+1)}\sqrt x \right]
e^{-x^2/x_0^2} \;\; , 
\end{eqnarray}
where 
\begin{eqnarray}
x_0&=&0.1,\;\; x_L=0.7 \;\; , \\
\alpha_A(\vec r)&=&0.1(A^{1/3}-1)\frac {4}{3}\sqrt {1-r^2/R_A^2} \;\; . 
\end{eqnarray}
For central $p+Au$ collisions, we should take the average over $r \leq r_0
(\sim 1.12 fm)$, where $r_0 \ll R_A (\sim A^{1/3} r_0)$.  Therefore
\begin{eqnarray}
S_A(x,\vec r) \simeq S_A(x,\vec 0) \;\; , 
\end{eqnarray}
and the integral over the transverse vector in Equation~(\ref{EQ:dncc}) gives
the following factor:  
\begin{eqnarray}
\int d^2r T_p(\vec r) T_{Au}(\vec r) S_{a/p}(x,\vec r) S_{Au}(x,\vec r) 
\simeq \frac {3A^{1/3}}{2\pi r_0^2} S_{Au}(x,\vec 0) \;\; . 
\label{EQ:A.3}
\end{eqnarray}
Note that there is no shadowing on the proton projectile in $p+Au$ collisions.

The other shadowing functions from Eskola \cite{shadowing} depend on $Q^2$ and
are also different for valence quarks, sea quarks and gluons.  The density
function is:
\begin{eqnarray}
\Gamma_{a/A}(x, Q^2,\vec r) &=& T_A(\vec r) f_{a/N}(x,Q^2) \nonumber \\
&& \times \left\{ 1-\frac {A T_A(\vec r)}{\int d^2 r T_A^2(\vec r)} \left[
1-S_{a/A}(x, Q^2) \right] \right\} \;\; . 
\end{eqnarray}
If we average over the impact parameter, we get 
\begin{eqnarray}
f_{a/A}(x,Q^2) & \equiv & \int d^2 r \Gamma_{a/A}(x, Q^2,\vec r) \nonumber \\
&=&A f_{a/N}(x,Q^2) S_{a/A}(x, Q^2) \;\; , 
\end{eqnarray}
which is precisely the definition of the impact-parameter-independent shadowing
function $S_{a/A}(x, Q^2)$.   
For central $p+Au$ collisions, the integral over the transverse vector yields
\begin{eqnarray}
&&\int d^2r T_p(\vec r) T_{Au}(\vec r) \left\{ 1\!-\!\frac {A T_{Au}(\vec
r)}{\int d^2 r T_{Au}^2(\vec r)} \left[ 1\!-\!S_{b/Au}(x,Q^2) \right] \right\} 
\nonumber \\[2ex]
&&\simeq \frac {3A^{1/3}}{2\pi r_0^2} \left[ \frac {4}{3} S_{b/Au}(x,
Q^2)\!-\!\frac {1}{3} \right] \;\; . 
\label{EQ:A.6}
\end{eqnarray}

From equations~(\ref{EQ:A.3}) and (\ref{EQ:A.6}), the overall shadowing factor
for central $p+Au$ collisions is  
\begin{eqnarray}
S_{Au}(x,\vec 0)
\end{eqnarray}
for HIJING's case, and 
\begin{eqnarray}
[4/3S_{b/Au}(x, Q^2)-1/3]
\end{eqnarray}
for Eskola's case. 